\documentclass[11pt,aps,a4paper,eqsecnum,amsmath,amssymb
              ,superscriptaddress,nofootinbib
              ]{revtex4-1}

\usepackage{graphicx}

\newcommand{\half}{\frac{1}{2}}

\begin{document}
\title{Phase Diagram of an Integrable Alternating $U_q[sl(2|1)]$ Superspin
  Chain }

\author{Holger Frahm}
\affiliation{%
Institut f\"ur Theoretische Physik, Leibniz Universit\"at Hannover,
Appelstra\ss{}e 2, 30167 Hannover, Germany}

\author{M\'arcio J. Martins}
\affiliation{%
Departamento de F\'isica, Universidade Federal de S\~ao Carlos,
C.P. 676, 13565-905 S\~ao Carlos (SP), Brazil}

\date{21. February 2012}

\begin{abstract}
  We construct a family of integrable vertex model based on the typical
  four-dimensional representations of the quantum group deformation of the Lie
  superalgebra $sl(2|1)$.  Upon alternation of such a representation with its
  dual this model gives rise to a mixed superspin Hamiltonian with local
  interactions depending on the representation parameter $\pm b$ and the
  deformation parameter ${\gamma}$.  As a subsector this model contains
  integrable vertex models with ordinary symmetries for twisted boundary
  conditions.  The thermodynamic limit and low energy properties of the mixed
  superspin chain are studied using a combination of analytical and numerical
  methods.  Based on these results we identify the phases realized in this
  system as a function of the parameters $b$ and $\gamma$.  The different
  phases are characterized by the operator content of the corresponding
  critical theory.  Only part of the spectrum of this effective theory can be
  understood in terms of the $U(1)$ symmetries related to the physical degrees
  of freedom corresponding to spin and charge.  The other modes lead to
  logarithmic finite-size corrections in the spectrum of the theory.
\end{abstract}

\maketitle

\section{Introduction}
Integrable quantum spin chains have long been a source of examples of systems
presenting rich critical behavior. The critical properties of quantum spin
chains based on simply laced Lie algebras are believed to be in the
universality class of Wess-Zumino-Witten models on the corresponding group
\cite{Affl86a,AfHa87}.  By way of contrast, the critical behavior of
integrable spin chains with both fermionic and bosonic degrees of freedom
involves a more subtle understanding. For instance, while superspin chains
invariant by the orthosympletic $osp(n|2m)$ symmetry are conformally invariant
\cite{MaNR98} those based on the $sl(n|m)$ superalgebra appears to be not even
relativistic \cite{Saleur00}.  In addition, the conformal properties of the
the spin chain with $osp(3|2)$ invariance was observed to be unusual thanks to
the presence of excitations with zero conformal weight \cite{MaNR98}. This
type of behaviour has subsequently been seen in the case of the a $sl(2|1)$
superspin chain that alternates fundamental and dual representations
\cite{EsFS05}. It was further noted that such degeneracy to the ground state
is dominated by a rather remarkable logarithmic finite-size corrections.  This
peculiar finite-size behaviour has been recently well elaborated from a
numerical perspective in the context of a staggered six-vertex model
\cite{IkJS08}. Here we shall argued that this model hides a twisted quantum
group algebra and its integrability can indeed be understood in terms of
bosonic and fermionic degrees of freedom.

The purpose of this paper is to introduce and investigate an integrable model
whose phase diagram presents all the above mentioned features at once.  It can
certainly be considered the prototype system mixing critical properties of
models based on ordinary Lie algebras and superalgebras.  The exactly solved
lattice system turns out to be the vertex model built up by alternating one of
the four dimensional representations of the $U_q[sl(2|1)]$ superalgebra with
its dual.  The fact that these representations can be labelled by a complex
number $b$ and $-b$ leads to a staggered vertex model with a free parameter.
Our paper is organized as follows: in the following section we present and
solve this alternating lattice model.  There we also expose certain underlying
spectral properties of this model which turn out to be essential to uncover
its many possible critical phases and discuss its relation to other systems
studied previously.  In Section~\ref{sec:hidden6v} it is argued that the model
hides integrable models based from ordinary symmetries with twisted boundary
conditions.  We then begin our analysis of the thermodynamics and critical
properties of the supersymmetric vertex model.  In Section~\ref{sec:mix-A} we
consider the case of staggering $0\le b<\half$.  Both in the antiferromagnetic
and the ferromagnetic regime the model is found to be in the universality
class of the $U_q[osp(2|2)]$ spin chain with central charge $c=-1$,
independent of the parameter $b$.  We continue with the discussion of the
critical behaviour of the ferromagnetic model for $b>\half$ in
Section~\ref{sec:phaseB}.  Here the low energy effective theory contains four
gapless modes, two of them can be identified with the physical spin and charge
degrees of freedom of the model.  Again, we find that the parameter $b$ is
irrelevant in the continuum limit.  Finally, we discuss the thermodynamical
limit of the antiferromagnetic model: in Section~\ref{sec:phaseC} the
behaviour for a special choice of the representation parameter satisfying a
self-duality condition is analyzed.  The operator content is found to be
similar to that of the ferromagnetic model.  The paper ends with a summary and
discussion of our results.

\section{The mixed $U_q[sl(2|1)]$ vertex model}
\label{sec:mixsl21q}
The four dimensional typical representation of the Lie superalgebra $sl(2|1)$
has the special feature that the eigenvalues of the azimuthal generator
associated to the fermionic degrees of freedom may be any complex number
\cite{ScNR77}. Therefore, its quantum group deformation $U_q[sl(2|1)]$ becomes
a rich two-parameter algebra and in turn provides a family of solutions of the
Yang-Baxter equation \cite{BGZD94,DGLZ95a,Maas95}.  We shall denote by $
R_{12}^{(b_1,b_2)}(\lambda)$ the respective $R$-matrix acting on the tensor
product $V_1^{(b_1)} \otimes V_2^{(b_2)}$ of two such different four
dimensional spaces. The upper labels emphasize the dependence of the
$R$-matrix on the extra complex variables $b_1$ and $b_2$ while $\lambda$ is
the usual spectral parameter. The general structure of the $R$-matrix in the
symmetrical grading $FBBF$ is given by \cite{Grun00},
\begin{equation}
  \label{Rmat}
{\cal R}_{12}^{(b_1,b_2)}(\lambda) = 
\sum_{j=1}^{4} {a}_{j} 
e_{j,j}^{(1)} \otimes e_{j,j}^{(2)}
+ \sum_{\stackrel{j,k=1}{j \neq k, 5-k}}^{4} 
b_{jk}e_{j,j}^{(1)} \otimes e_{k,k}^{(2)}  
+\sum_{\stackrel{j,k=1}{j \neq k,5-k}}^{4} 
c_{jk}e_{j,k}^{(1)} 
\otimes e_{k,j}^{(2)}  
+\sum_{{j,k=1}}^{4} 
d_{jk}e_{5-j,k}^{(1)} 
\otimes e_{j,5-k}^{(2)}  
\end{equation}
where $e_{j,k}^{(a)} \in \mathrm{End}(\mathbb{C}_{a}^{4})$ are the standard
$4\times 4$ Weyl matrices. 
%
The explicit expressions for the Boltzmann weights $a_{j}$, $b_{jk}$, $c_{jk}$
and $d_{jk}$ are presented in Appendix~\ref{app-weights}.

The above $R$-matrix satisfies the Yang-Baxter relation for any three general
distinct spaces $V_1^{(b_1)}$, $V_2^{(b_2)}$ and $V_3^{(b_3)}$, namely
\begin{equation}
  R_{12}^{(b_1,b_2)}(\lambda)
  R_{13}^{(b_1,b_3)}(\lambda+\mu)
  R_{23}^{(b_2,b_3)}(\mu)=
  R_{23}^{(b_2,b_3)}(\mu)
  R_{13}^{(b_1,b_3)}(\lambda+\mu)
  R_{12}^{(b_1,b_2)}(\lambda),
  \label{YBE}
\end{equation}

Considering the weights expressions (\ref{weiinit}) we see that
the extra variables $b_i$ enter the Yang-Baxter equation (\ref{YBE}) in a
similar manner as the spectral parameter however in a non-additive form.  An
immediate consequence is that in general there are two different types of Lax
operators obeying the Yang-Baxter algebra with the same $R$-matrix. Within the
quantum inverse scattering framework one can explore this freedom and
construct mixed integrable vertex model. For example, we can combine the
Boltzmann weights $R_{12}^{(b,b_1)}(\lambda)$ and $R_{12}^{(b,b_2)}(\lambda)$
in such way that the quantum space states alternate among the $b_1$ and $b_2$
representations while the auxiliary space is remains fixed at $b$. A schematic
representation of this system on a $2L \times 2L $ square lattice is depicted
in Figure~\ref{TFIX}.
\setlength{\unitlength}{3500sp}
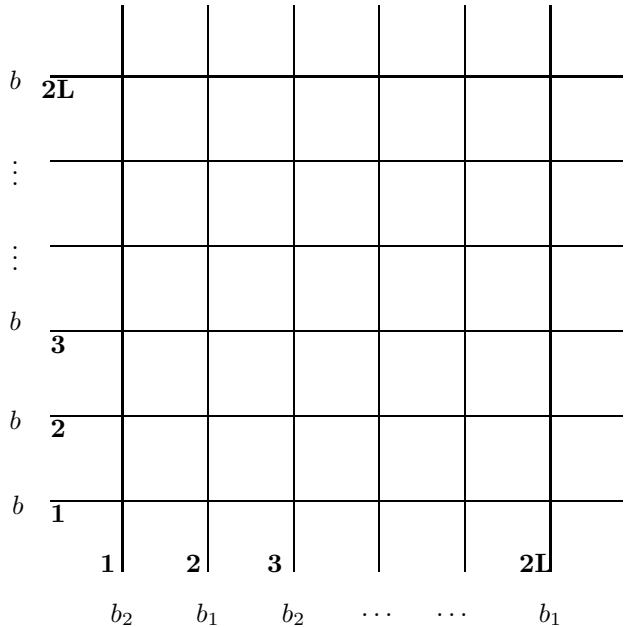
\begin{figure}[ht]
\begin{center}
\begin{picture}(6537,3912)(2000,-6361)
\put(3800,-2350){\makebox(0,0){\fontsize{10}{12}\selectfont }}
\put(4400,-2350){\makebox(0,0){\fontsize{10}{12}\selectfont }}
\put(5000,-2350){\makebox(0,0){\fontsize{10}{12}\selectfont }}
\put(5600,-2350){\makebox(0,0){\fontsize{10}{12}\selectfont }}
\put(6200,-2350){\makebox(0,0){\fontsize{10}{12}\selectfont }}
\put(6800,-2350){\makebox(0,0){\fontsize{10}{12}\selectfont }}
\put(3800,-6200){\makebox(0,0){\fontsize{10}{12}\selectfont \textbf{1}}}
\put(4400,-6200){\makebox(0,0){\fontsize{10}{12}\selectfont \textbf{2}}}
\put(5000,-6200){\makebox(0,0){\fontsize{10}{12}\selectfont \textbf{3} }}
\put(5600,-6200){\makebox(0,0){\fontsize{10}{12}\selectfont }}
\put(6200,-6200){\makebox(0,0){\fontsize{10}{12}\selectfont }}
\put(6800,-6200){\makebox(0,0){\fontsize{10}{12}\selectfont \textbf{2L}}}
\put(3450,-5850){\makebox(0,0){\fontsize{10}{12}\selectfont \textbf{1}}}
\put(3450,-5250){\makebox(0,0){\fontsize{10}{12}\selectfont \textbf{2}}}
\put(3450,-4650){\makebox(0,0){\fontsize{10}{12}\selectfont \textbf{3}}}
\put(3450,-4050){\makebox(0,0){\fontsize{10}{12}\selectfont }}
\put(3450,-3450){\makebox(0,0){\fontsize{10}{12}\selectfont }}
\put(3450,-2850){\makebox(0,0){\fontsize{10}{12}\selectfont \textbf{2L}}}
\put(7300,-5850){\makebox(0,0){\fontsize{10}{12}\selectfont }}
\put(7300,-5250){\makebox(0,0){\fontsize{10}{12}\selectfont }}
\put(7300,-4650){\makebox(0,0){\fontsize{10}{12}\selectfont }}
\put(7300,-4050){\makebox(0,0){\fontsize{10}{12}\selectfont }}
\put(7300,-3450){\makebox(0,0){\fontsize{10}{12}\selectfont }}
\put(7300,-2850){\makebox(0,0){\fontsize{10}{12}\selectfont }}
\put(3401,-2761){\line( 1, 0){4000}}
\put(3401,-3361){\line( 1, 0){4000}}
\put(3401,-3961){\line( 1, 0){4000}}
\put(3401,-4561){\line( 1, 0){4000}}
\put(3401,-5161){\line( 1, 0){4000}}
\put(3401,-5761){\line( 1, 0){4000}}
\put(3901,-6261){\line( 0, 1){4000}}
\put(4501,-6261){\line( 0, 1){4000}}
\put(5101,-6261){\line( 0, 1){4000}}
\put(6301,-6261){\line( 0, 1){4000}}
\put(6901,-6261){\line( 0, 1){4000}}
\put(5701,-6261){\line( 0, 1){4000}}
\put(3170,-5786){\makebox(0,0){$b$}}
\put(3150,-5186){\makebox(0,0){$b$}}
\put(3150,-4486){\makebox(0,0){$b$}}
\put(3150,-3986){\makebox(0,0){$\vdots$}}
\put(3150,-3386){\makebox(0,0){$\vdots$}}
\put(3150,-2786){\makebox(0,0){$b$}}
\put(3901,-6561){\makebox(0,0){$b_2$}}
\put(4501,-6561){\makebox(0,0){$b_1$}}
\put(5101,-6561){\makebox(0,0){$b_2$}}
\put(5701,-6561){\makebox(0,0){$\dots$}}
\put(6226,-6561){\makebox(0,0){$\dots$}}
\put(6901,-6561){\makebox(0,0){$b_1$}}
\end{picture} \par
\end{center}
\caption{The mixed vertex with fixed horizontal
$b$ representation.\label{TFIX}}
\end{figure}

The row-to-row transfer matrix  associated to the mixed vertex
model (\ref{TFIX}) can be formally written as the supertrace
\cite{Kulish85} over the auxiliary space ${\cal{A}} \sim V^{(b)} $ of the
following ordered product of operators,
\begin{equation}
\label{TRAMIX}
T^{(b,\{b_1,b_2\})}(\lambda) = 
\mathrm{Str}_{\mathcal{A}}\left[
\mathcal{R}^{b,b_1}_{\mathcal{A},2L}(\lambda)
\mathcal{R}^{b,b_2}_{\mathcal{A},2L-1}(\lambda)
\mathcal{R}^{b,b_1}_{\mathcal{A},2L-2}(\lambda) \cdots
\mathcal{R}^{b,b_2}_{\mathcal{A},1}(\lambda)
\right] \,.
\end{equation}

At this point we recall that due to the Yang-Baxter equations (\ref{YBE}) the
above transfer matrix commutes not only for arbitrary spectral parameters but
also for any values of variable $b$ labeling the horizontal space of
states. More precisely the transfer matrix (\ref{TRAMIX}) satisfies the
relation, namely
\begin{equation}
  \label{COMU}
  [T^{(b,\{b_1,b_2\})}(\lambda), 
  T^{(\bar{b},\{b_1,b_2\})}(\mu)]=0,\quad \forall b, \bar{b}
  \mathrm{~and~} \lambda, \mu. 
\end{equation}

The diagonalization of the transfer matrix (\ref{TRAMIX}) can be carried out
within the nested Bethe ansatz framework since $T^{(b,\{b_1,b_2\})}(\lambda)$
commutes with two distinct $U(1)$ symmetries.  A possible solution is to apply
the fusion procedure to obtain a recurrence relation for the eigenvalues of
the transfer matrix and combine it with some reasonable analyticity
assumptions to fix the corresponding Bethe equations
\cite{PfFr96,PfFr97,Grun00}. Yet another method is to explore directly the
commutation relations among the monodromy matrix elements in the four
dimensional representation. For instance, such constructive approach has been
applied to solve the isotropic limit $q=1$ of the plain transfer matrix
(\ref{TRAMIX}) where $b_1=b_2=b$ \cite{RaMa96}. Because these methods have
been already fully discussed in the literature we shall here present only the
final results for the eigenvalues of $T^{(b,\{b_1,b_2\})}(\lambda)$.
Considering the expressions for the Boltzmann weights (\ref{weiinit}) in any
of the aforementioned frameworks one can indeed compute the eigenvalues of the
transfer matrix (\ref{TRAMIX}).  We find that they can be conveniently written
in terms of the product of two terms,
\begin{equation}
\begin{aligned}
\label{GAMAGE}
 & \Lambda^{(b,\{b_1,b_2\})}(\lambda) \\
 & = \left[-F(\lambda, b, \{\lambda_\ell^{(1)}\})
          + \left(\frac{\sinh(\lambda-i\gamma (b-b_1)) \,
              \sinh(\lambda-i\gamma (b-b_2))}{ 
                  \sinh(\lambda-i\gamma (b+b_1+1))\, \sinh(\lambda-i\gamma
                  (b+b_2+1))} \right)^L
            G(\lambda, b, \{\lambda_\ell^{(1)}\}) \right] \\
  &\quad\times 
  \left[ F(\lambda, -b, \{\lambda_\ell^{(2)}\})
          - \left(\frac{\sinh(\lambda +i\gamma(b-b_1)) \, \sinh(\lambda+i\gamma ( b-b_2))}{
                  \sinh(\lambda+i\gamma(b+b_{1} - 1)) \, \sinh(\lambda+i\gamma (b+b_2-1))}
            \right)^L
            G(\lambda, -b, \{\lambda_\ell^{(2)}\}) \right] \,.
\end{aligned}
\end{equation}
The auxiliary functions 
$F(\lambda, b, \{\lambda_\ell^{(a)}\})$  and 
$G(\lambda, b, \{\lambda_\ell^{(a)}\})$  are given by,
\begin{equation}
\label{FUNGE}
\begin{aligned}
  F(\lambda, b, \{\lambda_\ell^{(a)}\}) &= \prod_{\ell=1}^{N_a}
          \frac{\sinh(\lambda_\ell^{(a)}-\lambda+i\gamma(b-1/2))}{
                \sinh(\lambda_\ell^{(a)}-\lambda-i\gamma(b-1/2))}\,,\\
  G(\lambda, b, \{\lambda_\ell^{(a)}\}) &= \prod_{\ell=1}^{N_a}
          \frac{\sinh(\lambda-\lambda_\ell^{(a)}-i\gamma(b+3/2))}{
                \sinh(\lambda-\lambda_\ell^{(a)}+i\gamma(b-1/2))}\,.
\end{aligned}
\end{equation}
while the set of rapidities $\lambda_j^{(1)}$ and $\lambda_j^{(2)}$ are
required to fulfill the following nested Bethe equations,
\begin{equation}
\label{BAGE}
\begin{aligned}
 \left\{
  \frac{\sinh(\lambda_j^{(1)} + i\gamma(b_1-\half))}{
        \sinh(\lambda_j^{(1)} - i\gamma(b_1-\half))}\,
  \frac{\sinh(\lambda_j^{(1)} + i\gamma(b_2-\half))}{
        \sinh(\lambda_j^{(1)} - i\gamma(b_2-\half))}
 \right\}^L &= 
  \prod_{k=1}^{N_2} \frac{\sinh(\lambda_j^{(1)}-\lambda_k^{(2)} -i\gamma)}{
                     \sinh(\lambda_j^{(1)}-\lambda_k^{(2)} +i\gamma)} \,,
  \quad j=1,\ldots,N_1\,,\\
 \left\{
  \frac{\sinh(\lambda_j^{(2)} - i\gamma(b_1+\half))}{
        \sinh(\lambda_j^{(2)} + i\gamma(b_1+\half))}\,
  \frac{\sinh(\lambda_j^{(2)} - i\gamma(b_2+\half))}{
        \sinh(\lambda_j^{(2)} + i\gamma(b_2+\half))}
 \right\}^L &= 
  \prod_{k=1}^{N_1} \frac{\sinh(\lambda_j^{(2)}-\lambda_k^{(1)} -i\gamma)}{
                     \sinh(\lambda_j^{(2)}-\lambda_k^{(1)} +i\gamma)} \,,
  \quad j=1,\ldots,N_2\,.
\end{aligned}
\end{equation}
We stress that the numbers $N_a$ of Bethe roots are directly related to the
two $U(1)$ symmetries of the model.  They determine the eigenvalues of the
$U_q[sl(2|1)]$ charge $B$ and spin-projection $S_3$ for the corresponding
Bethe state which are $B=(N_1-N_2)/2$ and $S_3=L-(N_1+N_2)/2$.  We also
observe that the Bethe equations depend only on the alternating
representations $b_1$ and $b_2$ of the quantum spaces.

In addition, we can use the commuting property (\ref{COMU}) to built up mixed
vertex models with alternation in both horizontal and vertical spaces of
states. Of particular interest are those whose transfer matrix commutes with
one-dimensional spin Hamiltonians possessing a finite number of local
interactions for any size $L$. The simplest such case occurs when we alternate
between a given representation $b$ and its dual counterpart $-b$. In
Figure~\ref{TMOD} we exhibit the graphical representation of this type of
double mixed vertex model. The corresponding transfer matrix is given in terms
of the following product of commuting operators,
\begin{equation}
\label{transmix}
T^{(mix)}(\lambda) = T^{(b,\{b,-b\})}(\lambda) T^{(-b,\{b,-b\})}(\lambda) \,.
\end{equation}
\setlength{\unitlength}{3500sp}
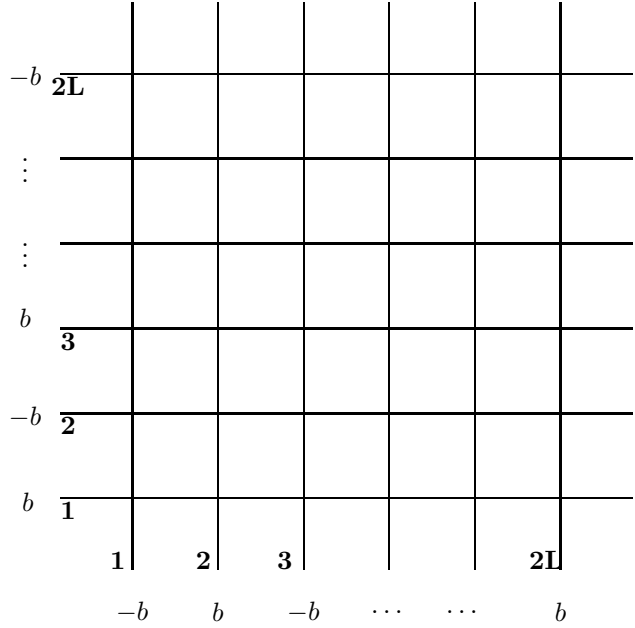
\begin{figure}[ht]
\begin{center}
\begin{picture}(6537,3912)(2000,-6361)
\put(3800,-2350){\makebox(0,0){\fontsize{10}{12}\selectfont }}
\put(4400,-2350){\makebox(0,0){\fontsize{10}{12}\selectfont }}
\put(5000,-2350){\makebox(0,0){\fontsize{10}{12}\selectfont }}
\put(5600,-2350){\makebox(0,0){\fontsize{10}{12}\selectfont }}
\put(6200,-2350){\makebox(0,0){\fontsize{10}{12}\selectfont }}
\put(6800,-2350){\makebox(0,0){\fontsize{10}{12}\selectfont }}
\put(3800,-6200){\makebox(0,0){\fontsize{10}{12}\selectfont \textbf{1}}}
\put(4400,-6200){\makebox(0,0){\fontsize{10}{12}\selectfont \textbf{2}}}
\put(5000,-6200){\makebox(0,0){\fontsize{10}{12}\selectfont \textbf{3} }}
\put(5600,-6200){\makebox(0,0){\fontsize{10}{12}\selectfont }}
\put(6200,-6200){\makebox(0,0){\fontsize{10}{12}\selectfont }}
\put(6800,-6200){\makebox(0,0){\fontsize{10}{12}\selectfont \textbf{2L}}}
\put(3450,-5850){\makebox(0,0){\fontsize{10}{12}\selectfont \textbf{1}}}
\put(3450,-5250){\makebox(0,0){\fontsize{10}{12}\selectfont \textbf{2}}}
\put(3450,-4650){\makebox(0,0){\fontsize{10}{12}\selectfont \textbf{3}}}
\put(3450,-4050){\makebox(0,0){\fontsize{10}{12}\selectfont }}
\put(3450,-3450){\makebox(0,0){\fontsize{10}{12}\selectfont }}
\put(3450,-2850){\makebox(0,0){\fontsize{10}{12}\selectfont \textbf{2L}}}
\put(7300,-5850){\makebox(0,0){\fontsize{10}{12}\selectfont }}
\put(7300,-5250){\makebox(0,0){\fontsize{10}{12}\selectfont }}
\put(7300,-4650){\makebox(0,0){\fontsize{10}{12}\selectfont }}
\put(7300,-4050){\makebox(0,0){\fontsize{10}{12}\selectfont }}
\put(7300,-3450){\makebox(0,0){\fontsize{10}{12}\selectfont }}
\put(7300,-2850){\makebox(0,0){\fontsize{10}{12}\selectfont }}
\put(3401,-2761){\line( 1, 0){4000}}
\put(3401,-3361){\line( 1, 0){4000}}
\put(3401,-3961){\line( 1, 0){4000}}
\put(3401,-4561){\line( 1, 0){4000}}
\put(3401,-5161){\line( 1, 0){4000}}
\put(3401,-5761){\line( 1, 0){4000}}
\put(3901,-6261){\line( 0, 1){4000}}
\put(4501,-6261){\line( 0, 1){4000}}
\put(5101,-6261){\line( 0, 1){4000}}
\put(6301,-6261){\line( 0, 1){4000}}
\put(6901,-6261){\line( 0, 1){4000}}
\put(5701,-6261){\line( 0, 1){4000}}
\put(3170,-5786){\makebox(0,0){$b$}}
\put(3150,-5186){\makebox(0,0){$-b$}}
\put(3150,-4486){\makebox(0,0){$b$}}
\put(3150,-3986){\makebox(0,0){$\vdots$}}
\put(3150,-3386){\makebox(0,0){$\vdots$}}
\put(3150,-2786){\makebox(0,0){$-b$}}
\put(3901,-6561){\makebox(0,0){$-b$}}
\put(4501,-6561){\makebox(0,0){$b$}}
\put(5101,-6561){\makebox(0,0){$-b$}}
\put(5701,-6561){\makebox(0,0){$\dots$}}
\put(6226,-6561){\makebox(0,0){$\dots$}}
\put(6901,-6561){\makebox(0,0){$b$}}
\end{picture} \par
\end{center}
\caption{The mixed vertex with alternation $\pm b$ 
in both horizontal and vertical
spaces.\label{TMOD}}
\end{figure}

Because we are dealing with a family of commuting operators the eigenvalues of
the transfer matrix $T^{(mix)}(\lambda)$ is just the product of the individual
eigenvalues,
\begin{equation}
  \label{GAMAMIX}
  \Lambda^{(mix)}(\lambda)= 
  \Lambda^{(b,\{b,-b\})}(\lambda) 
  \Lambda^{(-b,\{b,-b\})}(\lambda) 
\end{equation}
where $\Lambda^{(\pm b,\{b,-b\})}(\lambda)$ are easily computed from
Eqs.(\ref{GAMAGE}) and (\ref{FUNGE}).

By the same token, the corresponding Bethe equations are obtained from
Eqs.(\ref{BAGE}) by substituting $b_1=b$ and $b_2=-b$. We note that for this
choice of alternation the Bethe equations (\ref{BAGE}) turn out to be
invariant under the change of the sign of the parameter $b$. Since will be
referring to these equations in many distinct circumstances on the paper we
for sake of clarity shall quote them explicitly
\begin{equation}
\label{baebq}
\begin{aligned}
 \left\{
  \frac{\sinh(\lambda_j^{(1)} + i\gamma(b-\half))}{
        \sinh(\lambda_j^{(1)} - i\gamma(b-\half))}\,
  \frac{\sinh(\lambda_j^{(1)} - i\gamma(b+\half))}{
        \sinh(\lambda_j^{(1)} + i\gamma(b+\half))}
 \right\}^L &= 
  \prod_{k=1}^{N_2} \frac{\sinh(\lambda_j^{(1)}-\lambda_k^{(2)} -i\gamma)}{
                     \sinh(\lambda_j^{(1)}-\lambda_k^{(2)} +i\gamma)} \,,
  \quad j=1,\ldots,N_1\,,\\
 \left\{
  \frac{\sinh(\lambda_j^{(2)} + i\gamma(b-\half))}{
        \sinh(\lambda_j^{(2)} - i\gamma(b-\half))}\,
  \frac{\sinh(\lambda_j^{(2)} - i\gamma(b+\half))}{
        \sinh(\lambda_j^{(2)} + i\gamma(b+\half))}
 \right\}^L &= 
  \prod_{k=1}^{N_1} \frac{\sinh(\lambda_j^{(2)}-\lambda_k^{(1)} -i\gamma)}{
                     \sinh(\lambda_j^{(2)}-\lambda_k^{(1)} +i\gamma)} \,,
  \quad j=1,\ldots,N_2\,.
\end{aligned}
\end{equation}

We now observe that for $\lambda=0$ both $\mathcal{R}_{12}^{b,b}(\lambda)$ and
$\mathcal{R}_{12}^{-b,-b}(\lambda)$ turn out to be the graded permutator and
$T^{(mix)}(\lambda=0)$ become proportional to the two-sites translation
operator.  The respective local spin chain Hamiltonian is then constructed by
taking the logarithmic derivative of the transfer matrix (\ref{transmix}) at
$\lambda=0$, namely
\begin{equation}
  \label{Hmix}
  \mathcal{H}^{(mix)} = \left.i \frac{\partial}{\partial \lambda}
    \ln T^{(mix)}(\lambda) \right|_{\lambda=0}\,.
\end{equation}

The operator (\ref{Hmix}) defines an integrable superspin Hamiltonian with
local two- and three-spin interactions.  Considering Eqs.~(\ref{GAMAGE}),
(\ref{GAMAMIX}) we find that the eigenspectrum of ${H}^{(mix)}$ in a given
sector $N_1$ and $N_2$ is parameterized by the following expression
\begin{equation}
\label{Emix}
 E^{(mix)}_{N_1,N_2}(b,\gamma)= \sum_{a=1,2}\sum_{\ell=1}^{N_a}\left\{
 \frac{2\sin(\gamma(2b+1))}{\cos(\gamma(2b+1))-\cosh2\lambda_\ell^{(a)}} -
  \frac{2\sin(\gamma(2b-1))}{\cos(\gamma(2b-1))-\cosh2\lambda_\ell^{(a)}}  
 \right\}\,.
\end{equation}
where  $\{\lambda_\ell^{(a)}\}$, $a=1,2$ are solutions of the Bethe
equations (\ref{baebq}). 

We now discuss some properties concerning the eigenspectrum (\ref{baebq}) and
(\ref{Emix}) of the superspin chain which will be helpful in the analysis of
the thermodynamic limit behavior.  We first note that the spectrum of this
model remains unchanged under the replacements $2\gamma b\to \pi -2\gamma b $
and $\lambda_j^{(a)} \to \lambda_j^{(a)}+i\pi/2$. In other words we have a
remarkable spectral relation for two distinct values of $b$,
\begin{equation}
  \label{EIDEN1}
  E^{(mix)}_{N_1,N_2}(b,\gamma)= 
  E^{(mix)}_{N_1,N_2}(\pi/(2\gamma)-b,\gamma) \,.
\end{equation}
The identity (\ref{EIDEN1}) allows to restrict our study of the complete phase
diagram of the mixed superspin chain to the region $0 < \gamma b \le \pi/4$
for a given value of the anisotropy $\gamma$ lying in the regime $0<\gamma \le
\pi$.  From (\ref{EIDEN1}) we also see that the line $b\gamma=\pi/4$ is rather
special since the spectrum is mapped onto itself.  This implies that the
individual Bethe roots remain invariant under the shift $\lambda_j^{(a)}\to
\lambda_j^{(a)}+i\pi/2$ which reflects the presence of some discrete
$\mathbf{Z}_2$ invariance of the model on this line which we denote as
'self-dual line' of the model in the following.  We note here that a similar
fact has been observed before for a particular staggered six-vertex model
\cite{IkJS08}.

Furthermore, as is typical of spin chains derived from quantum group algebras,
one also expects that spectrum of $\mathcal{H}^{(mix)}$ at $\gamma$ and
$\pi-\gamma$ should be related to each other. Direct inspection of
Eqs.~(\ref{baebq}) and (\ref{Emix}) reveals this relation to be
\begin{equation}
  \label{EIDEN2}
  E^{(mix)}_{N_1,N_2}(b,\gamma)= 
  -E^{(mix)}_{N_1,N_2}(\gamma b/(\pi-\gamma),\pi-\gamma) 
\end{equation}
From Eq.(\ref{EIDEN2}) we conclude that the spectrum of $H^{(mix)}$ changes
sign under the replacement $\gamma\to\pi-\gamma$ while leaving the product
$b\gamma$ unchanged.  This symmetry is useful to study both the
antiferromagnetic mixed superspin chain (\ref{Hmix}) and the ferromagnetic one
with Hamiltonian $-\mathcal{H}^{(mix)}$ while considering the deformation
parameter on the region $0 <\gamma \le \pi/2$.

\begin{figure}
\includegraphics[width=0.75\textwidth]{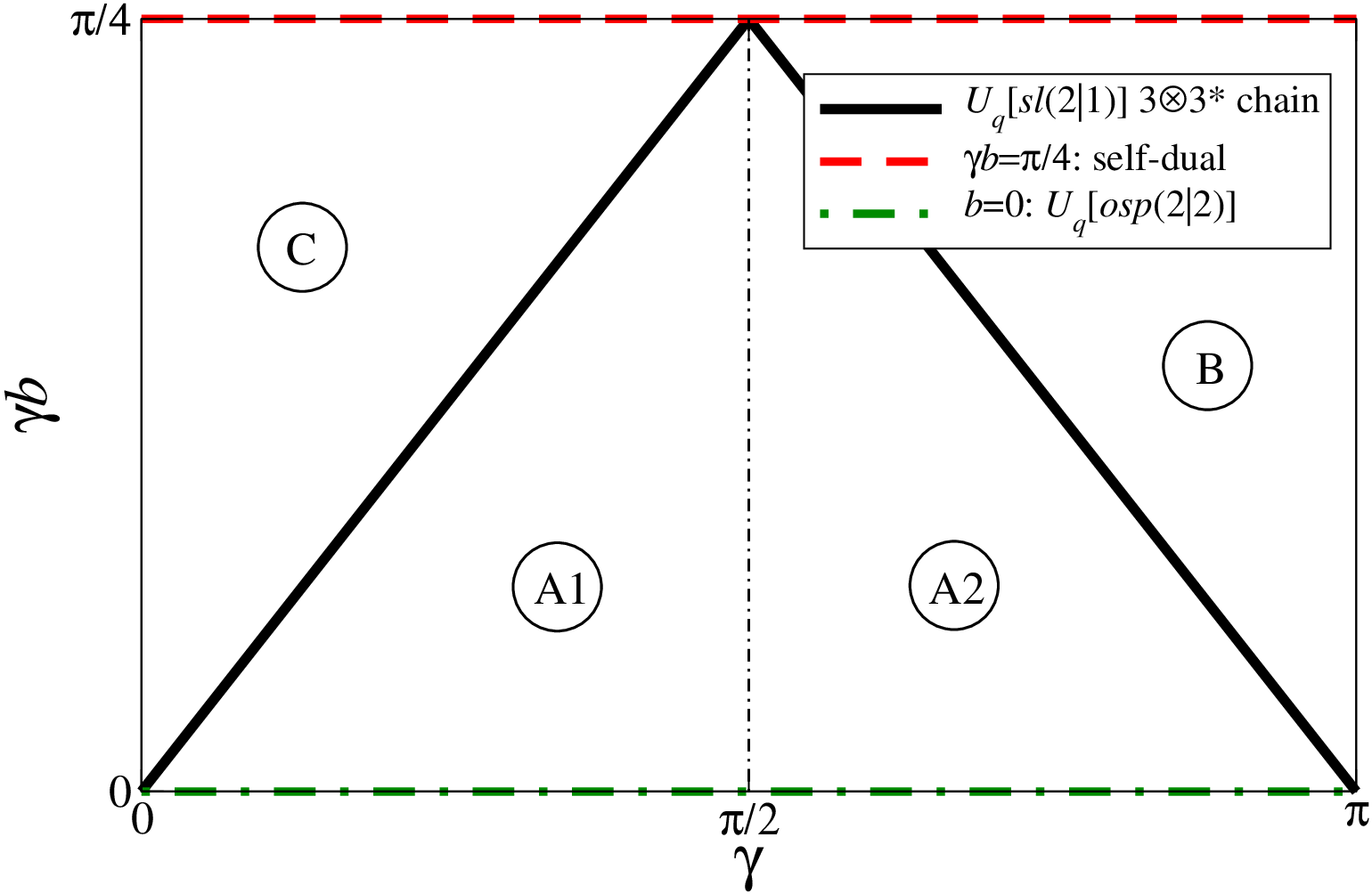}
\caption{Parameter space of the mixed superspin chain.\label{fig:phases}}
\end{figure}

In Figure~\ref{fig:phases} we summarize the region of parameters space one
should concentrate the analysis of the physical properties.

It turns out that for certain choices of the representation parameter $b$ the
mixed superspin model introduced above has been already studied in the
literature. In what follows we summarize the main previous results on this
system:
\begin{itemize}
\item For $b=0$ the alternating representations are the same and we have a
  homogeneous vertex model and a corresponding spin chain with only
  two-body interactions.  We see that the Bethe equations (\ref{baebq}) and
  the expressions (\ref{Emix}) for the eigenvalues simplify drastically.  In
  Ref.~\onlinecite{GaMa07} the critical properties of the corresponding
  \emph{antiferromagnetic} $U_q[osp(2|2)]$ superspin chain have been studied.
  The continuum theory was found to have central charge $c=-1$ with dimensions
  varying continuously with the deformation parameter $\gamma$.

\item For $b=\pm\half$ the four-dimensional representations used for the
  construction of the mixed superspin model (\ref{transmix}) degenerate into
  the \emph{atypical} three-dimensional ones, $3$ and $\bar{3}$, see also the
  discussion in Appendix~\ref{app-2site}.  The continuum model describing the
  low energy behaviour of the \emph{antiferromagnetic} $3\otimes
  \bar{3}$-superspin chain has been identified as a $c=0$ theory for
  $\gamma<\pi/4$ \cite{EsFS05,FrMa11}.\footnote{%
    Note that the parametrization of the deformation in
    Ref.~\onlinecite{FrMa11} differs from the one used here by a factor of
    $2$.}  The operator content has been attributed to one compact and one
  non-compact bosonic degree of freedom.  Both the compactification radius of
  the former and the spectral fine structure due to the latter depend on the
  deformation parameter $\gamma$.

  For \emph{ferromagmetic} exchange the spectrum of low-lying states of the
  $3\otimes \bar{3}$-superspin chain haw been found to be the same as for the
  $U_q[osp(2|2)]$ with central charge $c=-1$.  The scaling dimensions of this
  model exhibit exact spin charge separation, the excitations in both sectors
  are free bosons with compactification radii depending on $\gamma$.  In the
  isotropic limit, $\gamma\to0$, the magnetic part of the spectrum turns
  non-relativistic.
\end{itemize}

Considering the above informations we see that the phase diagram
Fig.~\ref{fig:phases} has been only marginally investigated so far.  In this
paper we provide results on the thermodynamics and critical properties of the
superspin chain which allow to characterize the model throughout its parameter
space.  Specifically, we shall investigate the model the four regions
\begin{itemize}
\item phase A1: the antiferromagnetic superspin chain for $b<\half$,
\item phase A2: the ferromagnetic superspin chain for $b<\half$,
\item phase B: the antiferromagnetic superspin chain for
  $\half<b\le\pi/4\gamma$. 
\item phase C: the antiferromagnetic superspin chain for
  $\half<b\le\pi/4\gamma $, and
\end{itemize}
Before turning to this problem let us first discuss another important feature
of the alternating vertex model (\ref{transmix}) that will be useful in the
analysis of its physical properties.  It turns out that this model hides
integrable systems based on ordinary symmetries with suitable twisted boundary
conditions.  We detail this property in next section.

%
\section{Hidden staggered six-vertex models }
\label{sec:hidden6v}
%
The interest in the study of staggered vertex models probably emerged with the
work of Temperley and Lieb where a remarkable equivalence between the
partition function of a staggered six-vertex model and that of a spin system
denominated $q$-state Potts model was proposed \cite{TeLi71}.  This
relationship was further elaborated for various types of lattices
\cite{BaKW76}.  In particularly it was argued that there exist two manifolds
of statistical weights for a staggered six-vertex model which are solvable by
Bethe ansatz methods \cite{Baxt82}.  Nowadays, the quantum inverse scattering
framework \cite{TaFa79} provides a general procedure to construct integrable
staggered vertex models from solutions of the Yang-Baxter equation.  Typical
examples are models whose transfer matrix are built up by staggering the
spectral parameters of an additive $R$-matrix between two or more different
values.

In what follows we shall argue that there is a one-to-one correspondence
between part of the spectrum of the mixed $U_q[sl(2|1)]$ vertex model and the
eigenvalues of the transfer matrix $T_{6v}(\lambda)$ of a staggered six-vertex
model with anti-periodic boundary conditions.  The latter can be written as
the product of commuting single-row transfer matrices
\begin{equation}
T_{6v}(\lambda)= T_{stg}(\lambda) T_{stg}(\lambda+2ib\gamma)\,
\label{TXXZ}
\end{equation}
where the operator $T_{stg}(\lambda)$ is the transfer matrix constructed from
the $R$-matrix of the six-vertex model in which the spectral parameters
alternates between $\lambda$ and $\lambda-2ib\gamma$.  The expression for
$T_{stg}(\lambda)$ is therefore given by
\begin{equation}
\label{TXXZstg}
T_{stg}(\lambda) = Tr_{\cal{A}} \left[ G_{\cal A}
\mathcal{R}^{(6v)}_{{\cal A} 2L}(\lambda) 
\mathcal{R}^{(6v)}_{{\cal A} 2L-1}(\lambda -2ib\gamma) 
\mathcal{R}^{(6v)}_{{\cal A} 2L-1}(\lambda) 
\cdots
\mathcal{R}^{(6v)}_{{\cal A} 1}(\lambda-2ib\gamma) \right]\,,
\end{equation}
with the standard $R$-matrix associated to the symmetrical six-vertex model
\begin{equation}
\mathcal{R}^{(6v)}(\lambda)=
\left(\begin{array}{cccc}
        1 & 0 & 0 & 0\\
        0 & 
\frac{\sinh(\lambda)}{\sinh(\lambda-i\gamma)} & 
\frac{\sinh(-i\gamma)}{\sinh(\lambda-i\gamma)} & 
0 \\
0 & 
\frac{\sinh(-i\gamma)}{\sinh(\lambda-i\gamma)} & 
\frac{\sinh(\lambda)}{\sinh(\lambda-i\gamma)} & 
0 \\
        0 & 0 & 0 & 1 \\
        \end{array}\right),
\label{Rsix-vertex}
\end{equation}
The matrix $G_{\cal A}$ in (\ref{TXXZstg}) encodes the freedom of choices of
toroidal boundary conditions compatible with integrability.  Here we assume
that it preserves the bulk $U(1)$ symmetry and therefore is diagonal,
\begin{equation}
G_{\cal A}=\left(\begin{array}{cc}
        1 & 0 \\
        0 & \mathrm{e}^{i\varphi} \\
        \end{array}\right),~~~~
0 \leq \varphi \leq \pi\,.
\label{bound}
\end{equation}

The transfer matrix (\ref{TXXZstg}) with boundary condition (\ref{bound}) can
be diagonalized by applying the so-called $ABCD$ algebraic Bethe ansatz method
\cite{TaFa79}.  Since this framework has been well discussed in the
literature, see for instance \cite{VladB}, we shall present only the main
results that are of interest here.  The underlying $U(1)$ invariance implies
that the Hilbert space can be split into disjoint sectors labeled by the
eigenvalues of the total azimuthal magnetization operator
$S_3=\half\sum_{j=1}^{2L} \sigma_j^z$.  The corresponding eigenvalues
$\Lambda_{N}(\lambda,\varphi)$ of $T_{stg}(\lambda)$ in the sector with
magnetization $L-N$ where $N=0,\cdots,L$ are given by,
\begin{equation}
\label{GAMAXXZstg}
\begin{aligned}
\Lambda_{N}(\lambda,\varphi;\{\lambda_j\}) &=
\prod_{j=1}^{N} \frac{\sinh(\lambda_{j}-\lambda-i\gamma/2)}
{\sinh(\lambda_{j} -\lambda +i\gamma/2)} \\
&+ \mathrm{e}^{i\varphi}
    \left[\frac{\sinh(\lambda-i2b\gamma)\sinh(\lambda)}{ 
                \sinh(\lambda-i\gamma)\sinh(\lambda-i\gamma-i2b\gamma)}\right]^{L}  
    \prod_{j=1}^{N} \frac{\sinh(\lambda-\lambda_{j}-i3\gamma/2)}{
                        \sinh(\lambda-\lambda_j-i\gamma/2)}  \,.
\end{aligned}
\end{equation}
The eigenstates are parameterized by the rapidities $\lambda_j$ which satisfy
the following Bethe ansatz equations,
\begin{equation}
\label{betheXXZa}
\left[\frac{\sinh(\lambda_{j}-i\gamma/2)\sinh(\lambda_j-i\gamma/2-i2b\gamma)}
{\sinh(\lambda_{j}+i\gamma/2)\sinh(\lambda_j+i\gamma/2 -i2b\gamma)}\right]^{L}=
\mathrm{e}^{i\varphi}
\prod_{\stackrel{k=1}{k \neq j}}^{N}  \frac{\sinh(\lambda_{j}-\lambda_{k}-i\gamma)}
{\sinh(\lambda_{j}-\lambda_{k}+i\gamma)},\quad j=1,\cdots,N .
\end{equation}
Using (\ref{TXXZ}) the transfer matrix eigenvalues
$\Lambda_{N}^{(6v)}(\lambda,\varphi)$ of the staggered six vertex model can be
written a product of two terms, namely
\begin{equation}
\Lambda_{N}^{(6v)}(\lambda,\varphi;\{\lambda_j\})=
\Lambda_{N}(\lambda,\varphi;\{\lambda_j\})
\Lambda_{N}(\lambda+i2b\gamma,\varphi;\{\lambda_j\})
\label{LAMTOTAL}
\end{equation}

We have now the main ingredients to identify part of the spectrum of the mixed
transfer matrix (\ref{transmix}) with the eigenvalues of the staggered
six-vertex model with anti-periodic boundary condition $\varphi=\pi$:
Facilitated by their symmetrical form the Bethe equations (\ref{baebq}) allow
for solutions in the zero charge sector ($N_1=N_2=N$) with coinciding roots,
i.e.\ $\{\lambda_j^{(1)}\} \equiv \{\lambda_j^{(2)}\}$.
Under these conditions is not difficult to see that Bethe ansatz equations
(\ref{baebq}) of the mixed $U_q[sl(2|1)]$ vertex model coincide with those
associated to the staggered six-vertex model (\ref{betheXXZa}) after the
identification $\lambda_j=\lambda_j^{(1)} +ib\gamma$.  Furthermore, by
expanding the products entering in Eq.(\ref{GAMAMIX}) and comparing them with
that given by expressions (\ref{GAMAXXZstg}), (\ref{LAMTOTAL}) one is able to
see the following direct correspondence,
\begin{equation}
\label{LAMrel}
  \Lambda_{N,N}^{(mix)}(\lambda;\{\lambda_j^{(1)}\},\{\lambda_j^{(1)}\}) \equiv
  \Lambda_{N}^{(6v)}(\lambda,\varphi=\pi;\{\lambda_j^{(1)}+ib\gamma\}) 
\end{equation}
where the rapidities are now solutions of
\begin{equation}
\label{betheXXZ}
\left[\frac{\sinh(\lambda_{j}^{(1)}+ib\gamma -i\gamma/2)}
           {\sinh(\lambda_{j}^{(1)}+ib\gamma +i\gamma/2)}\,
      \frac{\sinh(\lambda_j^{(1)}-ib\gamma -i\gamma/2)}
           {\sinh(\lambda_j^{(1)}-ib\gamma +i\gamma/2)}\right]^{L}=
  -\prod_{\stackrel{k=1}{k \neq j}}^{N}
  \frac{\sinh(\lambda_{j}^{(1)}-\lambda_{k}^{(1)}-i\gamma)}
       {\sinh(\lambda_{j}^{(1)}-\lambda_{k}^{(1)}+i\gamma)}\,,
  \quad j=1,\cdots,N .
\end{equation}
As a consequence of (\ref{LAMrel}) the corresponding energy eigenvalues
(\ref{Emix}) of the $U_q[sl(2|1)]$ superspin chain are related to those of the
spin chain associated to the staggered six-vertex model by
\begin{equation}
\label{ener-map}
 E_{N,N}^{(mix)} = 2 E_N^{(6v)}(\varphi=\pi)
\end{equation}
where
\begin{equation}
\label{ener-ft}
 E_N^{(6v)}(\varphi) = i\,\left. \frac{\partial}{\partial\lambda}
   \ln\,\Lambda_{N}^{(6v)}(\lambda,\varphi;\{\lambda_j^{(1)}+ib\gamma\})
 \right|_{\lambda=0}\,  
\end{equation}

We would like to conclude this section with the following remarks.  The
staggered six-vertex model has at least two particular lines in which certain
quantum group symmetries show up.  The first occurs at $b=1/2$ in which the
$R$-matrix (\ref{Rsix-vertex}) of the six-vertex model with rapidity $\lambda
=2i\gamma b$ becomes proportional to a projector. In this case the product of
neighbouring $R$-matrices in (\ref{TXXZstg}) behaves much like in the fusion
procedure for two spins-1/2 and the underlying quantum symmetry is that
present on the integrable XXZ spin-1 chain \cite{ZaFa80}.
The second peculiar line turns out to be again $\gamma b=\pi/4$ which is
equivalent to one of the integrable manifolds of the $q$-state Potts model
with antiferromagnetic couplings \cite{Baxt82}. This equivalence has been
recently re-elaborated in Ref.~\onlinecite{IkJS08} where arguments in favour
of a possible underlying quantum group symmetry were given however without any
precise proposal.  In what follows we shall argue that such an invariance is
directly related to the twisted quantum algebra $U_q[D_2^{(2)}]$ with $q=
\mathrm{e}^{2i\gamma}$.  In fact, we are going to show that the transfer
matrix spectrum of the staggered six-vertex model on the line $\gamma b=\pi/4$
is the same as that of the plain $U_q[D_2^2]$ vertex model with $L$ sites. In
order to do that we start by defining the latter transfer matrix,
\begin{equation}
  \label{TD22}
  T_{D_2^2}(\lambda) = Tr_{\cal{A}} \left[ \bar{G}_{\cal A}
    \mathcal{R}^{(D_2^2)}_{{\cal A} L}(\lambda) 
    \mathcal{R}^{(D_2^2)}_{{\cal A} L-1}(\lambda) 
    \cdots
    \mathcal{R}^{(D_2^2)}_{{\cal A} 1}(\lambda) \right],
\end{equation}
where $\mathcal{R}_{12}^{(D_2^2)}(\lambda)$ denotes the $R$-matrix of the
$U_q[D_2^2]$ vertex model.  This is a four state vertex model and the
explicitly form of the $R$-matrix can be found in the original work by Jimbo
\cite{Jimb86}.

The toroidal boundary condition is represented by the $c$-number matrix
$\bar{G}_{\cal A}$.  It turns out that the most general diagonal matrix
preserving the commutativity of the transfer matrix (\ref{TD22}) for different
values of the spectral parameter is,
\begin{equation}
  \bar{G}_{\cal A}=\left(\begin{array}{cccc}
      1 & 0 &0 &0\\
      0 & \mathrm{e}^{i\varphi}&0 &0 \\
      0 & 0& \mathrm{e}^{i\varphi}&0 \\
      0 & 0& 0 & \mathrm{e}^{2i\varphi}\\
    \end{array}\right),~~~~
  0 \leq \varphi \leq \pi.
  \label{boundd22}
\end{equation}

It is possible to find the eigenvalues ${\bar{\Lambda}}_{N}(\lambda,\varphi)$
of the transfer matrix (\ref{TD22}) by adapting the algebraic diagonalization
procedure devised in \cite{Mart99} to include the twisted boundary conditions
(\ref{boundd22}).  The index $N$ denotes the many possible $U(1)$ sectors that
are underlying the quantum $U_q[D_2^2]$ algebra.  Following such algebraic
approach we find that the eigenvalues of $T_{D_2^2}(\lambda)$ given in terms
of the Bethe rapidities $\{\mu_j\}$ can be written as,
\begin{eqnarray}
  {{\Lambda}}^{(D_2^2)}_{N}(\lambda,\varphi;\{\mu_j\})& =&
  \left[R_{11}^{11}(\lambda)\right]^{L}
  \prod_{j=1}^{N} \frac{R_{11}^{11}(\mu_{j}-\lambda)}
  {R_{12}^{12}(\mu_{j} -\lambda)} 
  +\mathrm{e}^{i\varphi}
  \left[\frac{R_{12}^{12}(\lambda)}{R_{11}^{11}(\lambda)}\right]^{L} \Lambda^{(16v)}(\lambda,\{\mu_j\})
  \nonumber\\
  &+&\mathrm{e}^{2i\varphi}
  \left[\frac{R_{41}^{41}(\lambda)}{R_{11}^{11}(\lambda)}\right]^{L} 
  \prod_{j=1}^{N} \frac{R_{12}^{12}(\lambda-\mu_{j})}
  {R_{41}^{41}(\lambda-\mu_{j})} 
  \label{EIND22}
\end{eqnarray}
where the elements $R_{ab}^{cd}$ can be read from the $U_q[D_2^2]$ $R$-matrix
(\cite{Jimb86}) by the relation $R(\lambda)= \displaystyle{
  \sum_{a,b,c,d=1}^{4}} R_{ab}^{cd}(\lambda) e_{ac} \otimes e_{bd}$.

To make a comparison with the eigenvalues of the staggered six-vertex model we
shall first normalized the $R$-matrix by setting $R_{11}^{11}(\lambda)=1$.
The expressions of the other needed $R$-matrix elements in Eq.~(\ref{EIND22})
can then be easily read from the original work by Jimbo \cite{Jimb86}.  We
next choose the spectral variable $x$ and the anisotropy $k$ used in this
later work to be $x=\mathrm{e}^{2\lambda}$ and
$k=\mathrm{e}^{2i\gamma}$. Considering these definitions the expressions of
the $R$-matrix elements entering Eq.(\ref{EIND22}) are,
\begin{equation}
  R_{12}^{12}(\lambda)= 
  \frac{\sinh(2\lambda)}{\sinh(2\lambda-2i\gamma)},~~~
  R_{41}^{41}(\lambda)= 
  \left[\frac{\sinh(2\lambda)}{\sinh(2\lambda-2i\gamma)}\right]^2
\end{equation}

The function $\Lambda^{(16v)}(\lambda,\{\mu_j\})$ corresponds to the
eigenvalue of the transfer matrix of an inhomogeneous sixteen-vertex that
naturally emerges in the nested algebraic Bethe ansatz formulation. As
explained in Ref.~\onlinecite{Mart99} such sixteen-vertex model is special and
its transfer matrix can be diagonalized without the recourse of a second Bethe
ansatz.  Considering the results of \cite{Mart99} one finds that the
expression for such eigenvalues in our notation is,
\begin{equation}
  \Lambda^{(16v)}(\lambda,\{\mu_j\})=
  \prod_{j=1}^{N} 
  \frac{\sinh(2\lambda-2\mu_j)+\epsilon_j\sinh(2i\gamma)}{\sinh(2\lambda-2\mu_j)}+
  \prod_{j=1}^{N} 
  \frac{\sinh(2\lambda-2\mu_j)-\epsilon_j\sinh(2i\gamma)}{\sinh(2\lambda-2\mu_j)}
\end{equation}
where $\epsilon_j=\pm$ is a discrete $Z_2$ variable entering in the
corresponding eigenvectors.

Putting all these results together and after few simplifications we find the
the final expression for the eigenvalue
${{\Lambda}}^{(D_2^2)}_{N}(\lambda,\varphi;\{\mu_j\})$ is given by,
\begin{eqnarray}
  \label{EIND22FI}
  {{\Lambda}}^{(D_2^2)}_{N}(\lambda,\varphi;\{\mu_j\}) &=&  
  \prod_{j=1}^{N} 
  \frac{\sinh(2\mu_j-2\lambda-2i\gamma)}{\sinh(2\mu_j-2\lambda)}+
  \mathrm{e}^{2i\varphi}
  \left[
    \frac{\sinh(2\lambda)}{\sinh(2\lambda-2i\gamma)}
  \right]^{2L} 
  \prod_{j=1}^{N} 
  \frac{\sinh(2\lambda-2\mu_j-2i\gamma)}{\sinh(2\lambda-2\mu_j)}
  \nonumber \\
  &+& 
  \mathrm{e}^{i\varphi}
  \left[
    \frac{\sinh(2\lambda)}{\sinh(2\lambda-2i\gamma)}
  \right]^{L} 
  \left \{ 
    \prod_{j=1}^{N} 
    \frac{2\sinh(\lambda-\mu_j-i\gamma)\cosh(\lambda-\mu_j+i\gamma)}{\sinh(2\lambda-2\mu_j)} \right.
  \nonumber \\
  &+& \left. \prod_{j=1}^{N} 
    \frac{2\sinh(\lambda-\mu_j+i\gamma)\cosh(\lambda-\mu_j-i\gamma)}{\sinh(2\lambda-2\mu_j)} \right \}
\end{eqnarray}
It is not difficult to see that Eq.(\ref{EIND22FI}) also factorizes into a
product of two terms.  By defining $\mu_j=\lambda_j+i\frac{\gamma}{2}$ we see
that such product form is exactly identified with the eigenvalues of the
staggered six-vertex model when $\gamma b=\pi/4$ given by
Eqs.(\ref{GAMAXXZstg}), (\ref{LAMTOTAL}).

We finally recall that the $D_2^2$ $R$-matrix can be obtained by Baxterizing
the dilute Birman-Wenzel-Murakami algebra associated to the $O(3)$ braid
representation \cite{Grim94}.  Following Ref.~\onlinecite{GaMa06} one can also
show that such $R$-matrix (now graded) can be alternatively be derived by
means of the dilute Baxterization of the $osp(1|2)$ superalgebra.  This means
that the staggered vertex model for $\gamma b=\pi/4$ hides both the fermionic
and bosonic degrees of freedom.  From this observation it follows directly that
the isotropic limit $\gamma\to\pi/2$ has indeed an $osp(2|2)$ symmetry as
first pointed out in Ref.~\onlinecite{IkJS10}.

\section{Thermodynamic limit and critical properties for $0\le b <\half$}
\label{sec:mix-A}
We are now going to discuss the thermodynamic limit and the low energy
properties of the mixed vertex model (\ref{transmix}).  Our discussion will be
based on the identification those root configurations solving the Bethe
equations (\ref{baebq}) which correspond to the low lying eigenstates of
(\ref{Hmix}) for given values of the parameters $b$ and $\gamma$ as obtained
by numerical diagonalization of small systems.

It is well established that in the thermodynamic limit $L\to\infty$ the
solutions of the Bethe equations (\ref{betheXXZ}) for the staggered six vertex
with $b\le\half$ are generically grouped into 'strings' consisting of $m$
complex rapidities $\lambda_{(m),j}$ characterized by a common real center
$\lambda_{(m)}$ and a parity $v_m=\pm1$:
\begin{equation}
\label{stringXXZ}
  \lambda_{(m),j} = \lambda_{(m)} +
  i\frac{\gamma}{2}\left({m+1}-2j\right)
  + i\frac{\pi}{4}\,(1-v_m)\,,\quad j=1,\ldots,m\,.
\end{equation}
The allowed values of $(m,v_m)$ depend on the anisotropy $\gamma$ in a rather
involved way \cite{TaSu72}.  For the mixed superspin chain the string
classification has been done for the case $b\equiv \half$ \cite{EsFS05,
  FrMa11}.  Away from this line we have to rely on the procedure outlined
above.  We find that for $b<\half$ most of the root configurations solving
(\ref{baebq}) corresponding to the ground state and low energy excitations can
be organized into strings of length $1$ of both parities: $(1,+)$, $(1,-)$.

\subsection{Phase A1: Anti-ferromagnetic regime for $0 \leq b < \half$ }
\label{sec:mix-afm}

Exact diagonalization of the Hamiltonian shows that the ground state of the
antiferromagnetic mixed superspin chain for $b\in[0,\half)$ lies in the
sectors $(N_1,N_2)=(L,L-1)$ and $(N_1,N_2)=(L-1,L)$ corresponding to charge
$B=\pm\half$ and spin $S_3=\half$ (the states with $S_3=-\half$ are obtained
by application of the global $\mathbb{Z}_2$-symmetry of the mixed chain, i.e.\
reversal of all spins), i.e.\ fourfold degenerate.  Furthermore, we find that
these states as well as many low lying excitations can be described by real
solutions to the Bethe equations (\ref{baebq}).  

Based on this observation we shall now study this state in the thermodynamic
limit $L\to\infty$ with $N_{a}/L\to1$:
taking the logarithm of the Bethe equations (\ref{baebq}) we obtain
\begin{equation}
\label{baelogA1}
\begin{aligned}
   L \left[\Phi(\lambda_j^{(1)},\gamma(b+\half)) 
   -\Phi(\lambda_j^{(1)},\gamma(b-\half))\right] &= 2\pi Q_j^{(1)} + \sum_{k=1}^{N_2}
      \Phi(\lambda_j^{(1)}-\lambda_k^{(2)},\gamma)\,, \qquad j=1,\ldots,N_1\\
   L \left[ \Phi(\lambda_j^{(2)},\gamma(b+\half)) 
   -\Phi(\lambda^{(2)},\gamma(b-\half))\right] &= 2\pi Q_j^{(2)} + \sum_{k=1}^{N_1}
      \Phi(\lambda_j^{(2)}-\lambda_k^{(1)},\gamma)\,, \qquad j=1,\ldots,N_2
\end{aligned}
\end{equation}
with
\begin{equation}
\label{logPhi}
    \Phi(x,y) = 2\arctan\left( \tanh(x) \cot(y)\right)\,.
\end{equation}
In (\ref{baelogA1}) the numbers $Q_j^{(a)}$ define the many possible branches
of the logarithm.  They have to be chosen integer or half-odd integer
depending on the parities of $N_{a}$ according to the rule
\begin{equation}
\label{qnosA1}
  Q_j^{(1)} \equiv \frac{N_2}{2} \mod 1\,,\qquad
  Q_j^{(2)} \equiv \frac{N_1}{2} \mod 1\,.
\end{equation}
The root configuration corresponding to the ground state in the sector
$(L,L-1)$ is defined by consecutive values for these quantum numbers, i.e.\
$Q_j^{(1)} = -(L-1)/2,-(L-3)/2,\ldots,(L-1)/2$ and $Q_j^{(2)} =
-L/2+1,-L/2+2,\ldots,L/2-1$.

\subsubsection*{Thermodynamic limit}
Following Yang and Yang \cite{YaYa69} we introduce counting functions
\begin{equation}
\label{countA1}
\begin{aligned}
  z^{(1)}(\lambda) &= 
      \half\left( \Phi(\lambda,\gamma(b+\half))
            - \Phi(\lambda,\gamma(b-\half)) \right)
      - \frac{1}{2L}\sum_{k=1}^{N_2} \Phi(\lambda-\lambda_k^{(2)},\gamma)\,,\\
  z^{(2)}(\lambda) &=
      \half\left( \Phi(\lambda,\gamma(b+\half))
            - \Phi(\lambda,\gamma(b-\half)) \right)
      - \frac{1}{2L}\sum_{k=1}^{N_1} \Phi(\lambda-\lambda_k^{(1)},\gamma)\,.
\end{aligned}
\end{equation}
By definition, we have $z^{(a)}(\lambda_j^{(a)}) = \pi Q_j^{(a)}/L$.  
In the thermodynamic limit the roots of the Bethe equations fill the entire
real axis with densities
\begin{equation}
  2\pi \rho^{(a)}(\lambda)
    = \frac{dz^{(a)}(\lambda)}{d\lambda}\,,\qquad a=1,2\,.
\end{equation}
They satisfy coupled linear integral equations obtained by taking derivatives
of Eqs.~(\ref{countA1})
\begin{equation}
\label{iglrhoA1}
\begin{aligned}
  \rho^{(1)}(\lambda) &= \frac{1}{4\pi}\left(\Phi'(\lambda,\gamma(b+\half))
            - \Phi'(\lambda,\gamma(b-\half))\right)
     -\int_{-\infty}^\infty\mathrm{d}\mu K_1(\lambda-\mu) \rho^{(2)}(\mu)\,,\\ 
  \rho^{(2)}(\lambda) &= \frac{1}{4\pi}\left(\Phi'(\lambda,\gamma(b+\half))
            - \Phi'(\lambda,\gamma(b-\half))\right)
     -\int_{-\infty}^\infty\mathrm{d}\mu K_1(\lambda-\mu) \rho^{(1)}(\mu)\,,
\end{aligned}
\end{equation}
with
\begin{equation}
\label{kernelK1}
  K_1 (\lambda) = \frac{1}{2\pi} \Phi'(\lambda,\gamma)
                = \frac{1}{\pi}\,\frac{\sin(2\gamma)}{\cosh2\lambda -
                  \cos2\gamma}\,.
\end{equation}
Eqs.~(\ref{iglrhoA1}) can be solved by Fourier transformation giving
\begin{equation}
\label{densA1}
\rho^{(1)}(\lambda)=\rho^{(2)}(\lambda)= \frac{1}{{\gamma}}\, 
\frac{\cos(\pi b) \cosh\left(\frac{\pi \lambda}{\gamma}\right)} 
  {\cosh\left(\frac{2 \pi \lambda}{\gamma}\right) + \cos(2\pi b)}\,. 
\end{equation}
Using this expression we can compute the energy density ${e}_{\infty}^{(A1)} =
E/L$ of the antiferromagnetic ground state from the infinite volume limit of
Eq.~(\ref{Emix}):
\begin{equation}
\label{einfA1}
  {e}_{\infty}^{(A1)} = -4 \int_{-\infty}^{\infty} \mathrm{d}\omega\,
    \frac{\cosh^2( {b \gamma}\omega) \sinh({(\pi-\gamma)\omega }/2)}
      {\sinh(\pi\omega/2) \cosh(\gamma\omega/2)}
\qquad\mathrm{for}\quad 0 \leq {\gamma} < \pi/2~~~\mathrm{and}~~0 \leq b <\half.
\end{equation}
Low-lying excitations over this ground state are described by modifications of
the configuration of Bethe roots, e.g.\ by alternative choices for the
integers $Q_j^{(a)}$.  The excitations have a linear dispersion which is found
to be $\epsilon^{(a)}(\lambda) \sim {v}_{A1}^{(mix)} p^{(a)}(\lambda)$
by standard methods \cite{HUBBARD}.  In the present model we obtain
\begin{equation}
  v_{A1}^{(mix)}=\left.
     \frac{1}{4\pi \rho^{(a)}(\lambda)}\,
     \frac{\partial{\epsilon}^{(a)}(\lambda)}{\partial\lambda}
  \right|_{\lambda=\infty}= \frac{\pi}{\gamma} \,,
\end{equation}
independent of the representation parameter $b$.

\subsubsection*{Analysis of the finite-size spectrum -- anti-ferromagnetic regime}
Refining the root density approach used above to compute the thermodynamic
properties of the antiferromagnetic ground state the finite-size corrections
to the low-lying energy levels can be computed
\cite{VeWo85,Vega88,WoEc87b,Suzu88,HUBBARD}.  In agreement with the
predictions \cite{BlCN86,Affl86} of conformal field theory they are found to
be of the form
\begin{equation}
\label{fseA1}
E(L,{\gamma})  -L {e}_{\infty}^{(A1)} =
  \frac{2\pi {v}_{A1}}{L} \left[ -\frac{1}{6} +
  X_{n_{1},n_{2}}^{m_1,m_{2}}({\gamma}) \right] 
+ o\left( L^{-1} \right),
\end{equation}
where the scaling dimensions $X_{n_{1},n_{2}}^{m_1,m_{2}}({\gamma})$ are given
by
\begin{equation}
\label{dimsA1}
\begin{aligned}
  X_{n_{1},n_{2}}^{m_1,m_{2}}({\gamma}) =& 
      \frac{\pi-\gamma}{4\pi}\, ( n_1+n_2)^2
    + \frac{\pi}{4(\pi-\gamma)}\, ( m_1+m_2)^2\\
    &+ \frac{\gamma}{4\pi}\, ( n_1-n_2)^2
    + \frac{\pi}{4\gamma}\, ( m_1-m_2)^2\,.
\end{aligned}
\end{equation}
Here the integers $n_a$, $a=1,2$, are related to the numbers of Bethe roots on
each level by $N_a=L-n_a$ and therefore determine the conserved $U(1)$ charge
and spin of the excitation to be $B=(n_1-n_2)/2$ and $S_3=(n_1+n_2)/2$.  The
macroscopic momentum (vorticity) of the excitations is determined by the
indices $m_a$.  As a consequence of the selection rule (\ref{qnosA1}) they
take integer (half-odd integer) values depending on the parity of $n_1\pm
n_2$:
\begin{equation}
\label{vortA1}
\begin{aligned}
  &\bullet \quad \mathrm{for~} n_1\pm n_2 \mathrm{~odd~}
    &\rightarrow \qquad & m_1,m_{2} = 0, \pm 1 ,\pm 2, \dots\\
  &\bullet \quad \mathrm{for~} n_1\pm n_2 \mathrm{~even~}\quad
    &\rightarrow \qquad &m_1,m_{2} = \pm \frac{1}{2}, \pm \frac{3}{2} 
            ,\pm \frac{5}{2} , \dots\,\, .
\end{aligned}
\end{equation}
Note that the scaling dimensions (\ref{dimsA1}) always decompose into two
parts depending on the charge and spin quantum numbers of the excitation
separately.  In the continuum theory the excitations are free bosons with
compactification radii $R_c^2=2\gamma/\pi=2-R_s^2$, respectively.

\subsubsection*{Numerical results}
To verify our expression (\ref{dimsA1}) for the scaling dimensions with the
selection rule (\ref{vortA1}) we have identified some of the corresponding
configurations of Bethe roots and solved the Bethe equations (\ref{baebq})
numerically for lattice sizes up to $L=100$.  From the numerical data for the
energy eigenvalues (\ref{Emix}) we then compute the sequence
\begin{equation}
\label{estimA1}
X(L,\gamma) = \frac{L}{2\pi {v}_{A1}}\left( E(L,{\gamma}) -
  L{e}_{\infty}^{(A1)} \right)  + \frac{1}{6}
\end{equation}
which in the thermodynamic limit is expected to extrapolate to the dimensions
(\ref{dimsA1}).  In Table~\ref{one1A} we present the finite-size estimates
for the energy of the antiferromagnetic ground state $E_0(L)$ for
$\gamma=2\pi/5$ and $2\pi/9$ and $b=0.1, 0.2, 0.3, 0.4$.  Extrapolation of the
data gives the predicted value $X_{1,0}^{0,0}=1/4$, independent of the
deformation parameter $\gamma$ and the representation parameter $b$.  The
corresponding finite-size scaling of the ground state energy is 
\begin{equation}
\label{fs0AF}
  E_0(L,{\gamma})  -L {e}_{\infty}^{(A1)} =
  \frac{2\pi {v}_{A1}}{L}\,\frac{1}{12} + o\left( L^{-1} \right)
  = \frac{\pi {v}_{A1}}{6L} + o\left( L^{-1} \right)
\end{equation}
corresponding to a central charge $c=-1$ of the continuum theory.  This
coincides with the result from \cite{GaMa07} for the $U_q[osp(2|2)]$ chain at
$b=0$.  

Further support for the proposed critical theory is given by corresponding
analysis of the finite-size behaviour of excitation energies: 
\begin{itemize}
\item a state corresponding to the $X_{0,0}^{\half,\half}=\frac{1}{4}
  \left(1-\gamma/\pi\right)^{-1}$ is described by $(N_1,N_2)=(L,L-1)$ real Bethe
  roots distributed symmetrically around the origin plus a single root
  $\lambda^{(2)}=i\pi/2$ on the second level.   The finite-size data and their
  extrapolation for this state are presented in Table~\ref{two1A}.

\item the dimension $X_{1,-1}^{\half,\half}=(\gamma/\pi) + \frac{1}{4}
  \left(1-\gamma/\pi\right)^{-1}$ is found in the sector $(N_1,N_2)=(L+1,L-1)$
  and given by $(L,L-1)$ real roots distributed symmetrically around the
  origin plus a single root $\lambda^{(1)}=i\pi/2$ on the first level.  The
  numerical data are given in Table~\ref{three1A}.

\item the configuration of Bethe roots for the state corresponding to
  $X_{1,0}^{1,0}=\frac{1}{4} + (\pi/4\gamma) + \frac{1}{4}
  \left(1-\gamma/\pi\right)^{-1}$ consists of $(L-1,L-1)$ real rapidities and
  a single additional root on the line $\mathrm{Im}(\lambda^{(a)}) = \pi/2$ in
  one level.  Numerical results are found in Table~\ref{four1A}.

\item The configuration of roots for $X_{1,1}^{\half,\half}=1 + (\gamma/\pi) +
  \frac{1}{4} \left(1-\gamma/\pi\right)^{-1}$ has two special features:
  considering the $(L,L-1)$ sector we find that there is one root
  $\lambda^{(2)}$ which is situated at $\infty$.  Within the root density
  formalism this leads to a phase shift of $\gamma$ in the logarithmic
  equations (\ref{baelogA1}) for the remaining $(L,L-2)$ roots (see
  \cite{FrMa11}).  The configuration of these roots changes at $\gamma=\pi/3$
  from all real to $(L-1,L-2)$ real and one first level root with imaginary
  part $\pi/2$ (such a situation has also been found in the Bethe ansatz
  solution of the twisted XXZ chain \cite{YuFo92}).  The finite-size data for
  (\ref{estimA1}) are shown in Table~\ref{five1A}.

\item Finally, the state with $X_{1,0}^{1,1}= \frac{1}{4} +
  \left(1-\gamma/\pi\right)^{-1}$ in the zero charge sector $(L-1,L-1)$ is
  given by a solution of the Bethe equations (\ref{betheXXZ}) corresponding to
  the degenerated XXZ sector of the mixed superspin chain.  The finite-size
  estimates and the extrapolation for the dimension are given in
  Table~\ref{six1A}. 
\end{itemize}

Summarizing the results of this section we have found that the critical
behaviour of the antiferromagnetic mixed superspin chain does not depend on
the staggering as long as $|b|<\half$.  It is described by a $c=-1$ conformal
field theory identical to the one obtained before in the context of the
antiferromagnetic $U_q[osp(2|2)]$ chain, i.e.\ for $b=0$ \cite{GaMa07}.

\subsection{Phase A2: ferromagnetic regime for $0 \leq b \leq 1/2$ }
\label{sec:mix-fm}
Let us now turn to the ferromagnetic regime of the mixed superspin chain.  As
discussed above, we can use the spectral relation
$\mathrm{spec}(H^{(mix)}(\gamma)) = \mathrm{spec}(-H^{(mix)}(\pi-\gamma))$ to
discuss this regime in the interval of anisotropies $0\le\gamma\le\pi/2$ using
the opposite sign for the energy eigenvalues (\ref{Emix}).  As a consequence,
the classification of solutions to the Bethe equations (\ref{baebq}) remains
unchanged.

\subsubsection*{Thermodynamic limit}
From exact diagonalization of the Hamiltonian for small system sizes we find
that the low energy states can be identified with configurations where the
Bethe roots with $\mathrm{Im}(\lambda_j^{(a)})=\pi/2$.  As in the
antiferromagnetic case the ground state in this regime is four-fold degenerate
and has charge $B=\pm\half$ and spin $S_3=\pm\half$ corresponding to the
sectors $(N_1,N_2)=(L,L-1)$ and $(L-1,L)$ of the Bethe equations.
Reparamaterisation of  the rapidities as $\lambda_j^{(a)} =
\mu_j^{(a)}+i\pi/2$ and taking the logarithm of (\ref{baebq}) we obtain
\begin{equation}
\label{baelogA2}
\begin{aligned}
   L \left[\Psi(\mu_j^{(1)},\gamma(b+1/2)) 
   -\Psi(\mu_j^{(1)},\gamma(b-1/2))\right] &= 2\pi R_j^{(1)} - \sum_{k=1}^{N_2}
      \Phi(\mu_j^{(1)}-\mu_k^{(2)},\gamma)\,, \qquad j=1,\ldots,N_1\\
   L \left[ \Psi(\mu_j^{(2)},\gamma(b+1/2)) 
   -\Psi(\mu_j^{(2)},\gamma(b-1/2))\right] &= 2\pi R_j^{(2)} - \sum_{k=1}^{N_1}
      \Phi(\mu_j^{(2)}-\mu_k^{(1)},\gamma)\,, \qquad j=1,\ldots,N_2
\end{aligned}
\end{equation}
where
\begin{equation}
\label{logPsi}
  \Psi(x,y) = 2\arctan\left( \tanh(x) \tan(y)\right)
\end{equation}
and $\Phi(x,y)$ has been defined in Eq.~(\ref{logPhi}).
%
Again, the numbers $R_j^{(a)}$ define the branches of the logarithm and have
to be chosen integer or half-odd integer depending on the parities of $N_{a}$
according to the rule
\begin{equation}
\label{qnosA2}
  R_j^{(1)} \equiv \frac{N_2}{2} \mod 1\,,\qquad
  R_j^{(2)} \equiv \frac{N_1}{2} \mod 1\,.
\end{equation}
To analyze the thermodynamic limit in this parameter region we proceed as
above: for $L\to\infty$ the roots $\mu_j^{(a)}$ fill the real axis with
densities
\begin{equation}
\label{densA2}
\sigma^{(a)}(\mu)= \frac{1}{(\pi-{\gamma})} 
\frac{\cos\left[\frac{\pi b{\gamma}}{\pi-{\gamma}}\right] 
  \cosh\left[\frac{\pi \mu}{\pi-{\gamma}}\right]} 
{\cosh\left[\frac{2 \pi \mu}{\pi-{\gamma}}\right] +
  \cos\left[\frac{2\pi {b \gamma}}{\pi-{\gamma}}\right]},
\qquad\mathrm{for}~~a=1,2. 
\end{equation}
As before we can now compute the energy density of the ferromagnetic ground
state form (\ref{Emix}) with the result
\begin{equation}
\label{einfA2}
{e}_{\infty}^{(A2)} = -4 \int_{-\infty}^{\infty} 
\mathrm{d} \omega \frac{
\cosh[\omega {b \gamma}]^2 \sinh[\omega {\gamma}/2]}{\sinh[\omega \pi/2] 
\cosh[\omega(\pi-{\gamma})/2]}
\qquad\mathrm{for}\quad 0 < {\gamma} \leq \pi/2~~~\mathrm{and}~~0 \leq b \leq 1/2.
\end{equation}
The low-lying excitations have a linear dispersion $\epsilon^{(a)}(\mu) \sim
{v}_{A2}^{(mix)} p^{(a)}(\mu)$ with Fermi velocity
\begin{equation}
  {v}_{A2}^{(mix)}= \left.
     \frac{\partial_\mu{\epsilon}^{(a)}(\mu)}{4\pi \sigma^{(a)}(\mu)} 
  \right|_{\mu=\infty}= \frac{\pi}{\pi-{\gamma}} \,.
\end{equation}

\subsubsection*{Analysis of the finite-size spectrum -- ferromagnetic regime}
The computation of finite-size corrections to the low-lying energy levels over
the ferromagnetic ground state within the root density formalism is completely
analogous to the antiferromagnetic regime above.  We find that the scaling
dimensions are 
\begin{equation}
\label{dimsA2}
\begin{aligned}
  X_{n_{1},n_{2}}^{m_1,m_{2}}({\gamma}) =& 
      \frac{\pi-\gamma}{4\pi}\, ( n_1-n_2)^2
    + \frac{\pi}{4(\pi-\gamma)}\, ( m_1-m_2)^2\\
    &+ \frac{\gamma}{4\pi}\, ( n_1+n_2)^2
    + \frac{\pi}{4\gamma}\, ( m_1+m_2)^2\,.
\end{aligned}
\end{equation}
Again the integers $n_a$, $a=1,2$, determine the conserved $U(1)$ charge and
spin of the corresponding excitation while the vorticity of the state is a
given by the $m_a$.  The latter are integer (half-odd integer) depending on
the parity of $n_1\pm n_2$:
\begin{equation}
\label{vortA2}
\begin{aligned}
  &\bullet \quad \mathrm{for~} n_1\pm n_2 \mathrm{~odd~}
    &\rightarrow \qquad & m_1,m_{2} = 0, \pm 1 ,\pm 2, \dots\\
  &\bullet \quad \mathrm{for~} n_1\pm n_2 \mathrm{~even~}\quad
    &\rightarrow \qquad &m_1,m_{2} = \pm \frac{1}{2}, \pm \frac{3}{2} 
            ,\pm \frac{5}{2} , \dots\,\, .
\end{aligned}
\end{equation}
As in the low-energy spectrum of the antiferromagnetic mixed superspin chain
we find an exact  separation of spin and charge degrees of freedom.  The
compactification radii, however, are interchanged.

\subsubsection*{Numerical results}
To support these results for the scaling dimensions (\ref{dimsA2}) with
(\ref{vortA2}) we have identified the Bethe configurations corresponding to
several states and studied the $L$-dependence of the corresponding energies
using the estimators
\begin{equation}
\label{estimA2}
X(L) = \frac{L}{2\pi {v}_{A2}}\left( E(L,{\gamma}) -
  L {e}_{\infty}^{(A2)} \right)  + \frac{1}{6}
\end{equation}
which in the thermodynamic limit is expected to extrapolate to the dimensions
(\ref{dimsA2}):
\begin{itemize}
\item the ferromagnetic ground state is described by $(N_1,N_2)=(L,L-1)$ roots
  on the line $\mathrm{Im}(\lambda_j^{(a)})=\pi/2$, with density approaching
  Eq.~(\ref{densA2}) in the thermodynamic limit.  The finite-size data
  (\ref{estimA2}) extrapolate to
  $X_{0,1}^{0,0}(\gamma)=X_{1,0}^{0,0}(\gamma)\equiv 1/4$ independently of
  $\gamma$ and the representation parameter $b$ (see Table~\ref{one1B}).  As a
  consequence the ground state energy scales like
  \begin{equation}
  \label{fs0F}
   E_0(L,\gamma)-L{e}_\infty^{(A2)} = \frac{\pi {v}_{A2}}{6L} + o(L^{-1})
  \end{equation}
  corresponding to a low energy effective theory with central charge $c=-1$.

\item the state giving the scaling dimension
  $X_{0,0}^{\half,-\half}=\frac{1}{4}(1-\gamma/\pi)^{-1}$ is found in the
  sector with $L$ roots with $\mathrm{Im}(\lambda_j^{(a)})=\pi/2$ for each
  level.  Two configuration of this type exist which are mapped onto each
  other by reflection at the imaginary axis.  They have one root on each level
  at $\lambda^{(1)}=-\lambda^{(2)} =\pm\infty$.  The numerical data for the
  scaling dimensions are shown in Table~\ref{two1B}. 

\item a similar configuration in the sector $(L-1,L-1)$ and without the roots
  at $\pm\infty$ gives rise to the dimension
  $X_{1,1}^{\half,-\half}=(\gamma/\pi) + \frac{1}{4}(1-\gamma/\pi)^{-1}$, see
  Table~\ref{three1B}. 

\item the scaling dimension $X_{1,1}^{\half,\half}=(\gamma/\pi) +
  (\pi/4\gamma)$ is in the degenerated XXZ sector of the mixed superspin chain
  and corresponds to a solution of (\ref{betheXXZ}) with $N=L-1$.   For
  $\gamma>\pi/4$ all of the roots have imaginary part $\pi/2$, at
  $\gamma=\pi/4$ the first of these roots jumps to the real axis.  The finite
  size extrapolation of $X_{1,1}^{\half,\half}$ is shown in Table~\ref{four1B}.

\item the configuration of Bethe roots corresponding to
  $X_{1,0}^{1,-1}=\frac{1}{4}+(1-\gamma/\pi)^{-1}$ is in the $(L,L-1)$ sector
  with $(L-1,L-1)$ roots on the line $\mathrm{Im}(\lambda^{(a)}=\pi/2$ where
  one of the roots on the second level is at $\infty$.  In addition there
  exists a single real root $\lambda^{(1)}$.  The finite-size data are in
  Table~\ref{five1B}.

\item The state leading to $X_{0,0}^{\half,\half}=(\pi/4\gamma)$ is again in
  the degenerate XXZ sector determined by the equations (\ref{betheXXZ}).  It
  consists of $L-1$ roots with imaginary part $\pi/2$ and a single real root
  for $\gamma>\pi/4$.  Below $\gamma=\pi/4$ the configuration changes into
  $L-2$ roots $\mathrm{Im}(\lambda)=\pi/2$ and a pair of complex conjugate
  roots forming a 2-string.  Lowering $\gamma$ further it is expected that
  longer strings are formed, similar as observed in \cite{FrMa11}.  The energy
  of the state depends on $\gamma$ in a continuous way, the extrapolation of
  the finite-size data can be found in Table~\ref{six1B}.
\end{itemize}

As in the antiferromagnetic regime the scaling dimensions obtained both from
our analysis of the thermodynamic limit and by solving the Bethe equations for
the mixed superspin chain in this phase do not depend on the staggering $b$.
In fact, the scaling dimensions (\ref{dimsA2}) coincide with what has been
found previously for the mixed superspin chain constructed from alternating
three-dimensional quark and antiquark representations of $U_q[sl(2|1)]$ which
is related to the present model in the limit $b\to\half$ \cite{FrMa11}: we
conclude that the critical theory of the mixed model is the same as that for
the critical $U_q[osp(2|2)]$ spin chain for \emph{all} $|b|\le\half$ and has
an effective central charge $c=-1$.


\section{The \emph{ferro}magnetic mixed superspin chain for $b>\half$}
\label{sec:phaseB}
Numerically diagonalizing the Hamiltonian (\ref{Hmix}) for parameters
$0<\gamma<\pi/2$ and $\half<b\le\pi/4\gamma$ we find the ground state of the
ferromagnetic model, i.e.\ with Hamiltonian $-\mathcal{H}^{(mix)}$, in the
sectors $(N_1,N_2) = (L\pm1,L\mp1)$, i.e.\ two degenerate singlets with charge
$B=\pm1$ and spin $S_3=0$, respectively.
For the self-dual model, $\gamma b=\pi/4$, the degeneracy of this ground state
is doubled for even $L$ as a consequence of the additional discrete
$\mathbb{Z}_2$ symmetry on this line.
For the low energy states we have identified the corresponding root
configurations solving the Bethe equations (\ref{baebq}): they are organized
into strings (\ref{stringXXZ}) of length $1$ with both parities, i.e.
$(1,\pm)$-strings.  Hence we consider solutions
\begin{equation}
\label{confB}
  \left\{\lambda_j^{(a)} \right\}_{j=1}^{N_a} \equiv 
  \left\{\lambda_j^{(a)} \right\}_{j=1}^{N_a^+} \cup 
  \left\{ \mu_j^{(a)}+i\frac{\pi}{2} \right\}_{j=1}^{N_a^-}\,,\quad a=1,2
\end{equation}
of the Bethe equations with $N_a = N_a^+ + N_a^-$ real parameters
$\lambda_j^{(a)}$ and $\mu_j^{(a)}$.  Taking the logarithm of
(\ref{baebq}) we obtain
\begin{equation}
\label{baelogB}
\begin{aligned}
  L\left( \Phi(\lambda_j^{(1)},\gamma(b+\half)) \right.
        &\left.- \Phi(\lambda_j^{(1)},\gamma(b-\half)) \right) = 2\pi Q^{(1)}_j\\
        &
  + \sum_{k=1}^{N_2^+} \Phi(\lambda_j^{(1)}-\lambda_k^{(2)},\gamma)
  - \sum_{k=1}^{N_2^-} \Psi(\lambda_j^{(1)}-\mu_k^{(2)},\gamma)\,,
 \quad j=1,\ldots,N_1^{+}\,,
\\
 -L\left( \Psi(\mu_j^{(1)},\gamma(b+\half)) \right.
        &\left.- \Psi(\mu_j^{(1)},\gamma(b-\half)) \right) = 2\pi R^{(1)}_j\\
        &
  - \sum_{k=1}^{N_2^+} \Psi(\mu_j^{(1)}-\lambda_k^{(2)},\gamma)
  + \sum_{k=1}^{N_2^-} \Phi(\mu_j^{(1)}-\mu_k^{(2)},\gamma)\,,
 \quad j=1,\ldots,N_1^{-}\,,
\\
  L\left( \Phi(\lambda_j^{(2)},\gamma(b+\half)) \right.
        &\left.- \Phi(\lambda_j^{(2)},\gamma(b-\half)) \right) = 2\pi Q^{(2)}_j\\
        &
  + \sum_{k=1}^{N_1^+} \Phi(\lambda_j^{(2)}-\lambda_k^{(1)},\gamma)
  - \sum_{k=1}^{N_1^-} \Psi(\lambda_j^{(2)}-\mu_k^{(1)},\gamma)\,,
 \quad j=1,\ldots,N_2^{+}\,,
\\
 -L\left( \Psi(\mu_j^{(2)},\gamma(b+\half)) \right.
        &\left.- \Psi(\mu_j^{(2)},\gamma(b-\half)) \right) = 2\pi R^{(2)}_j\\
        &
  - \sum_{k=1}^{N_1^+} \Psi(\mu_j^{(2)}-\lambda_k^{(1)},\gamma)
  + \sum_{k=1}^{N_1^-} \Phi(\mu_j^{(2)}-\mu_k^{(1)},\gamma)\,,
 \quad j=1,\ldots,N_2^{-}\,,
\end{aligned}
\end{equation}
where $\Phi(x,y)$
and $\Psi(x,y)$
have been introduced in Eqs.~(\ref{logPhi}) and (\ref{logPsi}) above.  The
quantum numbers $Q_j^{(a)}$, $R_j^{(a)}$ arise from specifying the branch of
the logarithm and uniquely characterize an eigenstate of the system.  They
have to be chosen integer or half-odd integer according to the parities of the
numbers $N_a^\pm$:
\begin{equation}
\label{qnosB}
\begin{aligned}
  &Q_j^{(1)} \equiv \frac{N_{2}^+}{2} \mod 1 \,, \quad 
   R_j^{(1)} \equiv \frac{N_{2}^-}{2} \mod 1 \,,\\
  &Q_j^{(2)} \equiv \frac{N_{1}^+}{2} \mod 1 \,, \quad 
   R_j^{(2)} \equiv \frac{N_{1}^-}{2} \mod 1 \,.
\end{aligned}
\end{equation}

\subsection{Thermodynamic limit}
We introduce counting functions for the relevant root configurations
(\ref{confB})
\begin{equation}
\label{countB}
\begin{aligned}
  z^{(1)}(\lambda) &= 
      \half\left( \Phi(\lambda,\gamma(b+\half))
            - \Phi(\lambda,\gamma(b-\half)) \right)
      - \frac{1}{2L}\sum_{k=1}^{N_2^+} \Phi(\lambda-\lambda_k^{(2)},\gamma)
      + \frac{1}{2L}\sum_{k=1}^{N_2^-} \Psi(\lambda-\mu_k^{(2)},\gamma)\,,\\
  y^{(1)}(\mu) &=
    -\half\left( \Psi(\mu,\gamma(b+\half)) 
               - \Psi(\mu,\gamma(b-\half)) \right) 
  + \frac{1}{2L}\sum_{k=1}^{N_2^+} \Psi(\mu-\lambda_k^{(2)},\gamma)
  - \frac{1}{2L}\sum_{k=1}^{N_2^-} \Phi(\mu-\mu_k^{(2)},\gamma)\,,
\end{aligned}
\end{equation}
and, similarly, $z^{(2)}(\lambda)$, $y^{(2)}(\mu)$.  Evaluating the counting
functions at a root of Eq.~(\ref{baelogB}) yields the corresponding quantum
number up to a factor of $\pi/L$, e.g.\ $z^{(a)}(\lambda_j^{(a)}) = \pi
Q_j^{(a)}/L$.  Taking the thermodynamic limit $L\to\infty$ with fixed ratios
$N_a^\pm/L$ the derivatives of the counting functions define the densities of
roots $\rho^{(a)}$, $\sigma^{(a)}$ and 'holes' $\rho_h^{(a)}$,
$\sigma_h^{(a)}$.
\begin{equation}
\begin{aligned}
  2\pi \left( \rho^{(a)}(\lambda) + \rho_h^{(a)}(\lambda)\right)
    &= -\frac{dz^{(a)}(\lambda)}{d\lambda}\,,\\
  2\pi \left( \sigma^{(a)}(\mu) + \sigma_h^{(a)}(\mu)\right)
    &= -\frac{dy^{(a)}(\mu)}{d\mu}\,.
\end{aligned}
\end{equation}
The signs are chosen such that the bare densities, i.e.\ the hole densities in
the reference state $N_a^\pm=0$
\begin{equation}
\label{baredB}
\begin{aligned}
  \rho_0(\lambda) &= -\frac{1}{4\pi} \left(
    \Phi'(\lambda,\gamma(b+\half)) - \Phi'(\lambda,\gamma(b-\half)) \right)\\
  &= -\frac{1}{2\pi}\,\left(
    \frac{\sin2\gamma(b+\half)}{\cosh2\lambda -\cos2\gamma(b+\half)}
   -\frac{\sin2\gamma(b-\half)}{\cosh2\lambda -\cos2\gamma(b-\half)} \right)
  \,,\\
  \sigma_0(\lambda) &=  \frac{1}{4\pi} \left( \Psi'(\lambda,\gamma(b+\half))
    - \Psi'(\lambda,\gamma(b-\half)) \right)\\
  &= \frac{1}{2\pi}\,\left(
    \frac{\sin2\gamma(b+\half)}{\cosh2\lambda +\cos2\gamma(b+\half)}
   -\frac{\sin2\gamma(b-\half)}{\cosh2\lambda +\cos2\gamma(b-\half)} \right)
\,,
\end{aligned}
\end{equation}
are positive near the origin.  Note, however, that $\rho_0(\lambda)$
changes sign at $\cosh(2\lambda_0) = \cos\gamma/\cos(2\gamma b)$.  On the line
$\gamma b=\pi/4$, where the model is self-dual under the transformation
(\ref{EIDEN1}), $\lambda_0=\pm\infty$ such that $\rho_0(\lambda)$ is
non-negative for real $\lambda$.  Away from this line, however, $\lambda_0$ is
finite which will require special attention in the analysis of the
thermodynamic limit within the root density formalism below.
For now, we proceed as in the previous sections and arrive at integral
equations for the densities
\begin{equation}
\label{iglrhoB}
  \left(
   \begin{array}{c} \rho^{(1)}(\lambda)\\\sigma^{(1)}(\lambda)\\
                    \rho^{(2)}(\lambda)\\\sigma^{(2)}(\lambda)\end{array}
      \right) = 
  \left(
   \begin{array}{c} \rho_0(\lambda)\\\sigma_0(\lambda)\\
                    \rho_0(\lambda)\\\sigma_0(\lambda)\end{array} \right)   
   + \int_{-\infty}^{\infty}\mathrm{d}\mu\,
 \mathbb{K}(\lambda-\mu)
  \left(
   \begin{array}{c} \rho^{(1)}(\mu)\\\sigma^{(1)}(\mu)\\
                    \rho^{(2)}(\mu)\\\sigma^{(2)}(\mu)\end{array}
      \right)
\end{equation}
with kernel matrix
\begin{equation}
\label{kernelB}
  \mathbb{K}(\lambda) = \left(
   \begin{array}{cccc}
     0 & 0 & K_1(\lambda) & -K_2(\lambda) \\
     0 & 0 & -K_2(\lambda) & K_1(\lambda) \\
     K_1(\lambda) & -K_2(\lambda) & 0 & 0 \\
     -K_2(\lambda) & K_1(\lambda) & 0 & 0 \\
   \end{array}
   \right)\,.
\end{equation}
The function $K_1(\lambda)$ has been given before in (\ref{kernelK1}) and
\begin{equation}
\label{kernelK2}
  K_2 (\lambda) = \frac{1}{2\pi} \Psi'(\lambda,\gamma)
                = \frac{1}{\pi}\,\frac{\sin(2\gamma)}{\cosh2\lambda +
                  \cos2\gamma}\,.
\end{equation}
Solving the integral equations (\ref{iglrhoB}) by Fourier transformation
one obtains
\begin{equation}
  \label{densB}
  \begin{aligned}
    \rho^{(a)}(\lambda) &= 
       \frac{\sin\frac{\pi\gamma(2b-1)}{\pi-2\gamma}}{2(\pi-2\gamma)}\,
       \left(\cosh\frac{2\pi \lambda}{\pi-2\gamma} -
         \cos\frac{\pi\gamma(2b-1)}{\pi-2\gamma}\right)^{-1}\,,\\ 
    \sigma^{(a)}(\lambda) &=
       \frac{\sin\frac{\pi\gamma(2b-1)}{\pi-2\gamma}}{2(\pi-2\gamma)}\,
       \left(\cosh\frac{2\pi \lambda}{\pi-2\gamma} +
         \cos\frac{\pi\gamma(2b-1)}{\pi-2\gamma}\right)^{-1}\,,
  \end{aligned}
\end{equation}
for $a=1,2$, corresponding to total densities
\begin{equation}
\label{denstotB}
  \frac{N_a^+}{2L} = \int_{-\infty}^\infty\mathrm{d}\lambda\, \rho^{(a)}(\lambda)
  = \half\frac{\pi-2\gamma b-\gamma}{\pi-2\gamma} 
  = \half - \frac{N_a^-}{2L} \,.
\end{equation}
On the self-dual line $\gamma b=\pi/4$ the densities $\rho^{(a)}$ and
$\sigma^{(a)})$ coincide.  In the limit $b\to\half$ the densities of
$(1,-)$-strings vanish, indicating the transition to the ground state of phase
A2 (see also Ref.~\onlinecite{FrMa11}).

Note that the densities (\ref{densB}) are positive for real $\lambda$ in the
entire phase B, hence a consistent description of the thermodynamic limit of
this state within the root density formalism is possible.  Using
(\ref{densB}) in (\ref{Emix}) we obtain the energy density of this eigenstate
of the ferromagnetic superspin chain:
\begin{equation}
  \label{einfB}
  \begin{aligned}
  \epsilon_\infty^{(B)} &\equiv
  \lim_{L\to\infty} \frac{1}{L} E^{(mix)}_{N_1^\pm,N_2^\pm} = 
  -4\pi\sum_{a=1,2} \left[
    \int_{-\infty}^{\infty}\mathrm{d}\lambda\,
    \rho_0(\lambda) \rho^{(a)}(\lambda)
    + \int_{-\infty}^{\infty}\mathrm{d}\mu\,
    \sigma_0(\mu)\sigma^{(a)}(\mu)
  \right]\\
  &= -4 \int_{-\infty}^\infty\mathrm{d}\omega\,
    \frac{\sinh(\gamma\omega/2)      \left(
          \sinh((\pi-\gamma)\omega/2)\cosh((\pi/2-2\gamma b)\omega)
            -\sinh(\gamma\omega/2)
      \right)
   }{\sinh(\pi\omega/2)\sinh((\pi-2\gamma)\omega/2)}\,.
  \end{aligned}
\end{equation}
Starting from this state we find that there are low energy excitations with
linear dispersion.  Their Fermi velocity is
\begin{equation}
\label{fermivB}
  v_{B} = \frac{2\pi}{\pi-2\gamma}\,.
\end{equation}

\subsection{Phase B: critical theory of the self-dual model}
As noted above the description of the thermodynamic limit above is guaranteed
to yield the ground state for parameters satisfying the self-duality condition
$\gamma b=\pi/4$.  Therefore we begin our discussion of the low energy
behaviour for this case.

\subsubsection*{Finite-size spectrum}
The finite-size corrections for the low energy spectrum of the superspin chain
on the self-dual line can again be studied based on the root density approach.
Unlike in the cases discussed in the previous sections, however, particular
combinations of quantum numbers for the vorticities of the four low energy
modes lead to a singular contribution to the finite-size energy.  This can be
traced back to a singularity of the integral kernel in (\ref{iglrhoB}), i.e.\
the fact that one of the eigenvalues of $(1-\widetilde{\mathbb{K}}(\omega=0))$
vanishes.
This situation is similar to that found previously for the mixed superspin
chain based on the atypical $b=\pm\half$ representations of $U_q[sl(2|1)]$
\cite{EsFS05,FrMa11} as well as for the ferromagnetic staggered six-vertex
model (\ref{TXXZ}) with $\gamma b=\pi/4$ \cite{JaSa06,IkJS08}.
As advocated in the aforementioned works this subtlety can be taken care of by
proper regularization the integral operator and leads to strong logarithmic
finite-size corrections for the contributions from the singular mode.
Put into the present context these considerations leads us to the following
proposal for the general form of the low energy spectrum
\begin{equation}
  \label{fseB}
  E(L,\gamma)-L\varepsilon_\infty^{(B)}(\gamma)
  = \frac{2\pi v_{B}}{L}\left[ -\frac{1}{3}
    + X_{n_1^+,n_1^-,n_2^+,n_2^-}^{m_1^+,m_1^-,m_2^+,m_2^-}(\gamma) \right] + o(L^{-1})\,.
\end{equation}
The scaling dimensions depend on the quantum numbers of the four massless
modes with Fermi velocities (\ref{fermivB}) in this phase:
\begin{equation}
  \label{dimsB}
  \begin{aligned}
  X_{n_1^+,n_1^-,n_2^+,n_2^-}^{m_1^+,m_1^-,m_2^+,m_2^-}(\gamma) &=
    \frac{\gamma}{4\pi} \left(n_1^++n_1^-+n_2^++n_2^-\right)^2
  + \frac{\pi}{16\gamma} \left(m_1^++m_1^-+m_2^++m_2^-\right)^2\\
&  + \frac{\pi-2\gamma}{8\pi} \left(n_1^++n_1^--n_2^+-n_2^-\right)^2
  + \frac{\pi}{8(\pi-2\gamma)} \left(m_1^++m_1^--m_2^+-m_2^-\right)^2\\
& +  \frac{1}{8}\left(n_1^+-n_1^--n_2^++n_2^-\right)^2
  + \frac{1}{8}\left(m_1^+-m_1^--m_2^++m_2^-\right)^2\\
& + \frac{K(L)}{8} \left(n_1^+-n_1^-+n_2^+-n_2^-\right)^2
  + \frac{1}{8K(L)} \left(m_1^+-m_1^-+m_2^+-m_2^-\right)^2\,.
 \end{aligned}
\end{equation}
The excitations are labeled by $n_a^\pm = (L/2) - N_a^\pm$, i.e.\ the
difference between the number of Bethe roots in (\ref{confB}) and the ground
state densities (\ref{denstotB}) on the line $b\gamma=\pi/4$ in the
thermodynamic limit.  By definition the $n_a^\pm$ are integers (half-odd
integers) for lattices with even (odd) $L$.
From the selection rule (\ref{qnosB}) we conclude that the vorticities of an
excitation with charges $(n_1^+,n_1^-,n_2^+,n_2^-)$ are integer or half-odd
integer values according to the rule
\begin{equation}
\label{vortB}
  m_a^\pm = \half\left(n_1^\pm-n_2^\pm+1\right)\mod 1\,,\quad
  a=1,2.
\end{equation}
The difference $n_1^\pm-n_2^\pm\equiv-(N_1^\pm-N_2^\pm)$ is always integer,
therefore the vorticities can take integer or half-odd integer values
depending on the numbers $n_a^\pm$.

Two of the gapless modes contributing to (\ref{dimsB}) can be identified with
the $U(1)$ symmetries corresponding to spin and charge of the mixed superspin
chain: the effective theories for these modes are those of free bosons with
compactification radii $R_s^2=4\gamma/\pi = 2-R_c^2$.
The existence of the third compact boson with radius $(R_3)^2=2$ does not
follow directly from the symmetries of the mixed superspin chain.

The coupling constant $K(L)$ of the fourth mode reflects the effect of the
regularization of the integral operator on the finite lattice: an analytical
derivation of its $L$-dependence does not exist so far.  Based on numerical
evidence $K(L)$ is expected to display a logarithmic dependence on the lattice
size $L$ and to vanish as $L\to\infty$.
Later on we shall present our own numerical results supporting such
finite-size logarithmic corrections for the superspin chain.  Here we draw on
previous studies of the staggered ferromagnetic six-vertex model
\cite{IkJS08}: as discussed in Section~\ref{sec:hidden6v} the scaling
dimensions (\ref{dimsB}) with $n_1^\pm=n_2^\pm=n^\pm$,
$m_1^\pm=m_2^\pm=m^\pm$ appear in the XXZ subsector of the spectrum.  Taking
into account (\ref{ener-map}) the finite-size spectrum of the staggered XXZ
model is
\begin{equation}
  \label{dimsBxxz}
  \begin{aligned}
    E^{(6v)}(L,\gamma)-\frac{L}{2}\varepsilon_\infty^{(B)}(\gamma) &=
    \frac{2\pi v_{B}}{L}\left[ -\frac{1}{6}
      + \widetilde{X}_{n^+,n^-}^{m^+,m^-}(\gamma) \right] + o(L^{-1})\,,\\
    \widetilde{X}_{n^+,n^-}^{m^+,m^-}(\gamma) =&\, \frac{\gamma}{2\pi}
    \left(n^++n^-\right)^2 + \frac{\pi}{8\gamma} \left(m^++m^-\right)^2\\%
    & + \frac{K(L)}{4} \left(n^+-n^-\right)^2 + \frac{1}{4K(L)}
    \left(m^+-m^-\right)^2\,.
 \end{aligned}
\end{equation}
Note that both the charge mode \emph{and} the boson with self-dual radius
$R_3^2=2$ disappear from the spectrum in this sector.
The scaling dimensions $\widetilde{X}_{n^+,n^-}^{m^+,m^-}(\gamma)$ have been
obtained directly for the staggered six vertex model previously \cite{JaSa06}.

Finally let us remark on the effect of boundary conditions on the finite-size
spectrum: for antiperiodic boundary conditions, i.e.\ twist $\varphi=\pi$ in
(\ref{bound}), relevant for the staggered six vertex model hidden inside the
superspin chain the vorticities are constrained by $m^\pm=\varphi/2\pi\mod 1$,
hence they take half odd integer values.

\subsubsection*{Numerical results}
\begin{figure}
\includegraphics[width=0.75\textwidth]{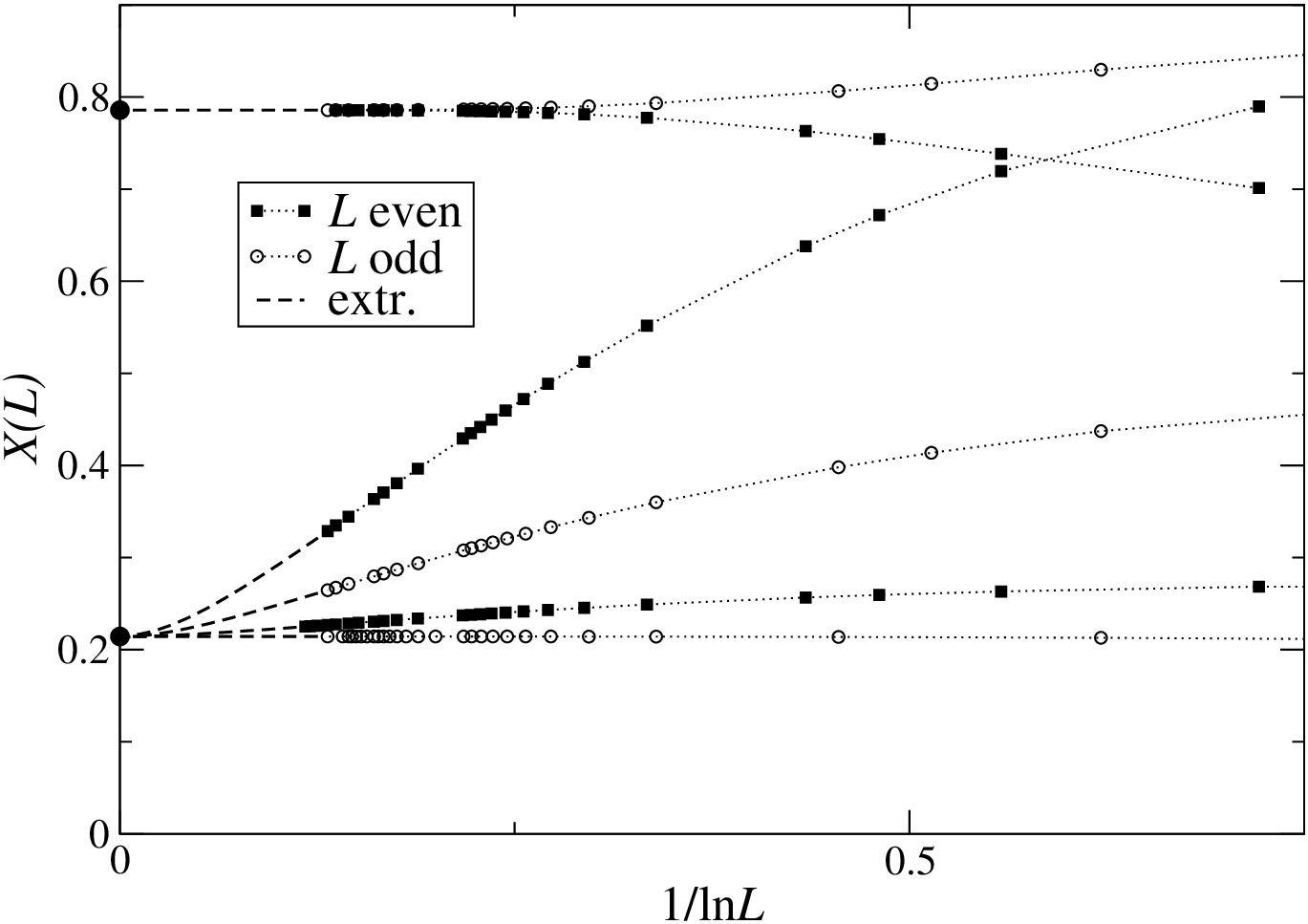}
\caption{Finite-size estimators for the scaling dimensions of some low lying
  states of the mixed superspin 
  chain for $\gamma=2\pi/7$ and $\gamma b=\pi/4$.  The lowest states
  extrapolate to $\half-\frac{\gamma}{\pi}=\frac{3}{14}$ and
  differ by multiples of $1/(\ln L)^2$.  The other states extrapolate to
  $\frac{1}{2} + \frac{\gamma}{\pi}=\frac{11}{14}$ without logarithmic
  corrections.  Dashed lines are rational function extrapolations of the
  numerical data to $L\to\infty$.
  \label{fig:dimBPotts}}
\end{figure}

As mentioned at the beginning of this section one of the ground states of the
mixed superspin chain is in the sector $(N_1,N_2)=(L+1,L-1)$ corresponding to
charge $B=+1$ and zero magnetization.
Following the selection rules for the $n_a^\pm$, $m_a^\pm$ given above we find
the lowest energy state for odd $L$ in this sector to be given by $\mathbf{n}
\equiv (n_1^+,n_1^-,n_2^+,n_2^-) = (-\half,-\half,\half,\half)$ and
$\mathbf{m}=(0,0,0,0)$.  According to (\ref{fseB}) the ground state energy
scales as
\begin{equation}
\label{gsB}
  E_0(L,\gamma)-L\varepsilon_\infty^{(B)}(\gamma)
  = -\frac{\pi v_{B}}{6L}\left[ 2\frac{6\gamma-\pi}{\pi} \right] + o(L^{-1})\,,
\end{equation}
giving an effective central charge $c_\mathrm{eff}= 2(6\gamma-\pi)/\pi$.  Note
that $c_{\mathrm{eff}}=0$ for $\gamma=\pi/6$.  There are, however, subleading
finite-size corrections to (\ref{gsB}).
Similarly, the ground state for even $L$ corresponds to the choice
$\mathbf{n}=(-1,0,0,+1)$ and $\mathbf{m}=(0,0,0,0)$ which results in the same
effective central charge, although there are logarithmic corrections of the
finite-size gap.  More generally, the finite-size energies of states within
the same spin and charge sector and furthermore $n_1^+-n_1^- = n_2^+-n_2^-$
differ by multiples of $K(L)$ only.  For example, the finite-size estimators
for the scaling dimensions $X(L) = L(E(L)-L\varepsilon_\infty^{(B)})/(2\pi
v_{B})+1/3$ of the following configurations (note that the $N_a^\pm$ are
integers, so the configurations can be realized for $L$ even \emph{or} odd
only!)
\begin{center}
\begin{tabular}{|cccc|cc|c|}
\hline
  $N_1^+$ & $N_1^-$ & $N_2^+$ &  $N_2^-$ & $\mathbf{n}$ & $\mathbf{m}$ & 
             $X_{\mathbf{n}}^{\mathbf{m}}$ \\
\hline
  ~$(L+1)/2$~ & ~$(L+1)/2$~ & ~$(L-1)/2$~ & ~$(L-1)/2$~ & 
             $(-\half,-\half, \half, \half)$ & $(0,0,0,0)$ &
             $\half - \frac{\gamma}{\pi}$ \\
  $(L+2)/2$ & $L/2$ & $L/2$ & $(L-2)/2$ & 
             $(-1,0,0,1)$ &  $(0,0,0,0)$ &
             $\half - \frac{\gamma}{\pi} +K(L) $ \\
  $(L+3)/2$ & $(L-1)/2$ & $(L+1)/2$ & $(L-3)/2$ & 
             $(-\frac{3}{2},\half,-\half,\frac{3}{2})$ &  $(0,0,0,0)$ &
             $\half - \frac{\gamma}{\pi} +4K(L) $ \\
  $(L+4)/2$ & $(L-2)/2$ & $(L+2)/2$ & $(L-4)/2$ & 
             $(-2,1,-1,2)$ &  $(0,0,0,0)$ &
             $\half - \frac{\gamma}{\pi} +9K(L) $ \\
\hline
  $L/2$ & $(L-2)/2$ & $(L-2)/2$ & $L/2$ & 
             $(0,1, 1,0)$ & $(0,0,0,0)$ &
             $\half +\frac{\gamma}{\pi}$ \\
  $(L-1)/2$ & $(L-1)/2$ & $(L-1)/2$ & $(L-1)/2$ & 
             $(\half,\half,\half,\half)$ & $(\half,-\half,-\half,\half)$ &
             $\half + \frac{\gamma}{\pi}$ \\
\hline
\end{tabular}
\end{center}
extrapolate to the same values $\half-\frac{\gamma}{\pi}$ and $\frac{1}{2}+
\frac{\gamma}{\pi}$, respectively.  The splitting of these levels for finite
$L$ and in particular the fine structure of gaps between the first four levels
reflecting the $L$-dependence of $K(L)$ is shown for $\gamma=2\pi/7$ in Figure
\ref{fig:dimBPotts}.  It is clearly seen that this fine structure in the
multiplet $X_{\mathbf{n}}^{\mathbf{n}}\to\half-\frac{\gamma}{\pi}$ is very
different from the corrections to scaling depending on the parity of the
system size as observed between the two states shown for the multiplet
$X_{\mathbf{n}}^{\mathbf{n}}\to\half+\frac{\gamma}{\pi}$.
\subsubsection*{The logarithmic fine structure}
Our numerical results show that the low lying levels of the mixed superspin
chain can be combined into groups with the same finite-size behaviour in the
thermodynamic limit.  For large but finite chains, however, these degeneracies
are lifted and gaps $L\Delta E\sim 1/(\ln L)^2$ appear.
Such logarithmic corrections to the scaling are usually understood as a
consequence of the presence of marginal operators in the spectrum of a
conformal field theory \cite{Card86a}.  In the present model there are several
such operators: first, the fact that charge and spin excitations are described
by bosons which are dual to each other, $R_c^2+R_s^2=2$, gives rise to
composite operators which have scaling dimension (\ref{dimsC}) $X=2$.  As has
been discussed for phases A1 and A2 above, such operators are also present in
the low energy sector of the model for $b<\half$ where no signs of logarithmic
corrections to scaling have been observed.  Here, however, the mode with
vanishing coupling constant appears together with a compact self-dual boson
with $(R_3)^2=2$ independent of the deformation parameter $\gamma$.  The
presence of these two modes, which are not related to the physical $U(1)$
symmetries of the $U_q[sl(2|1)]$ mixed superspin chain, may be responsible for
both the massive degeneracies of the low-lying levels in the thermodynamic
limit and the presence of logarithmic corrections in the finite-size spectrum
for large but finite systems.

An alternative scenario has been put forward by Ikhlef \emph{et al.}
\cite{IkJS08} in the context of the ferromagnetic staggered six-vertex model:
based on numerical evidence and supported by arguments from an RG analysis of
a supersphere sigma model they interpret the finite-size spectrum
(\ref{dimsBxxz}) as a signature of the emergence of a non-compact boson in
the effective field theory of the model.  As a consequence the coupling
constant should approach its long-distance value as
\begin{equation}
  \label{fslog}
  K(L) \simeq A\,[B+\ln L]^{-p} \,
\end{equation}
where $A$ and $B$ are functions of the system parameters and the exponent can
take values $p=1,2$.  The numerical data of Ref.~\onlinecite{IkJS08} for the
ferromagnetic staggered XXZ model with $\gamma<\pi/2$ and $b=\pi/4\gamma$ as
well as our own results are consistent with this proposal for $p=2$ and
$A(\gamma)=5\gamma/(\pi-2\gamma)$.  Eq.~(\ref{fslog}) also agrees with what
has been observed in the low energy spectrum of the antiferromagnetic
superspin chain based on alternating quark and antiquark representations of
$U_q[sl(2|1)]$ \cite{EsFS05,FrMa11}, although the amplitude $A$ displays a
different $\gamma$-dependence in that model.

\subsection{Phase B: critical theory for $\half<b<\pi/4\gamma$}
As mentioned above, the fact that the bare densities $\rho_0(\lambda)$ of the
$(1,+)$-strings (\ref{baredB}) are not positive definite for
$\half<b<\pi/4\gamma$ requires special attention.  To find the ground state
within the configurations (\ref{confB}) the energy density has to be
minimized by varying the support of the root densities $\rho^{(a)}$ (or
$\sigma^{(a)}$).  In the integral equations (\ref{iglrhoB}) this amounts to
changing the boundaries of integration to finite values.

\begin{figure}
\includegraphics[width=0.75\textwidth]{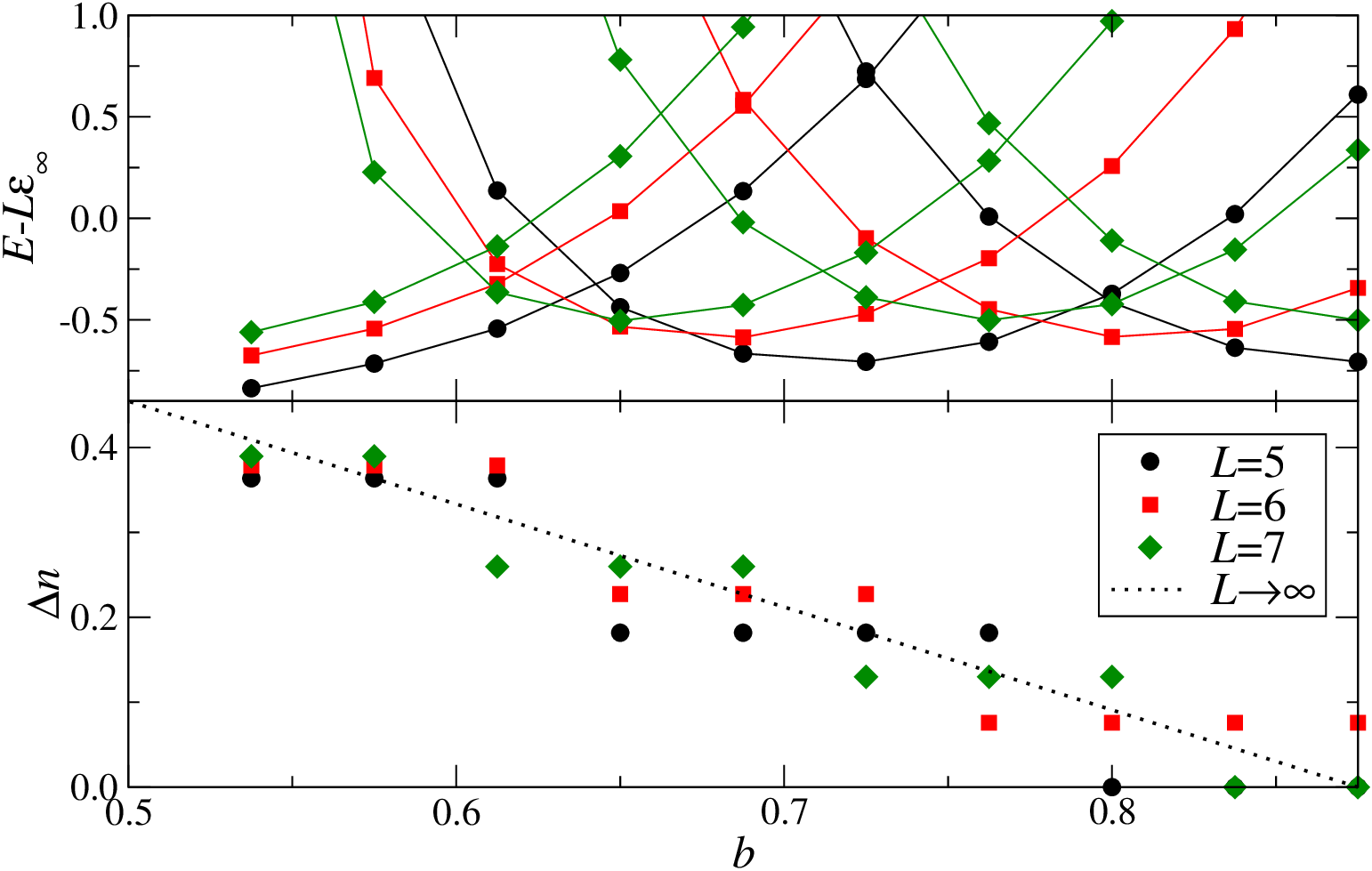}
\caption{Crossings of lowest levels in the $(N_1,N_2)=(L+1,L-1)$ sectors of
  the mixed superspin chain(upper panel) and differences in
  ground state string densities $\Delta n = (N_a^+-N_a^-)/2L$ (lower panel) as
  function of the staggering parameter $b$ for $L=5,6,7$.  The dotted line is
  the result (\ref{denstotB}) from the thermodynamic limit.
  \label{fig:Bcross}}
\end{figure}

Here we argue, however, that (\ref{densB}) does in fact describe the true
thermodynamic ground state of the mixed superspin chain based on the following
observations:
\begin{enumerate}
\item solving the integral equations (\ref{iglrhoB}) with finite boundaries
  numerically we find that (\ref{einfB}) is a lower bound for the
  energy density.
\item the ground state energies obtained by exact diagonalization of the
  superspin chain Hamiltonian up to $L=4$, i.e.\ $8$ sites, are reproduced
  from the Bethe ansatz.  The root configurations are consistent with the
  thermodynamic results.
\item the level crossings among the low energy states obtained from the Bethe
  ansatz for configurations with given numbers $N_a^\pm$ of roots support the
  result (\ref{denstotB}), see Fig.\ref{fig:Bcross}.
\end{enumerate}
%

Based on these findings we conjecture that the critical exponents are given by
(\ref{dimsB}) throughout the B phase.  As $b$ is varied away from the
self-dual line $\gamma b=\pi/4$, however, the numbers $n_a^\pm$ have to be
measured relatively to the state with densities given by (\ref{denstotB}),
i.e.\
\begin{equation}
\label{spinB}
  n_a^+ = L\left(\frac{\pi-\gamma(2b+1)}{\pi-2\gamma}\right) - N_a^+\,,
  \quad
  n_a^- = L\left(\frac{\gamma(2b-1)}{\pi-2\gamma}\right) - N_a^-\,.
\end{equation}
The selection rule (\ref{vortB}) for the vorticities $m_a^\pm$ remains
unchanged.  The quantum numbers $n_a^\pm$, on the other hand, take values
depending on the staggering $b$.
In the scaling dimensions (\ref{dimsB}), however, this affects \emph{only}
the contribution proportional to $K(L)$: according to (\ref{spinB}) the
numbers of spin and charge excitations, i.e.\ $n_1^++n_1^-+n_2^++n_2^-$ and
$n_1^++n_1^--n_2^+-n_2^-$, as well as the occupation number of the third
compact boson $n_1^+-n_1^--n_2^++n_2^-$ continue to be integers.

\begin{figure}
\includegraphics[width=0.75\textwidth]{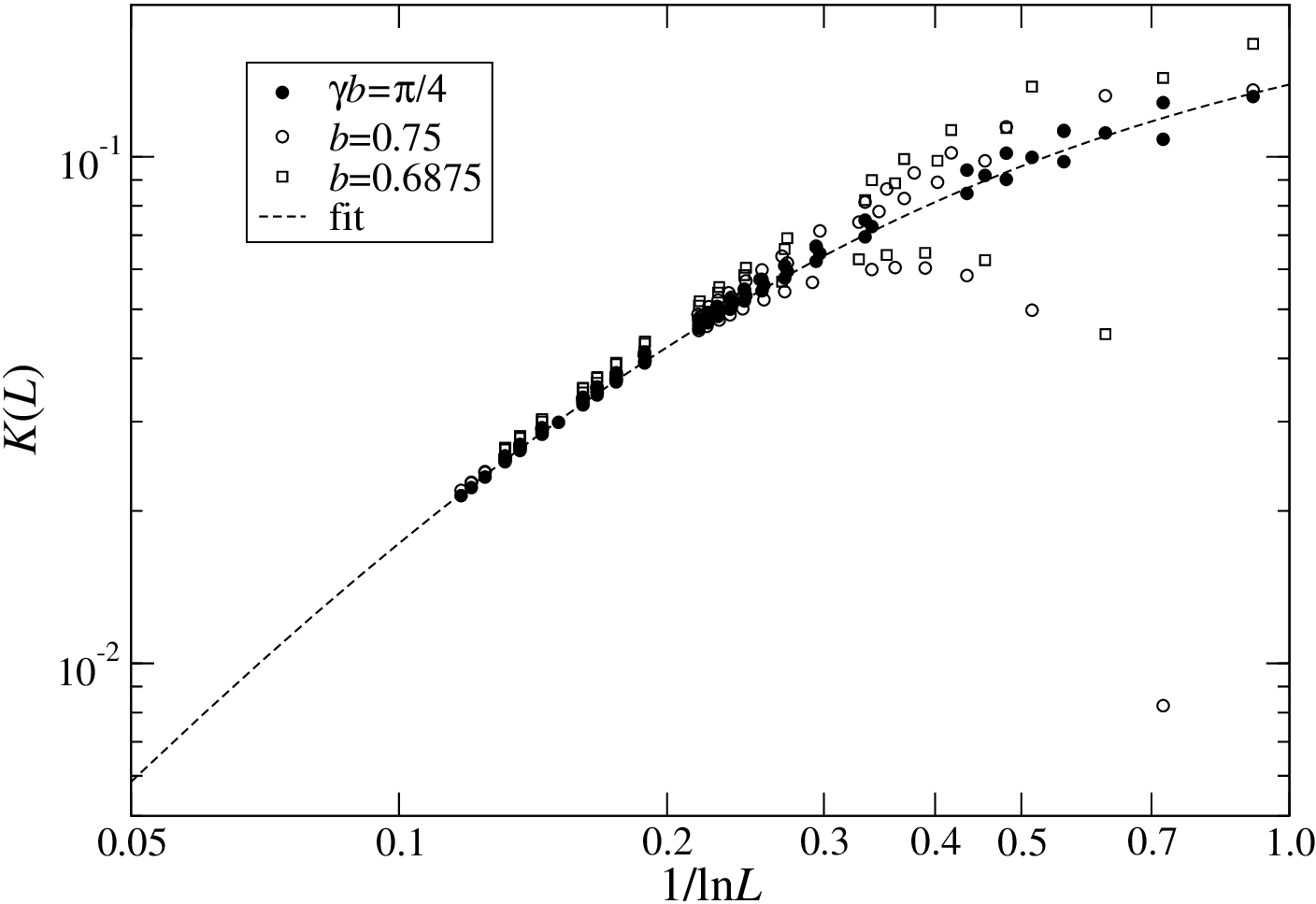}
\caption{ Coupling constant $K(L)$ extracted from the
  finite-size behaviour (\ref{dimsB}) of the lowest states in the
  $(N_1,N_2)=(L+1,L-1)$ sectors of the mixed superspin chain for
  $\gamma=2\pi/7$ and $b=0.75$ and $0.6875$.
  The dashed line is a fit of (\ref{fslog}) with $p=2$ and
  $A(\gamma)=5\gamma/(\pi-2\gamma)$ to the data for the self-dual case
  $b=\pi/4\gamma$. 
  \label{fig:Belogb}}
\end{figure}

To verify this conjecture we have studied the finite-size scaling of the
ground states for $\gamma=2\pi/7$ and $b=(\pi+\gamma)/6\gamma=3/4$ and
$b=(\pi+2\gamma)/8\gamma = 11/16$.  As before one of the ground states is in
the $(N_1,N_2)=(L+1,L-1)$ 
sector but according to (\ref{denstotB}) the ratios $N_a^+/N_a^-$ are $2$ and
$3$, respectively.  Depending on the system size $L$ the root configurations
and corresponding conformal dimensions according to (\ref{dimsB}) are for
$b=3/4$
\begin{center}
\begin{tabular}{|cccc|cc|c|}
\hline
  $N_1^+$ & $N_1^-$ & $N_2^+$ &  $N_2^-$ & $\mathbf{n}$ & $\mathbf{m}$ & 
             $X_{\mathbf{n}}^{\mathbf{m}}$ \\
\hline
 ~$(2L+3)/3$~ & ~$L/3$~ & ~$2L/3$~ & ~$(L-3)/3$~ & 
             $(-1,0,0, 1)$ & $(0,0,0,0)$ &
             $\half - \frac{\gamma}{\pi}+K(L)$ \\
  $(2L+1)/3$ & $(L+2)/3$~ & ~$(2L-2)/3$~ & ~$(L-1)/3$~ & 
             $(-\frac{1}{3},-\frac{2}{3},\frac{2}{3},\frac{1}{3})$ & $(0,0,0,0)$ &
             $\half - \frac{\gamma}{\pi}+\frac{1}{9}K(L)$ \\
  $(2L+2)/3$ & $(L+1)/3$~ & ~$(2L-1)/3$~ & ~$(L-2)/3$~ & 
             $(-\frac{2}{3},-\frac{1}{3},\frac{1}{3},\frac{2}{3})$ & $(0,0,0,0)$ &
             $\half - \frac{\gamma}{\pi}+\frac{1}{9}K(L)$ \\
\hline
\end{tabular}
\end{center}
and for $b=11/16$:
\begin{center}
\begin{tabular}{|cccc|cc|c|}
\hline
  $N_1^+$ & $N_1^-$ & $N_2^+$ &  $N_2^-$ & $\mathbf{n}$ & $\mathbf{m}$ & 
             $X_{\mathbf{n}}^{\mathbf{m}}$ \\
\hline
 ~$(3L+4)/4$~ & ~$L/4$~ & ~$3L/4$~ & ~$(L-4)/4$~ & 
             $(-1,0,0, 1)$ & $(0,0,0,0)$ &
             $\half - \frac{\gamma}{\pi}+K(L)$ \\
 $(3L+1)/4$ & ~$(L+3)/4$~ & $(3L-3)/4$ & $(L-1)/4$ & 
             $(-\frac{1}{4},-\frac{3}{4},\frac{3}{4}, \frac{1}{4})$ & $(0,0,0,0)$ &
             $\half - \frac{\gamma}{\pi}+\frac{1}{4}K(L)$ \\
 $(3L+2)/4$ & ~$(L+2)/4$~ & $(3L-2)/4$ & $(L-2)/4$ & 
             $(-\half,-\half,\half,\half)$ & $(0,0,0,0)$ &
             $\half - \frac{\gamma}{\pi}$ \\
 $(3L+3)/4$ & ~$(L+1)/4$~ & $(3L-1)/4$ & $(L-3)/4$ & 
             $(-\frac{3}{4},-\frac{1}{4},\frac{1}{4}, \frac{3}{4})$ & $(0,0,0,0)$ &
             $\half - \frac{\gamma}{\pi}+\frac{1}{4}K(L)$ \\
\hline
\end{tabular}
\end{center}
In Figure~\ref{fig:Belogb} our results for the coupling constant $K(L)$ as
extracted from the numerical data for these parameters are presented together
with the ones obtained from the ground state energies of self-dual model (see
Fig.~\ref{fig:dimBPotts}) and a fit of the conjecture (\ref{fslog}) with
$p=2$ and $A(\gamma)=5\gamma/(\pi-2\gamma)$: from these data we conclude that
the amplitude $A(\gamma,b)$ of the coupling constant is independent of the
staggering parameter $b$.

This would imply that the parameter $b$ is irrelevant as far as the
compactified bosonic degrees of freedom are concerned, while it acts as some
kind of twist in the mode with vanishing coupling constant.

Approaching the limit $b\to\half+0$ from above the densities
$\sigma^{(a)}(\lambda)$ of $(1,-)$ strings vanish as does the corresponding
Fermi velocity.  On the line $b=\half$ the critical theory has central charge
$c=-1$ with the operator content given in Section \ref{sec:mix-fm} and also in
Ref.~\onlinecite{FrMa11} before.

\section{The \emph{anti-ferro}magnetic mixed superspin chain \newline on the
  self-dual line}
\label{sec:phaseC}
Our numerical analysis of the Hamiltonian (\ref{Hmix}) for small lattice sizes
indicates that the ground state of the anti-ferromagnetic superspin chain sits
in the sectors $(N_1,N_2)= (L\pm1,L\pm1)$ corresponding to charge zero and
spin $S_3=\pm1$ for parameters $0<\gamma<\pi/2$ and $\gamma b=\pi/4$, i.e. on
the self-dual line.  For $L$ odd the spectrum displays this double degeneracy
while we observe a four-fold degenerate ground state for $L$ even.
For these parameters we have identified the root configurations corresponding
to low lying states solving the Bethe equations (\ref{baebq}): apart from
strings (\ref{stringXXZ}) we find roots located on the lines $Im(\lambda)= \pm
\gamma b$ which in fact dominate the low energy states.  
As an example, the ground state configuration of the anti-ferromagnetic mixed
superspin chain on the self-dual line is given in terms of composites
combining roots from both levels of the Bethe ansatz as
\begin{equation}
  \label{bstrings}
  \lambda_m^{(1)} =  \left(\lambda_m^{(2)}\right)^* 
  = \lambda_m \pm i\gamma b\,, 
\end{equation}
with real center $\lambda_m\in\mathbb{R}$.  
%

These configurations continue to exist away from the self-dual line.  It turns
out, however, that for finite chains and generic $\half< b<\pi/4\gamma$ the
imaginary parts of these roots deviate from the asymptotic values $\pm\gamma
b$ by corrections which vanish as the system size $L\to\infty$.  In the limit
$b\to\half$ the '$b$-strings' (\ref{bstrings}) turn into the strange 2-string
configurations describing the low-lying states of the anti-ferromagnetic
$3\otimes\bar{3}$ superspin chain before \cite{EsFS05,FrMa11}.
Away from $b=\half$, however, the numerical data indicate that these states
are not in the low energy sector of the model any more.  Instead there appears
to be a crossover to different $b$-dependent ground states.

Therefore we are going to concentrate our investigation of the
anti-ferromagnetic superspin chain (\ref{Hmix}) for $b>\half$ on the self-dual
case: here the additional root configurations discussed above become exact,
i.e.\ with imaginary parts $\equiv\pm i\pi/4$ independent of the system size
$L$.  This allows to describe most of the low energy states by the structure
\begin{equation}
\label{confC}
\left\{\lambda_j^{(a)} \right\}_{j=1}^{N_a} \equiv 
\left\{ \mu_j^{(a,+)} - i\frac{\pi}{4} \right\}_{j=1}^{N_a^+} \cup 
\left\{ \mu_j^{(a,-)} + i\frac{\pi}{4} \right\}_{j=1}^{N_a^-}  
\,,\quad a=1,2,
\quad\mathrm{for}~~b\gamma=\pi/4
\end{equation}
where $\mu_j^{(a,\pm)} \in \mathbb{R} $ and $N_a=N_a^+ + N_a^{-}$ for $a=1,2$.
We shall now use this fact to employ the analytical tools of the previous
sections to start our investigations of the thermodynamic limit and
finite-size properties of the anti-ferromagnetic mixed superspin chain on the
self-dual line.  It is convenient to shift the rapidities $\mu_j^{(a,\pm)}$ as
follows
\begin{equation}
\label{shiftC}
\mu_j^{(a,\pm)}= \xi_j^{(a,\pm)} +i \pi/4\,,\quad a=1,2 \,.  
\end{equation}
Substituting (\ref{shiftC}) into the Bethe equations (\ref{baebq}) and taking
their logarithms we find that the resulting equations for $\xi_j^{(a,\pm)}$
are
\begin{equation}
\label{baelogC}
\begin{aligned}
L\Phi(2\xi_j^{(1,\pm)},\gamma) 
&=2\pi Q^{(1,\pm)}_j
  + \sum_{k=1}^{N_2^{\pm}} \Phi(\xi_j^{(1,\pm)}-\xi_k^{(2,\pm)},\gamma)
  - \sum_{k=1}^{N_2^{\mp}} \Psi(\lambda_j^{(1,\pm)}-\mu_k^{(2,\mp)},\gamma)\,,
\quad j=1,\ldots,N_1^{\pm}\,,
\\
L\Phi(2\xi_j^{(2,\pm)},\gamma) 
&=2\pi Q^{(2,\pm)}_j
  + \sum_{k=1}^{N_1^{\pm}} \Phi(\xi_j^{(2,\pm)}-\xi_k^{(1,\pm)},\gamma)
  - \sum_{k=1}^{N_1^{\mp}} \Psi(\lambda_j^{(2,\pm)}-\mu_k^{(1,\mp)},\gamma)\,,
\quad j=1,\ldots,N_2^{\pm}\,.
\end{aligned}
\end{equation}
The functions $\Phi(x,y)$ and $\Psi(x,y)$ have been defined in
Eqs.~(\ref{logPhi}) and (\ref{logPsi}) above.
The quantum numbers $Q_j^{(a,\pm)}$ define the many possible branches of the
logarithm and uniquely characterize an eigenstate of the system.  They have to
be chosen integer or half-odd integer according to the parities of the numbers
$N_a^\pm$:
\begin{equation}
\label{qnosC}
\begin{aligned}
 &Q_j^{(1,\pm)} \equiv \frac{L+N_{2}^{\pm}}{2} \mod 1 \,, \quad 
 &Q_j^{(2,\pm)} \equiv \frac{L+N_{1}^{\pm}}{2} \mod 1 \,. \quad 
\end{aligned}
\end{equation}
Note that as in phase A2 there exist root configurations with one or more
roots located at $\pm\infty$ (see also Appendix~\ref{app-2site}).  To deal
with such configurations one can follow the above procedure to obtain
equations for the finite roots \emph{after} taking into account the infinite
one(s) explicitely.

\subsection{Thermodynamic limit}
As before we begin our analysis of the thermodynamic limit by introducing
counting functions for the rapidities $\xi_j^{(a,\pm)}$
\begin{equation}
\label{countC}
\begin{aligned}
z^{(1,\pm)}(\xi) &= 
      \frac{1}{2}\Phi(2\xi,\gamma)
      - \frac{1}{2L}\sum_{k=1}^{N_2^{\pm}} \Phi(\xi-\xi_k^{(2,\pm)},\gamma)
      + \frac{1}{2L}\sum_{k=1}^{N_2^{\mp}} \Psi(\xi-\xi_k^{(2,\mp)},\gamma)\,,\\
z^{(2,\pm)}(\lambda) &= 
      \frac{1}{2}\Phi(2\xi,\gamma)
      - \frac{1}{2L}\sum_{k=1}^{N_1^{\pm}} \Phi(\xi-\xi_k^{(1,\pm)},\gamma)
      + \frac{1}{2L}\sum_{k=1}^{N_1^{\mp}} \Psi(\xi-\xi_k^{(1,\mp)},\gamma)\,,\\
\end{aligned}
\end{equation}
When $L \to \infty $ the roots $\xi_j^{(a,\pm)}$ tend towards a continuous
distribution and the derivatives of the counting functions define the
corresponding densities of particles $\sigma^{(a,\pm)}(\xi)$ and holes
$\sigma_h^{(a,\pm)}(\xi)$, namely
\begin{equation}
\begin{aligned}
\label{densiC}
2\pi \left( \sigma^{(a,\pm)}(\xi) + \sigma_h^{(a,\pm)}(\xi)\right)
&= \frac{dz^{(a,\pm)}(\xi)}{d\xi}\,.
\end{aligned}
\end{equation}
For the ground state, the counting functions are dominated by the particle
densities and Eqs.~(\ref{baelogC}) turn into the following coupled integral
relations for the densities $\sigma^{(a,\pm)(\xi)}$,
\begin{equation}
\label{iglrhoC}
  \left(
   \begin{array}{c} \sigma^{(1,+)}(\xi)\\\sigma^{(1,-)}(\xi)\\
                    \sigma^{(2,+)}(\xi)\\\sigma^{(2,-)}(\xi)\end{array}
      \right) = \frac{1}{4\pi}
  \left(
   \begin{array}{c} \Phi'(2\xi,\gamma)\\\Phi'(2\xi,\gamma)\\
                    \Phi'(2\xi,\gamma)\\\Phi'(2\xi,\gamma) 
\end{array} \right)   
   - \int_{-\infty}^{\infty}\mathrm{d}\mu\,
 \mathbb{K}(\xi-\mu)
  \left(
   \begin{array}{c} \sigma^{(1,+)}(\mu)\\\sigma^{(1,-)}(\mu)\\
                    \sigma^{(2,+)}(\mu)\\\sigma^{(2,-)}(\mu)\end{array}
      \right)
\end{equation}
where the kernel matrix $\mathbb{K}(\lambda)$ has been defined in
Eq.~(\ref{kernelB}) above.
As a consequence of the fact that Eqs.~(\ref{iglrhoC}) are invariant under the
exchanges $1 \leftrightarrow 2$ and $+ \leftrightarrow -$ all the densities
$\sigma^{(a,\pm)}(\xi)$ are the same.  The remaining scalar integral equation
can be solved by standard Fourier methods giving
\begin{equation}
  \label{densCPotts}
  \sigma^{(a,\pm)}(\xi)= \frac{1}{4 \gamma \cosh(\frac{\pi}{\gamma} \xi)}\,,
\quad a=1,2\,.
\end{equation}

Using (\ref{densCPotts}) in (\ref{Emix}) we obtain the energy density of this
eigenstate of the anti-ferromagnetic superspin chain on the self-dual line
$\gamma b= \pi/4$
\begin{equation}
  \label{einfCPotts}
  \epsilon_\infty^{(C)}= 
  -2\int_{-\infty}^\infty\mathrm{d}\omega\,
    \frac{\sinh((\pi-2\gamma)\omega/4)}{
          \sinh(\pi\omega/4)\cosh(\gamma \omega/2)} \,.
\end{equation}
Starting from this state we find that there are four branches of low energy
excitations with linear dispersion.  Their Fermi velocity is
\begin{equation}
\label{fermivC}
  v_{C} = \frac{\pi}{\gamma}\,.
\end{equation}

\subsection{Critical theory of the self-dual anti-ferromagnetic superspin
  chain}
\subsubsection*{Finite-size spectrum}
Similar as for phase B we can employ the standard techniques to compute the
finite-size scaling behaviour of the low-energy excitations and find that the
energy gaps are given by
\begin{equation}
  \label{fseC}
  E(L,\gamma)-L\varepsilon_\infty^{(C)}(\gamma)
  = \frac{2\pi v_{C}}{L}\left[ -\frac{1}{3}
    + X_{n_1^+,n_1^-,n_2^+,n_2^-}^{m_1^+,m_1^-,m_2^+,m_2^-}(\gamma) \right] + o(L^{-1})\,.
\end{equation}
where the scaling dimensions 
$X_{n_1^+,n_1^-,n_2^+,n_2^-}^{m_1^+,m_1^-,m_2^+,m_2^-}(\gamma)$ associated to the
four possible gapless modes are  given by,
\begin{equation}
  \label{dimsC}
  \begin{aligned}
  X_{n_1^+,n_1^-,n_2^+,n_2^-}^{m_1^+,m_1^-,m_2^+,m_2^-}(\gamma) &=
    \frac{\gamma}{4\pi} \left(n_1^++n_1^--n_2^+-n_2^-\right)^2
  + \frac{\pi}{16\gamma} \left(m_1^++m_1^--m_2^+-m_2^-\right)^2\\
&  + \frac{\pi-2\gamma}{8\pi} \left(n_1^++n_1^-+n_2^++n_2^-\right)^2
  + \frac{\pi}{8(\pi-2\gamma)} \left(m_1^++m_1^-+m_2^++m_2^-\right)^2\\
& +  \frac{1}{8}\left(n_1^+-n_1^-+n_2^+-n_2^-\right)^2
  + \frac{1}{8}\left(m_1^+-m_1^-+m_2^+-m_2^-\right)^2\\
& + \frac{K(L)}{8} \left(n_1^+-n_1^--n_2^++n_2^-\right)^2
  + \frac{1}{8K(L)} \left(m_1^+-m_1^--m_2^++m_2^-\right)^2\,.
 \end{aligned}
\end{equation}
Excitations are labeled again by charge indices $n_a^{\pm}=L/2-N_a^{\pm}$,
$a=1,2$ and corresponding vorticities $m_a^{\pm}$.  From the constraint
(\ref{qnosC}) that they are linked to the parities of $n_a^{\pm}$ by
\begin{equation}
\label{vortC}
  m_a^\pm = \half\left(n_1^\pm+n_2^\pm+1\right)\mod 1\,,\quad
  a=1,2\,.
\end{equation}
Similar to the discussion of the finite-size spectrum in phase B three of the
elementary critical modes contributing to (\ref{dimsC}) can be identified as
free bosons: the modes corresponding to spin and charge excitations have
compactification radii depending on the deformation parameter as $R_c^2=
4\gamma/\pi = 2-R_s^2$, i.e.\ they are interchanged as compared to what has
been found in the ferromagnetic regime B.  As in that phase there is a third
compact boson which cannot be identified with the $U(1)$ charges of the
superspin chain, its radius takes the self-dual value $(R_3)^2= 2$ independent
of $\gamma$.
Again, the finite-size behaviour of the fourth mode cannot be studied within
this approach: as a consequence of the singular kernel of the Bethe ansatz
integral equations one only finds that the corresponding coupling constant
$K(L)$ vanishes in the thermodynamic limit.  This implies that states with
$n_1^+-n_1^--n_2^++n_2^-=0$ are degenerate to order $L^{-1}$ and only states
with $m_1^+-m_1^--m_2^++m_2^-=0$ appear in the low energy spectrum.

A preliminary verification of the above proposal (\ref{dimsC}) can be easily
done by considering the subsector associated to the staggered six-vertex
model.  In this case we have the same charge and vorticity indices for both
levels $a=1,2$.  Taking into account the relation (\ref{ener-map}) between the
spectra of the mixed superspin chain and the staggered XXZ model and setting
$n_1^{\pm}=n_2^{\pm}=n^{\pm}$, $m_1^{\pm}=m_2^{\pm}=m^{\pm}$ in (\ref{dimsC})
we find
\begin{equation}
  \label{dimsCxxz}
  \begin{aligned}
  E^{(6v)}(L,\gamma)&-\frac{L}{2}\varepsilon_\infty^{(C)}(\gamma) =
    \frac{2\pi v_{C}}{L}\left[ -\frac{1}{6}
      + \widetilde{X}_{n^+,n^-}^{m^+,m^-}(\gamma) \right] + o(L^{-1})\,,\\
  \widetilde{X}_{n^+,n^-}^{m^+,m^-}(\gamma) &= \frac{(\pi-2\gamma)}{4\pi}
    \left(n^++n^-\right)^2 + \frac{\pi}{4(\pi-2\gamma)} \left(m^++m^-\right)^2\\
    &\quad+ \frac{1}{4} \left(n^+-n^-\right)^2 + \frac{1}{4}
    \left(m^+-m^-\right)^2\,.
 \end{aligned}
\end{equation}
From (\ref{dimsCxxz}) we see that the modes associated to the amplitude
$K(L)$ do not contribute to the spectrum of the staggered six-vertex model.
The conformal dimensions $\widetilde{X}_{n^+,n^-}^{m^+,m^-}(\gamma)$ are in
accordance with those proposed recently in Ref.~\onlinecite{Ikhl11}.

\subsubsection*{Numerical results}
\begin{figure}
\includegraphics[width=0.75\textwidth]{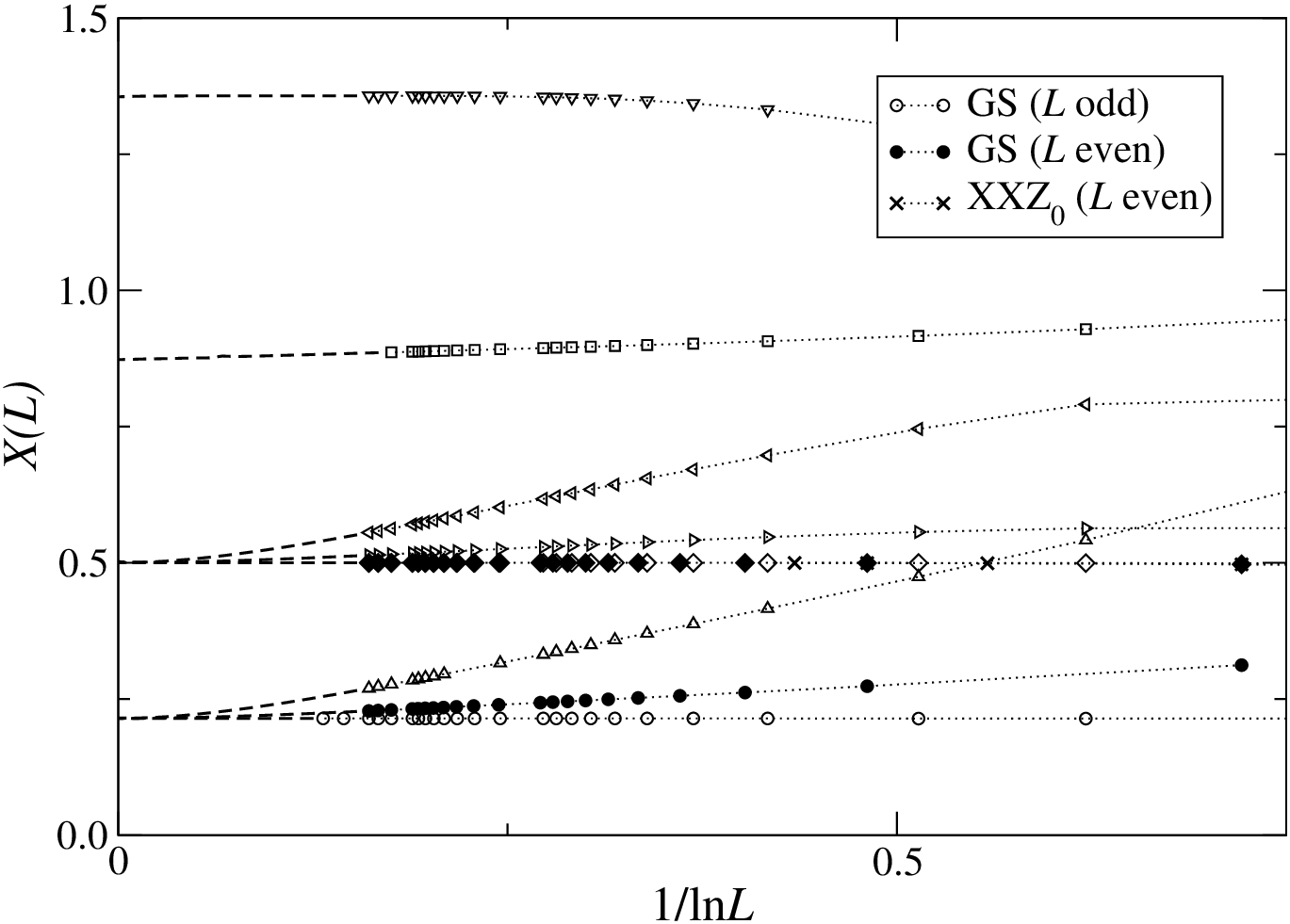}
\caption{Finite-size estimators $X(L) = L(E(L)-L\varepsilon_\infty^{(C)})
  /(2\pi v_{C}) + 1/3$ for the scaling dimensions of some low lying states
  of the anti-ferromagnetic mixed superspin chain for $\gamma=2\pi/7$ on the
  self-dual line $\gamma b=\pi/4$.  Dimensions appearing for odd (even) $L$
  are shown as open (filled) symbols.  The scaling dimensions of the lowest
  states extrapolate to $\half - \frac{\gamma}{\pi}=\frac{3}{14}$ and $\half$
  but differ by multiples of $K(L)$.  The other levels displayed in the figure
  extrapolate to $\frac{\pi}{4\gamma}= \frac{7}{8}$ and
  $\frac{5}{2}-\frac{4\gamma}{\pi}=\frac{19}{14}$, 
  respectively. Dashed lines are rational function
  extrapolations of the numerical data to $L\to\infty$.
  \label{fig:dimCPotts}}
\end{figure}

From the exact diagonalization of the mixed superspin chain for small $L$ we
have identified the ground state to be located in the sectors
$(N_1,N_2)=(L\pm1,L\pm1)$.  Following the selection rules for the $n_a^\pm$,
$m_a^\pm$ given above, the lowest level is given by
$\mathbf{n}\equiv(n_1^+,n_1^-,n_2^+,n_2^-) = (\half,\half,\half,\half)$ for
$L$ odd and $\mathbf{n}= (0,1,1,0)$ for $L$ even while
$\mathbf{m}\equiv(m_1^+,m_1^-,m_2^+,m_2^-) =(0,0,0,0)$ in both cases.
According to (\ref{fseC}) the ground state energy scales as
\begin{equation}
\label{gsC}
\begin{aligned}
  E_0(L,\gamma) -L\varepsilon_\infty^{(C)}(\gamma)
  &= -\frac{\pi v_{C}}{6L}\left[ 2\frac{6\gamma-\pi}{\pi} \right] +
  o(L^{-1})\,, \quad\mathrm{for~}L\mathrm{~odd}\,\\
  &= -\frac{\pi v_{C}}{6L}\left[ 2\frac{6\gamma-\pi}{\pi}\right]
     +\frac{2\pi v_{C}}{L}\left[\half K(L)\right] +
  o(L^{-1})\,, \quad\mathrm{for~}L\mathrm{~even}\,.
\end{aligned}
\end{equation}
Note that the ground state of the odd $L$ superspin chains is also in the XXZ
subsector of this model.
As mentioned above, $K(L)\to0$ for $L\to\infty$, therefore we find an
effective central charge $c_\mathrm{eff}= 2(6\gamma-\pi)/\pi$ in both cases.
Upon fine-tuning of the deformation parameter to $\gamma=\pi/6$ the ground
state energy of the mixed superspin chain is exactly
$L\varepsilon_\infty^{(C)}(\gamma)$ without subleading corrections, i.e.\
$c_\mathrm{eff}=0$.
It turns out that there is a family of states in this sector with finite-size
gap to the ground states (\ref{gsC}) vanishing as multiples of $K(L)$: the
lowest energy in the sector $N_1^+=N_2^-=(L-1+\ell)/2$ and
$N_1^+=N_2^-=(L-1-\ell)/2$ with $\ell=0,\pm1,\pm2,\ldots$ is degenerate with
the ground state (note that $N_a^\pm$ are integers, therefore only
configurations with even (odd) $\ell$ can be realized for odd (even) $L$).  In
Figure \ref{fig:dimCPotts} the splitting of the levels with
\begin{center}
\begin{tabular}{|cccc|cc|c|}
\hline
  $N_1^+$ & $N_1^-$ & $N_2^+$ &  $N_2^-$ & $\mathbf{n}$ & $\mathbf{m}$ & 
             $X_{\mathbf{n}}^{\mathbf{m}}$ \\
\hline
  ~$(L-1)/2$~ & ~$(L-1)/2$~ & ~$(L-1)/2$~ & ~$(L-1)/2$~ & 
             $(\half,\half, \half, \half)$ & $(0,0,0,0)$ &
             $\half - \frac{\gamma}{\pi}$ \\
  $L/2$ & $(L-2)/2$ & $(L-2)/2$ & $L/2$ & 
             $(0,1,1,0)$ &  $(0,0,0,0)$ &
             $\half - \frac{\gamma}{\pi} +\half K(L) $ \\
  $(L+1)/2$ & $(L-3)/2$ & $(L-3)/2$ & $(L+1)/2$ & 
             $(-\half,\frac{3}{2},\frac{3}{2},-\half)$ &  $(0,0,0,0)$ &
             $\half - \frac{\gamma}{\pi} +2K(L) $\\
\hline
\end{tabular}
\end{center}
as a function of the system size is shown for $\gamma=2\pi/7$, $\gamma
b=\pi/4$.
The gaps between the lowest levels vanish for $L\to\infty$, similarly as has
been found in phase B above.
Based on the proposal (\ref{fslog}) we can analyze the behaviour of the
coupling constant $K(L)$ appearing in the scaling dimensions of the mixed
superspin chain.
We have used the numerical data for the ground state energy of the even $L$
chains to extract $K(L)$.  Comparing $K(L)$ with $K(L/2)$ we obtain estimates
for the amplitude $A(\gamma)$ for various values of the deformation parameter.
As shown in Table \ref{tab:logestC} the proposal (\ref{fslog}) does describe
the data quite well with
\begin{equation}
  \label{ampC}
  A(\gamma,b=\pi/4\gamma) = 5\,\frac{\pi-2\gamma}{4\gamma}\,.
\end{equation}

A second group of states with scaling dimensions extrapolating to
$X(\gamma)=\half$ independent of $\gamma$ presented in Figure
\ref{fig:dimCPotts} is given by the following configurations:
\begin{center}
\begin{tabular}{|cccc|cc|c|}
\hline
  $N_1^+$ & $N_1^-$ & $N_2^+$ &  $N_2^-$ & $\mathbf{n}$ & $\mathbf{m}$ & 
             $X_{\mathbf{n}}^{\mathbf{m}}$ \\
\hline
  $L/2$ & $L/2$ & $(L-2)/2$ & $(L-2)/2$ & 
             $(0,0,1,1)$ &  $(0,0,0,0)$ &
             $\half$\\
  $L/2$ & $L/2$ & $L/2$ & $L/2$ & 
             $(0,0,0,0)$ & $(\half,-\half,\half,-\half)$ &
             $\half$ \\
  ~$(L+1)/2$~ & ~$(L-1)/2$~ & ~$(L-3)/2$~ & ~$(L-1)/2$~ & 
             $(-\half,\half, \frac{3}{2}, \half)$ & $(0,0,0,0)$ &
             $\half + \half K(L)$ \\
\hline
\end{tabular}


\end{center}
Among these, the state corresponding to
$X_{(0,0,0,0)}^{(\half,-\half,\half,-\half)}(\gamma)=\half$ is the lowest
state in the XXZ subsector of the model for even $L$, it has spin $S_3=0$.

For the mixed superspin chain two more states with this finite-size behaviour
appear.  They are given by configurations involving two infinite roots
$\lambda^{(1)}=-\lambda^{(2)}=\infty$ of the Bethe equations (\ref{baebq}) in
addition to the $N_a^\pm$ finite ones\footnote{%
  The existence of such configurations has been already established for the
  2-site system, see Appendix \ref{app-2site}, and for the low energy states
  of the system for $b<\half$.}:
\begin{center}
\begin{tabular}{|cccc|cc|c|}
\hline
  $N_1^+$ & $N_1^-$ & $N_2^+$ &  $N_2^-$ & $\mathbf{n}$ & $\mathbf{m}$ & 
             $X_{\mathbf{n}}^{\mathbf{m}}$ \\
\hline
  $(L-1)/2$ & $(L-1)/2$ & $(L-1)/2$ & $(L-1)/2$ & 
             $(0,0,0,0)$ & $(\half,-\half,\half,-\half)$ &
             $\half$\\
  $(L+1)/2$ & $(L-3)/2$ & $(L-3)/2$ & $(L+1)/2$ &
             $(-1,1,1,-1)$ & $(\half,-\half,\half,-\half)$ &
             $\half + 2 K(L)$ \\
\hline
\end{tabular}
\end{center}
The finite-size estimators of the scaling dimensions for these states at
$\gamma=2\pi/7$ are also included in Figure~\ref{fig:dimCPotts}.

We have identified two more states with larger scaling dimensions: the
configuration of the state with dimension
$X_{(0,0,0,0)}^{(\half,\half,-\half,-\half)}(\gamma)$ contains two infinite
roots $\lambda^{(1)}=-\lambda^{(2)}=\infty$, the one with dimension
$X_{(\half,\frac{3}{2},\half,\frac{3}{2})}^{(0,0,0,0)}(\gamma)$ is an
excitation within the XXZ subsector of the model.  The corresponding
configurations of finite roots are given by
\begin{center}
\begin{tabular}{|cccc|cc|c|}
\hline
  $N_1^+$ & $N_1^-$ & $N_2^+$ &  $N_2^-$ & $\mathbf{n}$ & $\mathbf{m}$ & 
             $X_{\mathbf{n}}^{\mathbf{m}}$  \\
\hline
  $~(L-1)/2~$ & $~(L-1)/2~$ & $~(L-1)/2~$ & $~(L-1)/2~$ & 
             $(0,0,0,0)$ &  $(\half,\half,-\half,-\half)$ &
             $\frac{\pi}{4\gamma}$ \\
  $~(L-1)/2~$ & $~(L-3)/2~$ & $~(L-1)/2~$ & $~(L-3)/2~$ & 
             $(\half,\frac{3}{2},\half,\frac{3}{2})$ &  $(0,0,0,0)$ &
             $\frac{5}{2} - \frac{4\gamma}{\pi}$ \\
\hline
\end{tabular}

\end{center}

\begin{figure}
\includegraphics[width=0.75\textwidth]{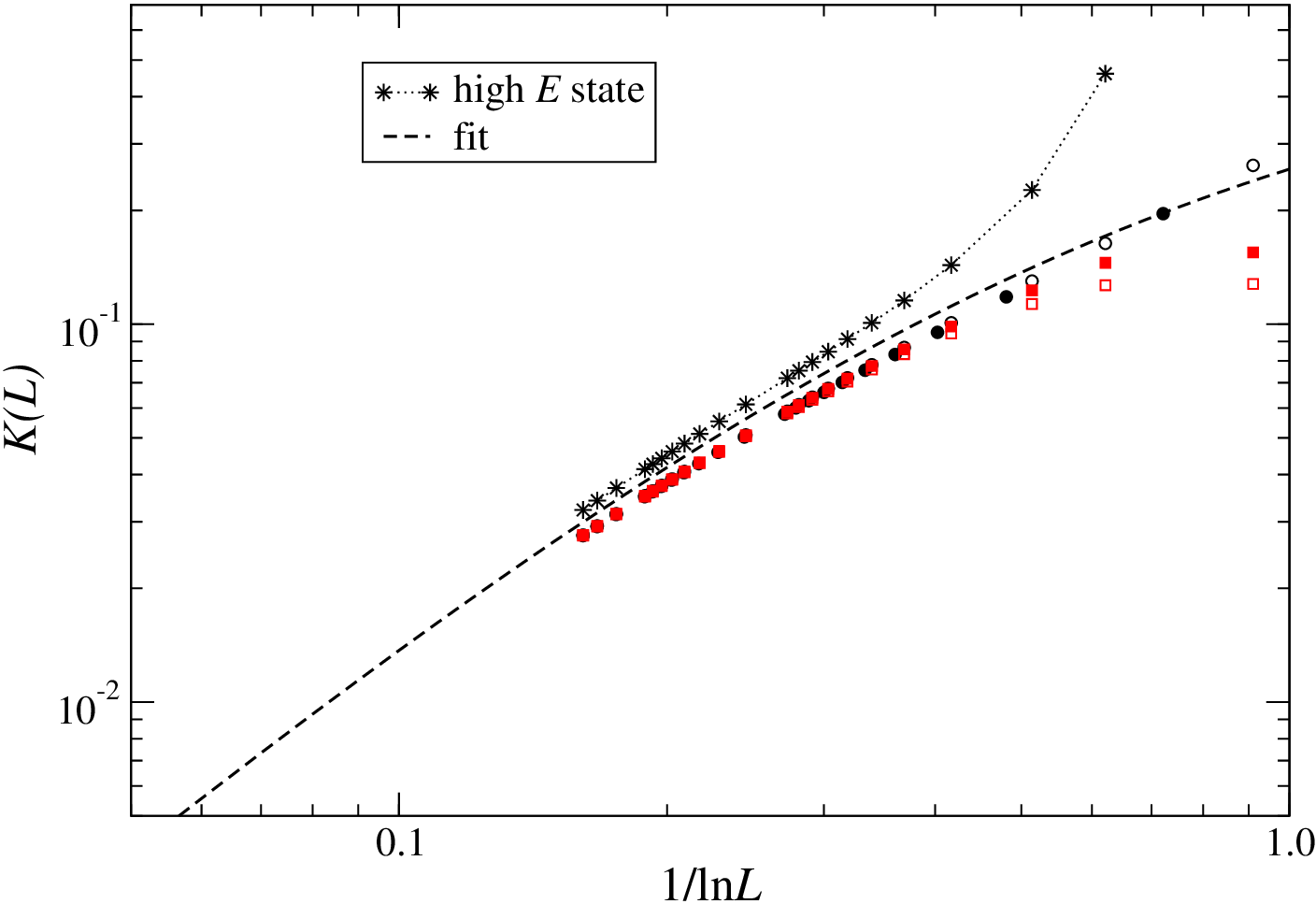}
\caption{Coupling constant $K(L)$ extracted from four lying levels of the
  finite-size spectrum with $X(\gamma)\to (\pi-2\gamma)/2\pi$ (black symbols)
  and for $X(\gamma)\to \half$ (red symbols) for $\gamma=2\pi/7$ and $b=7/8$.
  Data obtained from the high energy state with dimension (\ref{dimChighE})
  are denoted by '*'. The dashed line is a fit of (\ref{fslog}) with
  $A(\gamma=2\pi/7)=15/8$ to the numerical data.
  \label{fig:fineC}}
\end{figure}

To provide further evidence for the proposed $L$-dependence (\ref{fslog}) we
have extracted the coupling constant $K(L)$ from all gaps between the levels
in the groups extrapolating to $X(\gamma)=\half-\gamma/\pi$ and $\half$,
respectively.  In Figure \ref{fig:fineC} we present the resulting data for
$\gamma=2\pi/7$, $b=7/8$ together with a fit of (\ref{fslog}) using $p=2$ and
$A(\gamma=2\pi/7)=15/8$ according to (\ref{ampC}).
Additional support for this behaviour of the coupling constant comes from a
Bethe state with charge $B=\half$ and spin $S_3=2$ appearing in the spectrum
of the mixed superspin chain for odd $L$: choosing $N_1^+=N_1^-=N_2^+=(L-1)/2$
and $N_2^-=(L-3)/2$ and quantum numbers (\ref{qnosC})
\begin{equation}
  \begin{aligned}
    & Q_j^{(1,+)} = -\frac{L+1}{4}+j\,,\quad
      Q_j^{(1,-)} = -\frac{L+3}{4}+j\,,\\
    & Q_j^{(2,+)} = -\frac{L+1}{4}+j\,,\quad
      Q_j^{(2,-)} = -\frac{L-3}{4}+j\,,\\
  \end{aligned}
\end{equation}
($j$ takes values $1,\ldots,N_a^\pm$ in the corresponding sequences) we have
solved the Bethe equations numerically.  In the thermodynamic limit this state
disappears from the low energy spectrum.  Therefore it plays no role for the
effective field theory and it should not be expected that the finite-size
analysis employed above is applicable here.
Nevertheless the finite-size behaviour of this level can be described by
(\ref{fseC}) and (\ref{dimsC}) with a ``scaling dimension''
\begin{equation}
\label{dimChighE}
  X_{(\half,\half,\half,\frac{3}{2})}^{(0,-\half,0,\half)}(\gamma)
  = \frac{5\pi-4\gamma}{4\pi} + \frac{1}{8}K(L) +\frac{1}{8K(L)} \,.
\end{equation}
Using this prediction we can extract $K(L)$ from our numerical data for this
state.  Surprisingly, it agrees well with what has been obtained from the
logarithmic fine structure of the low-lying states before, see Figure
\ref{fig:fineC}.

\subsection{$\half\le b<\pi/4\gamma$ and connection to the $3\otimes\bar{3}$
  superspin chain}
As mentioned at the beginning of this section configurations involving roots
with $Im(\lambda_j^{(a)})\approx \pm\gamma b$ continue to exist away from the
self-dual line.  Restricting oneself to these configurations as in
(\ref{confC}), however, does not capture the low energy part of the spectrum
of the anti-ferromagnetic mixed superspin chain as $b\to\half$.  In fact,
comparing (\ref{baebq}) or (\ref{betheXXZ}) with the Bethe equations obtained
for spin chains based on general spin-$S$ representations of $U_q[sl(2)]$ the
phase factors appearing in the Bethe equations for the present model can be
identified with those arising from effective spins $(b+\half)$ and
$-(b-\half)$, respectively.  Hence, for $b>\half$ we have a system mixing
representations with different signs of the effective spin.  For such a case
an analysis of the thermodynamic limit based on some kind of string hypothesis
is not known.

For $b=\half$, however, some of the statistical weights (\ref{weiinit})
appearing in the $R$-matrix vanish.  The remaining non-degenerate subsector of
the mixed superspin chain (\ref{transmix}) at $b=\half$ coincides with the
spectrum of the antiferromagnetic superspin chain built from alternating
three-dimensional quark $[3]$ and antiquark $[\bar{3}]$ representations of
$U_q[sl(2|1)]$, see also Appendix \ref{app-2site}.  The low energy spectrum of
that model has been studied mostly in the zero charge sector, $N_1=N_2$, in
particular for root configurations consisting of the 'strange strings'
\cite{EsFS05,FrMa11}.
Under an adiabatic change of the staggering parameter from $b=\half$ to the
self-dual line these configurations evolve into the $b$-string configurations
(\ref{bstrings}) studied for the low energy states of the self-dual
anti-ferromagnetic mixed superspin chain (\ref{Hmix}) above.
As a consequence the resulting scaling dimensions of these states exhibit the
same dependence on the deformation parameter $\gamma$, i.e.\ (\ref{dimsC})
for $n_1^\pm=n_2^\mp$, although with a different amplitude of the coupling
constant $K(L)$.  The effective central charges of the two models, however,
differ: the effective central charge of the $3\otimes\bar{3}$ chain changes
from $c=0$ for $0\le\gamma<\pi/4$ to $c_{\mathrm{eff}}= 3-6(\pi-2\gamma)/\pi$
for $\pi/4<\gamma<\pi/2$.  This should be compared to the finite-size spectrum
of the present model which has $c_{\mathrm{eff}}=4-6(\pi-2\gamma)/\pi$
according to (\ref{fseC}).

This difference can be interpreted as indication that projecting out the
sector of the mixed superspin chain which degenerates as $b=\half$ removes not
only the states governing the low energy sector of the model for $b>\half$ but
also one of the gapless modes present in the critical theory for the self-dual
line.  A proof of this connection, however, would require a better
understanding of the antiferromagnetic mixed superspin chain for
$\half<b<\pi/4\gamma$ which is beyond of the scope of this paper.

\section{Summary \& Conclusion}
%
In this paper we have introduced a family of vertex models from alternating
four-dimensional typical representations of the superalgebra $U_q[sl(2|1)]$.
These models or, equivalently, the mixed superspin chains are integrable for a
range of two continuous parameters $q=\mathrm{e}^{2i\gamma}$ related to the
deformation $b$ labelling the typical representation.  We have investigated
the thermodynamic limit of the model and identified the low energy effective
theory which determine the critical behaviour of the model in the different
parameter regions.

For $|b|<\half$ we have identified the critical theory as a $c=-1$ conformal
field theory exhibiting exact separation of spin and charge degrees of
freedom.  This holds for both the antiferromagnetic and ferromagnetic
superspin chain or, using the identity (\ref{EIDEN2}) for the entire critical
range of the deformation parameter $0<\gamma<\pi$.  Note that as the isotropic
points $\gamma=0$ ($\gamma=\pi$) are approached, the dispersion of the charge
(spin) mode becomes non-relativistic.  This agrees with what has been found
before for the antiferromagnetic $U_q[osp(2|2)]$ chain and the ferromagnetic
$U_q[sl(2|1)]$ superspin chain with alternating three-dimensional atypical
representations \cite{GaMa07,FrMa11}.  It is also consistent with what is
known for the six-vertex model hidden inside the superspin chain: here the
alternation of representations in the supersymmetric model corresponds to a
staggering $\pm i\gamma b$ of the spectral parameter, see (\ref{TXXZ}).  The
critical properties of spin chains resulting from such a construction with
\emph{real} staggering (implying purely imaginary $b$) \cite{PoZv93,FrRo97}
are identical to those of the homogeneous XXZ spin chain
\cite{Hame85,WoEc87a,AlBB87}, i.e.\ those of a free boson compactified to a
circle with radius depending on the deformation parameter.

The case $b=\pm\half$ requires special attention: for this choice of the
representation parameter some of the vertex weights vanish and the Yang-Baxter
rquation (\ref{YBE}) provides the basis for the fusion procedure allowing to
construct vertex models based on higher-dimensional representations of the
underlying algebra.  For the present model this leads to a degeneration of the
spectrum leaving the superspin chain with alternating three-dimensional quark
and antiquark representations as the non-trivial sector.  For the
ferromagnetic superspin chain the critical behaviour on the line $b=\half$
coincides with what we have found here for $|b|<\half$ \cite{FrMa11}.
The zero-charge sector of the antiferromagnetic $3\otimes \bar{3}$ model has
also been studied previously: starting from the XXZ spin-$1$ chain
\cite{ZaFa80} hidden inside this sector at least part of the operator content
of the critical theory has been identified \cite{EsFS05,FrMa11}.  In addition
to the compactified boson and the Ising mode of the spin-$1$ chain indications
were found for the presence of another critical mode with unsual properties,
possibly signalling the presence of a boson with non-compact target space in
the spectrum of the antiferromagnetic superspin chain.

For values of the representation parameter between $b=\half$ and the self-dual
line $b=\pi/4\gamma$ we have analyzed the thermodynamic limit of the
ferromagnetic mixed superspin chain based on the exact solution and supported
by numerical analysis.  Similar as in the phases for $|b|<\half$ the operator
content does not depend on the representation parameter.  The low energy
spectrum, however, contains four gapless modes.  Only two of these can be
related to the physical degrees of freedom of the model, i.e.\ spin and
charge.  One of the remaining modes is a free boson compactified at the
self-dual radius $R^2=2$, or equivalently a $SU(2)_1$ WZW model, independent
of the deformation parameter.  The fourth mode has a coupling constant which
vanishes in the thermodynamic limit.  The spectral degeneracies are lifted for
lattices of finite length $L$ leading to a logarithmic fine structure in the
spectrum of scaling dimensions.  This behaviour is similar to what we have
discussed above for the antiferromagnetic $3\otimes \bar{3}$ superspin chain.
Variation of the value of the representation parameter $b$ acts on this mode
as a twist.  Interestingly, this mode with vanishing coupling constant has
been observed before in model appearing as the XXZ subsector of the
ferromagnetic mixed superspin chain for $b=\pi/4\gamma$ \cite{JaSa06,IkJS08}
which is related to the antiferromagnetic critical line of the Potts model.

For the antiferromagnetic mixed superspin chain with $\half<b\le\pi/4\gamma$
the root configurations solving the Bethe equations (\ref{baebq}) relevant for
the low energy states in the thermodynamic limit have been indentified
\emph{only} on the self-dual line $b=\pi/4\gamma$.  On this line the operator
content of the critical theory has been found to be related to that of the
ferromagnetic chain under the replacement $\gamma \leftrightarrow
(\pi/2)-\gamma$.  In particular, there are two critical modes in addition to
the ones related to spin and charge degrees of freedom.  In this regime,
however, the bosonic mode with radius $R^2=2$ replaces the noncompact one in
the XXZ subsector.  This is the reason why no logarithmic fine structure has
been observed in the anisotropic critical regime of the Potts model
\cite{Ikhl11}.

The mixed superspin chain built from alternating a four-dimensional typical
representation of the superalgebra $U_q[sl(2|1)]$ with its dual shows an
extremely rich phase diagram.  Within this model other systems with ordinary
symmetries and twisted boundary conditions are hidden which opens the
possibility to study the peculiar critical properties found in staggered
models in a larger context.  There are still several open problems in the
analysis of this model: most important, a basis for an analytical study of the
thermodynamic properties of the antiferromagnetic model inside phase C is
still lacking.  The ground states for $b=\half$ and $b=\pi/4\gamma$ can be
related to each other by adiabatic variation of the representation parameter.
Our numerical studies, however, provide strong evidence that the ground states
and low lying excitations for intermediate values of $b$ are not accessible
this way.  Another problem is the lack of an analytical derivation of the
finite-size behaviour of the spectrum related to the mode with vanishing
conformal weight.

The model considered here can be extended in several promising directions:
first the $\mathbb{Z}_2$ staggering of the vertex model can be generalized to
obtain models with larger unit cell.  In the enlarged parameter space more
phases are expected to appear with boundaries related to higher-level fusion
of the underlying single-row transfer matrices.  Another way to realize more
general phases is by building lattice models based on higher-dimensional
representations of the underlying algebra.  Already the homogeneous models
built on the $4S+1$-dimensional atypical representations of $sl(2|1)$ display
an interesting critical behaviour \cite{FrPT98,Frahm99}.  Alternation of
vertices carrying representations dual to each other and including deformation
should lead to a rich phase diagram extending what is known about the critical
behaviour of the higher spin XXZ models \cite{AlMa89a,AlMa89,FrYF90,FrYu90}.

\begin{acknowledgments}
  We thank Guiliano A.\,P.\ Ribeiro who participated in an early stage of this
  project.  MJM has been supported in part by CAPES/DAAD grant no.~4062/11-6.
  Additional funding for this project has been provided by the Deutsche
  Forschungsgemeinschaft, the Brazilian Foundation CNPq, and the Centre for
  Quantum Engineering and Space-Time Research (QUEST).
\end{acknowledgments}

\newpage
\appendix
\section{The Boltzmann weights}
\label{app-weights}
The explicit expressions for the Boltzmann weights $a_{j}$, $b_{jk}$, $c_{jk}$
and $d_{jk}$ appearing in the $R$-matrix (\ref{Rmat}) can be obtained from
Ref.~\onlinecite{Grun00} after correcting a few misprints.  Defining the
quantum group parameter by $q=\exp(2i\gamma)$ we find that their dependence on
the four dimensional representation values $b_1$ and $b_2$ are,
\begin{equation}
\label{weiinit}
\begin{aligned}
  a_1&=a_4=-1,\quad
  a_2=\frac{\sinh(i\gamma(1-b1-b2)+\lambda)}
  {\sinh(i\gamma(1-b1-b2)-\lambda)},
  \quad
  a_3=\frac{\sinh(i\gamma(1+b1+b2)+\lambda)}
  {\sinh(i\gamma(1+b1+b2)-\lambda)}\,,
  \\
  b_{12}&=b_{42}=\frac{\sinh(i\gamma(b1-b2)-\lambda)}
  {\sinh(i\gamma(b1+b2-1)+\lambda)},
  \quad
  b_{21}=b_{24}=\frac{\sinh(i\gamma(b2-b1)-\lambda)}
  {\sinh(i\gamma(b1+b2-1)+\lambda)}\,,
  \\
  b_{13}&=b_{43}=\frac{\sinh(i\gamma(b2-b1)-\lambda)}
  {\sinh(-i\gamma(b1+b2+1)+\lambda)},\quad
  b_{31}=b_{34}=\frac{\sinh(i\gamma(b1-b2)-\lambda)}
  {\sinh(-i\gamma(b1+b2+1)+\lambda)}\,,
  \\
  c_{12}&=\exp(-2\lambda)c_{21}=
  c_{42}=\exp(-2\lambda)c_{24}=\exp(-\lambda)
  \frac{\sqrt{\sinh(i\gamma(2b_1-1))\sinh(i\gamma(2b_2-1))}}{
        \sinh(\lambda+i\gamma(b_1+b_2-1))}\,,
  \\
  c_{13}&=\exp(2\lambda)c_{31}=
  c_{43}=\exp(2\lambda)c_{34}=\exp(\lambda)
  \frac{\sqrt{\sinh(i\gamma(2b_1+1))\sinh(i\gamma(2b_2+1))}}{
        \sinh(i\gamma(b_1+b_2+1)-\lambda)}\,,
\\
d_{41}&=d_{14}=\frac{
  \sinh(i\gamma(b1-b2)+\lambda)\sinh(i\gamma(b2-b1)+\lambda)}
{\sinh(i\gamma(b1+b2-1)+\lambda)\sinh(i\gamma(b1+b2+1)-\lambda)}\,,
\\
d_{32}&=\frac{
  \sinh(i\gamma(b1-b2)+\lambda)\sinh(i\gamma(b1-b2-2)+\lambda)}
{\sinh(i\gamma(b1+b2-1)+\lambda)\sinh(i\gamma(b1+b2+1)-\lambda)}\,,
\\
d_{23}&=\frac{
\sinh(i\gamma(b2-b1)+\lambda)\sinh(i\gamma(b2-b1-2)+\lambda)}
{\sinh(i\gamma(b1+b2-1)+\lambda)\sinh(i\gamma(b1+b2+1)-\lambda)}\,,
\\
d_{42}&=\exp(-2 \lambda -2i \gamma) 
d_{31}=-\exp(-2\lambda) d_{34}
=-\exp(-2i\gamma)d_{12} 
\\
&=\exp(-\lambda-i\gamma)
\frac{\sqrt{\sinh(i\gamma(2b_1-1))\sinh(i\gamma(2b_2+1))} \,\,
  \sinh(i\gamma(b1-b2)+\lambda)}{\sinh(i\gamma(b1+b2-1)+\lambda)
  \sinh(i\gamma(b1+b2+1)-\lambda)}\,,
\\
d_{43}&=-\exp(2 \lambda) d_{24}=-\exp(-2i\gamma) d_{13}
=\exp(2\lambda-2i\gamma)d_{21} 
 \\
&=-\exp(\lambda-i\gamma)
\frac{\sqrt{\sinh(i\gamma(2b_2-1))\sinh(i\gamma(2b_1+1))} \,\,
\sinh(i\gamma(b2-b1)+\lambda)}{\sinh(i\gamma(b1+b2-1)+\lambda)
\sinh(i\gamma(b1+b2+1)-\lambda)}\,,
\\
d_{33}&=\exp(4\lambda)d_{22}\\ 
     &=\exp(2\lambda)
     \frac{\sqrt{\sinh(i\gamma(2b_1+1))\sinh(i\gamma(2b_2+1))
         \sinh(i\gamma(2b_1-1)) \sinh(i\gamma(2b_2-1))}}
     {\sinh(i\gamma(b1+b2-1)+\lambda) \sinh(i\gamma(b1+b2+1)-\lambda)}\,,
\\
d_{44}&=\frac{\exp(-2\lambda)
  \sinh(2i\gamma)+\exp(-2i\gamma)\cosh(2i\gamma(b_1-b_2))
  -\cosh(2i\gamma(b_1+b_2))}{2 \sinh(i\gamma(b1+b2-1)+\lambda)
\sinh(i\gamma(b1+b2+1)-\lambda)}\,,
\\
d_{11}&=\frac{-\exp(2\lambda)
  \sinh(2i\gamma)+\exp(2i\gamma)\cosh(2i\gamma(b_1-b_2))
  -\cosh(2i\gamma(b_1+b_2))}{2 \sinh(i\gamma(b1+b2-1)+\lambda)
  \sinh(i\gamma(b1+b2+1)-\lambda)}\,.
\end{aligned}
\end{equation}

\newpage
\section{The two-site model}
\label{app-2site}
%
For $L=1$ the transfer matrices (\ref{TRAMIX}) act on a Hilbert space of
dimension $16$.  The eigenvalues $\Lambda^{(b,\{b,-b\})}(\lambda)$ and the
corresponding Bethe roots are obtained from solutions of the Bethe equations:

\begin{itemize}
\item $(N_1,N_2)=(0,0)$ -- the pseudo vacuum:
  \begin{equation}
    \label{eig1-00}
    \Lambda^{(b,\{b,-b\})}_{00}(\lambda) = \left(\frac{\sin\gamma}{
        \sinh(\lambda-i\gamma)}\right)^2\,
    \frac{\sinh(2\lambda-i\gamma(2b+1)) 
      \sinh(2\lambda+i\gamma(2b-1))}{
      \sinh(\lambda-i\gamma(2b+1))
      \sinh(\lambda+i\gamma(2b-1))}\,.
  \end{equation}
  This state is degenerate to the one with $B=0$, $S_3=-1$ due to the discrete
  spin inversion symmetry $S_3\leftrightarrow -S_3$ of the model.

\item $(N_1,N_2)=(1,0)$:\newline
  The Bethe equation has two solutions: $\lambda^{(1)} = 0$ giving the
  transfer matrix eigenvalue
  \begin{equation}
    \label{eig1-10a}
    \begin{aligned}
      \Lambda^{(b,\{b,-b\})}_{10,+}(\lambda) =& -4 \left(\frac{\sin\gamma}{
          \sinh(\lambda-i\gamma)}\right)^2\,
      \sin(\gamma(b+\half))\sin(\gamma(b-\half))\,\\
      &\quad\times \frac{\cosh(\lambda-i\gamma(b+\half))
        \cosh(\lambda+i\gamma(b-\half))
      }{
        \sinh(\lambda-i\gamma(2b+1))
        \sinh(\lambda+i\gamma(2b-1)) }\,,
    \end{aligned}
  \end{equation}
  and $\lambda^{(1)}=i\pi/2$, resulting in
  \begin{equation}
    \label{eig1-10b}
    \begin{aligned}
      \Lambda^{(b,\{b,-b\})}_{10,-}(\lambda) =& 4 \left(\frac{\sin\gamma}{
          \sinh(\lambda-i\gamma)}\right)^2\,
      \cos(\gamma(b+\half))\cos(\gamma(b-\half))\,\\
      &\quad\times \frac{\sinh(\lambda-i\gamma(b+\half))
        \sinh(\lambda+i\gamma(b-\half))
      }{
        \sinh(\lambda-i\gamma(2b+1))
        \sinh(\lambda+i\gamma(2b-1)) }\,.
    \end{aligned}
  \end{equation}
  Both of these eigenvalues are fourfold degenerate: first they are present in
  the $S_3=-\half$ sector due to spin inversion.  In addition, since the
  expressions (\ref{eig1-10a}) and (\ref{eig1-10b}) are invariant under
  $b\leftrightarrow -b$ charge conjugation, the same eigenvalues in the sector
  $(N_1,N_2)=(0,1)$ for $\lambda^{(2)} = 0$ and $i\pi/2$, respectively.

\item $(N_1,N_2)=(2,0)$:\newline
  The solution to the Bethe equations is $\lambda_1^{(1)}=0$,
  $\lambda_2^{(1)}=i\pi/2$ giving 
  \begin{equation}
    \label{eig1-20}
    \begin{aligned}
      \Lambda^{(b,\{b,-b\})}_{20}(\lambda) =& \half\, \left(\frac{\sin\gamma}{
          \sinh(\lambda-i\gamma)}\right)^2\,
      \frac{1}{
        \sinh(\lambda-i\gamma(2b+1))
        \sinh(\lambda+i\gamma(2b-1)) }\\
      &\times \left(2\sin^2\gamma +\sinh^2\lambda + \sinh^2(\lambda-i\gamma)
        +\sinh^2(\lambda-2i\gamma b) +\right.\\
      &\quad \left. \sinh^2(\lambda-i\gamma(2b+1))
        -\sinh^2[2\lambda-i\gamma(2b+1)]
      \right)\,.
    \end{aligned}
  \end{equation}
  This state has $S_3=0$.  For generic $b\ne0$ (\ref{eig1-20}) is not invariant
  under charge conjugation $b\leftrightarrow -b$, therefore the eigenvalue in
  the $(N_1,N_2)=(0,2)$ sector corresponding to the solution
  $\lambda_1^{(2)}=0$, $\lambda_2^{(2)}=i\pi/2$ of the Bethe equations is
  \begin{equation}
    \label{eig1-02}
    \Lambda^{(b,\{b,-b\})}_{02}(\lambda) = \Lambda^{(-b,\{b,-b\})}_{20}(\lambda)\,.
  \end{equation}

\item $(N_1,N_2)=(1,1)$:\newline
  In this sector we have three different eigenvalues corresponding to rather
  peculiar solutions of the Bethe equations:

  The first solution is actually a one parametric family defined by
  \begin{equation}
    \label{curve11}
    \cosh(\lambda^{(1)} + \lambda^{(2)}) \cos(\gamma)
    = \cosh( \lambda^{(1)} - \lambda^{(2)}) \cos(2\gamma b)\,.
  \end{equation}
  The eigenvalue corresponding to all solutions of (\ref{curve11}) is
  \begin{equation}
    \label{eig1-110}
    \begin{aligned}
      \Lambda^{(b,\{b,-b\})}_{11,0}(\lambda) =& -\left(\frac{\sin\gamma}{
          \sinh(\lambda-i\gamma)}\right)^2\,
      \frac{\sin(\gamma(2b+1))\,\sin(\gamma(2b-1))}{
        \sinh(\lambda-i\gamma(2b+1))
        \sinh(\lambda+i\gamma(2b-1))}\,.
    \end{aligned}
  \end{equation}
  It appears twice in the spectrum of $T^{(b)}(\lambda)$.

  Additional solutions to the Bethe equations in this sector involve at least
  one root at $\pm\infty$.  Choosing $\lambda^{(1)}=-\infty$ the other Bethe
  equations becomes
  \begin{equation}
    \frac{\sinh(\lambda^{(2)} + i\gamma(b-\half))}{
      \sinh(\lambda^{(2)} - i\gamma(b-\half))}\,
    \frac{\sinh(\lambda^{(2)} - i\gamma(b+\half))}{
      \sinh(\lambda^{(2)} + i\gamma(b+\half))}
    = \mathrm{e}^{-2i\gamma}\,.
  \end{equation}
  This equation has two solutions: the first one is contained in
  (\ref{curve11}) leading to the eigenvalue (\ref{eig1-110}) discussed above ,
  while for the other one $\lambda^{(2)}=+\infty$ leading to the eigenvalue
  \begin{equation}
    \label{eig1-11p}
    \begin{aligned}
      \Lambda^{(b,\{b,-b\})}_{11,+}(\lambda) =&
      -\frac{1}{4}\,\left(\frac{\sin\gamma}{ 
          \sinh(\lambda-i\gamma)}\right)^2\,
      \frac{1}{
        \sinh(\lambda-i\gamma(2b+1))
        \sinh(\lambda+i\gamma(2b-1))}\,\\
      &\quad \times \left[
        \mathrm{e}^{2\lambda-i\gamma}\left(1+\mathrm{e}^{-4i\gamma b} \right)
        -2\cos\gamma \right] \left[
        \mathrm{e}^{-2\lambda+i\gamma}\left(1+\mathrm{e}^{-4i\gamma b}\right)
        -2\cos\gamma \right]\,.
    \end{aligned}
  \end{equation}
  Although this eigenvalue is found in the zero charge sector of the model it
  is not invariant under charge conjugation $b\leftarrow -b$.  In fact,
  choosing the solution $\lambda^{(1)}=+\infty$ and $\lambda^{(2)}=-\infty$ to
  the Bethe equations we obtain a third eigenvalue in this sector, i.e.
  \begin{equation}
    \label{eig1-11m}
    \Lambda^{(b,\{b,-b\})}_{11,-}(\lambda) =  \Lambda^{(-b,\{b,-b\})}_{11,+}(\lambda)\,.
  \end{equation}

\end{itemize}

%
Finally, let us remark on the limit $b\to\pm\half$: in the rational
$sl(2|1)$-invariant case the four dimensional 'typical' representation
$[b,\half]$ degenerates into one of the three-dimensional atypical ones
$[\half]_\pm$ \cite{ScNR77}.  Similarly, the decomposition of the tensor
product of two such representations degenerates in the limit with
$b_1=-b_2\equiv b\to\pm\half$.  Instead of the $sl(2|1)$ octet $[0,1]$ and an
eight-dimensional indecomposable one finds an octet and an $sl(2|1)$ singlet:
\begin{equation}
  \begin{aligned}
    \left[b,\half\right] \otimes \left[-b,\half\right] &=
    \left[0,1\right] \oplus \left[0,\half,-\half,0\right]\,\\
    \Rightarrow\quad
    \left[\half\right]_+ \otimes \left[\half\right]_- &=
   \left[0,1\right]  \oplus \left[0\right]\,.
  \end{aligned}
\end{equation}
Periodic boundary conditions break the symmetry of the $q$-deformed vertex
model but still one observes degenerations arising in the limit
$b\to\pm\half$.  This shows up through singularities in the in the vertex
weights (\ref{weiinit}) of the $U_q[sl(2|1)]$ model.  They allow to obtain a
fusion relation from the Yang-Baxter equation for the monodromy matrix which
can be exploited to construct integrable models with higher spin.  Another
consequence of these singularities is that part of the Hilbert space of the
mixed vertex model decouples leaving the model built from the
three-dimensional quark and antiquark representations \cite{EsFS05,FrMa11} as
the non-trivial part.

For the two-site model considered here this leads to the vanishing of the
eigenvalues (\ref{eig1-10a}) and (\ref{eig1-110}) for any values of the
deformation parameter $\gamma$ while
\begin{equation}
  \Lambda_{20}^{(\pm\half,\{\half,-\half\})}(\lambda) = 
  \Lambda_{02}^{(\pm\half,\{\half,-\half\})}(\lambda) \equiv -4\sin^2\gamma
\end{equation}
become independent of the spectral parameter $\lambda$.  For the superspin
chain Hamiltonian (\ref{Hmix}) this implies the appearance of zero energy
eigenvalues.  Eliminating the decoupled states from the Hilbert space only one
of the singlets (\ref{eig1-110}) remains in the spectrum.  This state,
parameterized by Bethe roots $\lambda^{(1)}=\lambda^{(2)}=0$, is $c=0$ ground
state of the antiferromagnetic $3\otimes\bar{3}$-superspin chain discussed in
Refs.~\onlinecite{EsFS05,FrMa11} for $L=1$.

\newpage

\begin{thebibliography}{10}%
\makeatletter
\providecommand \@ifxundefined [1]{%
 \ifx #1\undefined \expandafter \@firstoftwo
 \else \expandafter \@secondoftwo
\fi
}%
\providecommand \@ifnum [1]{%
 \ifnum #1\expandafter \@firstoftwo
 \else \expandafter \@secondoftwo
\fi
}%
\providecommand \enquote [1]{``#1''}%
\providecommand \bibnamefont  [1]{#1}%
\providecommand \bibfnamefont [1]{#1}%
\providecommand \citenamefont [1]{#1}%
\providecommand\href[0]{\@sanitize\@href}%
\providecommand\@href[1]{\endgroup\@@startlink{#1}\endgroup\@@href}%
\providecommand\@@href[1]{#1\@@endlink}%
\providecommand \@sanitize [0]{\begingroup\catcode`\&12\catcode`\#12\relax}%
\@ifxundefined \pdfoutput {\@firstoftwo}{%
 \@ifnum{\z@=\pdfoutput}{\@firstoftwo}{\@secondoftwo}%
}{%
 \providecommand\@@startlink[1]{\leavevmode}%
 \providecommand\@@endlink[0]{}%
}{%
 \providecommand\@@startlink[1]{%
  \leavevmode
  \pdfstartlink
   attr{/Border[0 0 1 ]/H/I/C[0 1 1]}%
   user{/Subtype/Link/A<</Type/Action/S/URI/URI(#1)>>}%
  \relax
 }%
 \providecommand\@@endlink[0]{\pdfendlink}%
}%
\providecommand \url  [0]{\begingroup\@sanitize \@url }%
\providecommand \@url [1]{\endgroup\@href {#1}{\urlprefix}}%
\providecommand \urlprefix [0]{URL }%
\providecommand \Eprint[0]{\href }%
\@ifxundefined \urlstyle {%
  \providecommand \doi [1]{doi:\discretionary{}{}{}#1}%
}{%
  \providecommand \doi [0]{doi:\discretionary{}{}{}\begingroup
  \urlstyle{rm}\Url }%
}%
\providecommand \doibase [0]{http://dx.doi.org/}%
\providecommand \Doi[1]{\href{\doibase#1}}%
\providecommand \bibAnnote [3]{%
  \BibitemShut{#1}%
  \begin{quotation}\noindent
    \textsc{Key:}\ #2\\\textsc{Annotation:}\ #3%
  \end{quotation}%
}%
\providecommand \bibAnnoteFile [2]{%
  \IfFileExists{#2}{\bibAnnote {#1} {#2} {\input{#2}}}{}%
}%
\providecommand \typeout [0]{\immediate \write \m@ne }%
\providecommand \selectlanguage [0]{\@gobble}%
\providecommand \bibinfo [0]{\@secondoftwo}%
\providecommand \bibfield [0]{\@secondoftwo}%
\providecommand \translation [1]{[#1]}%
\providecommand \BibitemOpen[0]{}%
\providecommand \bibitemStop [0]{}%
\providecommand \bibitemNoStop [0]{.\EOS\space}%
\providecommand \EOS [0]{\spacefactor3000\relax}%
\providecommand \BibitemShut [1]{\csname bibitem#1\endcsname}%
\bibitem{Affl86a}%
  \BibitemOpen
  \bibfield{author}{%
  \bibinfo {author} {\bibfnamefont{I.}~\bibnamefont{Affleck}},\ }%
  \bibfield{journal}{%
  \bibinfo {journal} {Nucl. Phys. B}\ }%
  \textbf{\bibinfo {volume} {265}},\ \bibinfo {pages} {409} (\bibinfo {year}
  {1986})%
  \bibAnnoteFile{NoStop}{Affl86a}%
\bibitem{AfHa87}%
  \BibitemOpen
  \bibfield{author}{%
  \bibinfo {author} {\bibfnamefont{I.}~\bibnamefont{Affleck}}\ and\ \bibinfo
  {author} {\bibfnamefont{F.~D.~M.}\ \bibnamefont{Haldane}},\ }%
  \bibfield{journal}{%
  \bibinfo {journal} {Phys. Rev. B}\ }%
  \textbf{\bibinfo {volume} {36}},\ \bibinfo {pages} {5291} (\bibinfo {year}
  {1987})%
  \bibAnnoteFile{NoStop}{AfHa87}%
\bibitem{MaNR98}%
  \BibitemOpen
  \bibfield{author}{%
  \bibinfo {author} {\bibfnamefont{M.~J.}\ \bibnamefont{Martins}}, \bibinfo
  {author} {\bibfnamefont{B.}~\bibnamefont{Nienhuis}},\ and\ \bibinfo {author}
  {\bibfnamefont{R.}~\bibnamefont{Rietman}},\ }%
  \bibfield{journal}{%
  \bibinfo {journal} {Phys. Rev. Lett.}\ }%
  \textbf{\bibinfo {volume} {81}},\ \bibinfo {pages} {504} (\bibinfo {year}
  {1998}),\ \Eprint{http://arxiv.org/abs/cond-mat/9709051}{cond-mat/9709051}%
  \bibAnnoteFile{NoStop}{MaNR98}%
\bibitem{Saleur00}%
  \BibitemOpen
  \bibfield{author}{%
  \bibinfo {author} {\bibfnamefont{H.}~\bibnamefont{Saleur}},\ }%
  \bibfield{journal}{%
  \bibinfo {journal} {Nucl. Phys. B}\ }%
  \textbf{\bibinfo {volume} {578}},\ \bibinfo {pages} {552} (\bibinfo {year}
  {2000}),\ \Eprint{http://arxiv.org/abs/solv-int/9905007}{solv-int/9905007}%
  \bibAnnoteFile{NoStop}{Saleur00}%
\bibitem{EsFS05}%
  \BibitemOpen
  \bibfield{author}{%
  \bibinfo {author} {\bibfnamefont{F.~H.~L.}\ \bibnamefont{Essler}}, \bibinfo
  {author} {\bibfnamefont{H.}~\bibnamefont{Frahm}},\ and\ \bibinfo {author}
  {\bibfnamefont{H.}~\bibnamefont{Saleur}},\ }%
  \bibfield{journal}{%
  \bibinfo {journal} {Nucl. Phys. B}\ }%
  \textbf{\bibinfo {volume} {712 [FS]}},\ \bibinfo {pages} {513} (\bibinfo
  {year} {2005}),\
  \Eprint{http://arxiv.org/abs/cond-mat/0501197}{cond-mat/0501197}%
  \bibAnnoteFile{NoStop}{EsFS05}%
\bibitem{IkJS08}%
  \BibitemOpen
  \bibfield{author}{%
  \bibinfo {author} {\bibfnamefont{Y.}~\bibnamefont{Ikhlef}}, \bibinfo {author}
  {\bibfnamefont{J.~L.}\ \bibnamefont{Jacobsen}},\ and\ \bibinfo {author}
  {\bibfnamefont{H.}~\bibnamefont{Saleur}},\ }%
  \bibfield{journal}{%
  \bibinfo {journal} {Nucl. Phys. B}\ }%
  \textbf{\bibinfo {volume} {789}},\ \bibinfo {pages} {483} (\bibinfo {year}
  {2008}),\ \Eprint{http://arxiv.org/abs/cond-mat/0612037}{cond-mat/0612037}%
  \bibAnnoteFile{NoStop}{IkJS08}%
\bibitem{ScNR77}%
  \BibitemOpen
  \bibfield{author}{%
  \bibinfo {author} {\bibfnamefont{M.}~\bibnamefont{Scheunert}}, \bibinfo
  {author} {\bibfnamefont{W.}~\bibnamefont{Nahm}},\ and\ \bibinfo {author}
  {\bibfnamefont{V.}~\bibnamefont{Rittenberg}},\ }%
  \bibfield{journal}{%
  \bibinfo {journal} {J. Math. Phys.}\ }%
  \textbf{\bibinfo {volume} {18}},\ \bibinfo {pages} {155} (\bibinfo {year}
  {1977})%
  \bibAnnoteFile{NoStop}{ScNR77}%
\bibitem{BGZD94}%
  \BibitemOpen
  \bibfield{author}{%
  \bibinfo {author} {\bibfnamefont{A.~J.}\ \bibnamefont{Bracken}}, \bibinfo
  {author} {\bibfnamefont{M.~D.}\ \bibnamefont{Gould}}, \bibinfo {author}
  {\bibfnamefont{Y.-Z.}\ \bibnamefont{Zhang}},\ and\ \bibinfo {author}
  {\bibfnamefont{G.~W.}\ \bibnamefont{Delius}},\ }%
  \bibfield{journal}{%
  \bibinfo {journal} {J. Phys. A}\ }%
  \textbf{\bibinfo {volume} {27}},\ \bibinfo {pages} {6551} (\bibinfo {year}
  {1994}),\ \Eprint{http://arxiv.org/abs/hep-th/9405138}{hep-th/9405138}%
  \bibAnnoteFile{NoStop}{BGZD94}%
\bibitem{DGLZ95a}%
  \BibitemOpen
  \bibfield{author}{%
  \bibinfo {author} {\bibfnamefont{G.~W.}\ \bibnamefont{Delius}}, \bibinfo
  {author} {\bibfnamefont{M.~D.}\ \bibnamefont{Gould}}, \bibinfo {author}
  {\bibfnamefont{J.~R.}\ \bibnamefont{Links}},\ and\ \bibinfo {author}
  {\bibfnamefont{Y.-Z.}\ \bibnamefont{Zhang}},\ }%
  \bibfield{journal}{%
  \bibinfo {journal} {Int. J. Mod. Phys. A}\ }%
  \textbf{\bibinfo {volume} {10}},\ \bibinfo {pages} {3259} (\bibinfo {year}
  {1995}),\ \Eprint{http://arxiv.org/abs/hep-th/9408006}{hep-th/9408006}%
  \bibAnnoteFile{NoStop}{DGLZ95a}%
\bibitem{Maas95}%
  \BibitemOpen
  \bibfield{author}{%
  \bibinfo {author} {\bibfnamefont{Z.}~\bibnamefont{Maassarani}},\ }%
  \bibfield{journal}{%
  \bibinfo {journal} {J. Phys. A}\ }%
  \textbf{\bibinfo {volume} {28}},\ \bibinfo {pages} {1305} (\bibinfo {year}
  {1995}),\ \Eprint{http://arxiv.org/abs/hep-th/9407032}{hep-th/9407032}%
  \bibAnnoteFile{NoStop}{Maas95}%
\bibitem{Grun00}%
  \BibitemOpen
  \bibfield{author}{%
  \bibinfo {author} {\bibfnamefont{J.}~\bibnamefont{Gruneberg}},\ }%
  \bibfield{journal}{%
  \bibinfo {journal} {Nuclear Physics B}\ }%
  \textbf{\bibinfo {volume} {568}},\ \bibinfo {pages} {594} (\bibinfo {year}
  {2000})%
  \bibAnnoteFile{NoStop}{Grun00}%
\bibitem{Kulish85}%
  \BibitemOpen
  \bibfield{author}{%
  \bibinfo {author} {\bibfnamefont{P.~P.}\ \bibnamefont{Kulish}},\ }%
  \bibfield{journal}{%
  \bibinfo {journal} {J. Sov. Math.}\ }%
  \textbf{\bibinfo {volume} {35}},\ \bibinfo {pages} {2648} (\bibinfo {year}
  {1986}),\ \bibinfo {note} {[Zap. Nauch. Semin. LOMI {\bf 145}, 140 (1985)]}%
  \bibAnnoteFile{NoStop}{Kulish85}%
\bibitem{PfFr96}%
  \BibitemOpen
  \bibfield{author}{%
  \bibinfo {author} {\bibfnamefont{M.~P.}\ \bibnamefont{Pfannm{\"u}ller}}\ and\
  \bibinfo {author} {\bibfnamefont{H.}~\bibnamefont{Frahm}},\ }%
  \bibfield{journal}{%
  \bibinfo {journal} {Nucl. Phys. B}\ }%
  \textbf{\bibinfo {volume} {479}},\ \bibinfo {pages} {575} (\bibinfo {year}
  {1996}),\ \Eprint{http://arxiv.org/abs/cond-mat/9604082}{cond-mat/9604082}%
  \bibAnnoteFile{NoStop}{PfFr96}%
\bibitem{PfFr97}%
  \BibitemOpen
  \bibfield{author}{%
  \bibinfo {author} {\bibfnamefont{M.~P.}\ \bibnamefont{Pfannm{\"u}ller}}\ and\
  \bibinfo {author} {\bibfnamefont{H.}~\bibnamefont{Frahm}},\ }%
  \bibfield{journal}{%
  \bibinfo {journal} {J. Phys. A}\ }%
  \textbf{\bibinfo {volume} {30}},\ \bibinfo {pages} {L543} (\bibinfo {year}
  {1997})%
  \bibAnnoteFile{NoStop}{PfFr97}%
\bibitem{RaMa96}%
  \BibitemOpen
  \bibfield{author}{%
  \bibinfo {author} {\bibfnamefont{P.~B.}\ \bibnamefont{Ramos}}\ and\ \bibinfo
  {author} {\bibfnamefont{M.~J.}\ \bibnamefont{Martins}},\ }%
  \bibfield{journal}{%
  \bibinfo {journal} {Nucl. Phys. B}\ }%
  \textbf{\bibinfo {volume} {474 [FS]}},\ \bibinfo {pages} {678} (\bibinfo
  {year} {1996}),\
  \Eprint{http://arxiv.org/abs/hep-th/9604072}{hep-th/9604072}%
  \bibAnnoteFile{NoStop}{RaMa96}%
\bibitem{GaMa07}%
  \BibitemOpen
  \bibfield{author}{%
  \bibinfo {author} {\bibfnamefont{W.}~\bibnamefont{Galleas}}\ and\ \bibinfo
  {author} {\bibfnamefont{M.~J.}\ \bibnamefont{Martins}},\ }%
  \bibfield{journal}{%
  \bibinfo {journal} {Nucl. Phys. B}\ }%
  \textbf{\bibinfo {volume} {768}},\ \bibinfo {pages} {219} (\bibinfo {year}
  {2007}),\ \Eprint{http://arxiv.org/abs/hep-th/0612281}{hep-th/0612281}%
  \bibAnnoteFile{NoStop}{GaMa07}%
\bibitem{FrMa11}%
  \BibitemOpen
  \bibfield{author}{%
  \bibinfo {author} {\bibfnamefont{H.}~\bibnamefont{Frahm}}\ and\ \bibinfo
  {author} {\bibfnamefont{M.~J.}\ \bibnamefont{Martins}},\ }%
  \bibfield{journal}{%
  \bibinfo {journal} {Nucl. Phys. B}\ }%
  \textbf{\bibinfo {volume} {847}},\ \bibinfo {pages} {220} (\bibinfo {year}
  {2011}),\ \Eprint{http://arxiv.org/abs/1012.1753}{arXiv:1012.1753}%
  \bibAnnoteFile{NoStop}{FrMa11}%
\bibitem{TeLi71}%
  \BibitemOpen
  \bibfield{author}{%
  \bibinfo {author} {\bibfnamefont{H.~N.~V.}\ \bibnamefont{Temperley}}\ and\
  \bibinfo {author} {\bibfnamefont{E.~H.}\ \bibnamefont{Lieb}},\ }%
  \bibfield{journal}{%
  \bibinfo {journal} {Proc. R. Soc. Lond. A}\ }%
  \textbf{\bibinfo {volume} {332}},\ \bibinfo {pages} {251} (\bibinfo {year}
  {1971})%
  \bibAnnoteFile{NoStop}{TeLi71}%
\bibitem{BaKW76}%
  \BibitemOpen
  \bibfield{author}{%
  \bibinfo {author} {\bibfnamefont{R.~J.}\ \bibnamefont{Baxter}}, \bibinfo
  {author} {\bibfnamefont{S.~B.}\ \bibnamefont{Kelland}},\ and\ \bibinfo
  {author} {\bibfnamefont{F.~Y.}\ \bibnamefont{Wu}},\ }%
  \bibfield{journal}{%
  \bibinfo {journal} {J. Phys. A}\ }%
  \textbf{\bibinfo {volume} {9}},\ \bibinfo {pages} {397} (\bibinfo {year}
  {1976})%
  \bibAnnoteFile{NoStop}{BaKW76}%
\bibitem{Baxt82}%
  \BibitemOpen
  \bibfield{author}{%
  \bibinfo {author} {\bibfnamefont{R.~J.}\ \bibnamefont{Baxter}},\ }%
  \bibfield{journal}{%
  \bibinfo {journal} {Proc. R. Soc. Lond. A}\ }%
  \textbf{\bibinfo {volume} {383}},\ \bibinfo {pages} {43} (\bibinfo {year}
  {1982})%
  \bibAnnoteFile{NoStop}{Baxt82}%
\bibitem{TaFa79}%
  \BibitemOpen
  \bibfield{author}{%
  \bibinfo {author} {\bibfnamefont{L.~A.}\ \bibnamefont{Takhtajan}}\ and\
  \bibinfo {author} {\bibfnamefont{L.~D.}\ \bibnamefont{Faddeev}},\ }%
  \bibfield{journal}{%
  \bibinfo {journal} {Russ. Math. Survey}\ }%
  \textbf{\bibinfo {volume} {34}},\ \bibinfo {pages} {11} (\bibinfo {year}
  {1979})%
  \bibAnnoteFile{NoStop}{TaFa79}%
\bibitem{VladB}%
  \BibitemOpen
  \bibfield{author}{%
  \bibinfo {author} {\bibfnamefont{V.~E.}\ \bibnamefont{Korepin}}, \bibinfo
  {author} {\bibfnamefont{N.~M.}\ \bibnamefont{Bogoliubov}},\ and\ \bibinfo
  {author} {\bibfnamefont{A.~G.}\ \bibnamefont{Izergin}},\ }%
  \emph{\bibinfo {title} {{Quantum Inverse Scattering Method and Correlation
  Functions}}}\ (\bibinfo {publisher} {Cambridge University Press},\ \bibinfo
  {address} {Cambridge},\ \bibinfo {year} {1993})%
  \bibAnnoteFile{NoStop}{VladB}%
\bibitem{ZaFa80}%
  \BibitemOpen
  \bibfield{author}{%
  \bibinfo {author} {\bibfnamefont{A.~B.}\ \bibnamefont{Zamolodchikov}}\ and\
  \bibinfo {author} {\bibfnamefont{V.~A.}\ \bibnamefont{Fateev}},\ }%
  \bibfield{journal}{%
  \bibinfo {journal} {Sov. J. Nucl. Phys.}\ }%
  \textbf{\bibinfo {volume} {32}},\ \bibinfo {pages} {298} (\bibinfo {year}
  {1980})%
  \bibAnnoteFile{NoStop}{ZaFa80}%
\bibitem{Jimb86}%
  \BibitemOpen
  \bibfield{author}{%
  \bibinfo {author} {\bibfnamefont{M.}~\bibnamefont{Jimbo}},\ }%
  \bibfield{journal}{%
  \bibinfo {journal} {Comm. Math. Phys.}\ }%
  \textbf{\bibinfo {volume} {102}},\ \bibinfo {pages} {537} (\bibinfo {year}
  {1986})%
  \bibAnnoteFile{NoStop}{Jimb86}%
\bibitem{Mart99}%
  \BibitemOpen
  \bibfield{author}{%
  \bibinfo {author} {\bibfnamefont{M.~J.}\ \bibnamefont{Martins}},\ }%
  \bibfield{journal}{%
  \bibinfo {journal} {Phys. Rev. E}\ }%
  \textbf{\bibinfo {volume} {59}},\ \bibinfo {pages} {7220} (\bibinfo {year}
  {1999})%
  \bibAnnoteFile{NoStop}{Mart99}%
\bibitem{Grim94}%
  \BibitemOpen
  \bibfield{author}{%
  \bibinfo {author} {\bibfnamefont{U.}~\bibnamefont{Grimm}},\ }%
  \bibfield{journal}{%
  \bibinfo {journal} {J. Phys. A}\ }%
  \textbf{\bibinfo {volume} {27}},\ \bibinfo {pages} {5897} (\bibinfo {year}
  {1994})%
  \bibAnnoteFile{NoStop}{Grim94}%
\bibitem{GaMa06}%
  \BibitemOpen
  \bibfield{author}{%
  \bibinfo {author} {\bibfnamefont{W.}~\bibnamefont{Galleas}}\ and\ \bibinfo
  {author} {\bibfnamefont{M.~J.}\ \bibnamefont{Martins}},\ }%
  \bibfield{journal}{%
  \bibinfo {journal} {Nucl. Phys. B}\ }%
  \textbf{\bibinfo {volume} {732}},\ \bibinfo {pages} {444} (\bibinfo {year}
  {2006}),\ \Eprint{http://arxiv.org/abs/nlin/0509014}{nlin/0509014}%
  \bibAnnoteFile{NoStop}{GaMa06}%
\bibitem{IkJS10}%
  \BibitemOpen
  \bibfield{author}{%
  \bibinfo {author} {\bibfnamefont{Y.}~\bibnamefont{Ikhlef}}, \bibinfo {author}
  {\bibfnamefont{J.~L.}\ \bibnamefont{Jacobsen}},\ and\ \bibinfo {author}
  {\bibfnamefont{H.}~\bibnamefont{Saleur}},\ }%
  \bibfield{journal}{%
  \bibinfo {journal} {J. Phys. A}\ }%
  \textbf{\bibinfo {volume} {43}},\ \bibinfo {pages} {225201} (\bibinfo {year}
  {2010}),\ \Eprint{http://arxiv.org/abs/0911.3003}{arXiv:0911.3003}%
  \bibAnnoteFile{NoStop}{IkJS10}%
\bibitem{TaSu72}%
  \BibitemOpen
  \bibfield{author}{%
  \bibinfo {author} {\bibfnamefont{M.}~\bibnamefont{Takahashi}}\ and\ \bibinfo
  {author} {\bibfnamefont{M.}~\bibnamefont{Suzuki}},\ }%
  \bibfield{journal}{%
  \bibinfo {journal} {Prog. Theor. Phys.}\ }%
  \textbf{\bibinfo {volume} {48}},\ \bibinfo {pages} {2187} (\bibinfo {year}
  {1972})%
  \bibAnnoteFile{NoStop}{TaSu72}%
\bibitem{YaYa69}%
  \BibitemOpen
  \bibfield{author}{%
  \bibinfo {author} {\bibfnamefont{C.~N.}\ \bibnamefont{Yang}}\ and\ \bibinfo
  {author} {\bibfnamefont{C.~P.}\ \bibnamefont{Yang}},\ }%
  \bibfield{journal}{%
  \bibinfo {journal} {J. Math. Phys.}\ }%
  \textbf{\bibinfo {volume} {10}},\ \bibinfo {pages} {1115} (\bibinfo {year}
  {1969})%
  \bibAnnoteFile{NoStop}{YaYa69}%
\bibitem{HUBBARD}%
  \BibitemOpen
  \bibfield{author}{%
  \bibinfo {author} {\bibfnamefont{F.~H.~L.}\ \bibnamefont{Essler}}, \bibinfo
  {author} {\bibfnamefont{H.}~\bibnamefont{Frahm}}, \bibinfo {author}
  {\bibfnamefont{F.}~\bibnamefont{G{\"o}hmann}}, \bibinfo {author}
  {\bibfnamefont{A.}~\bibnamefont{Kl{\"u}mper}},\ and\ \bibinfo {author}
  {\bibfnamefont{V.~E.}\ \bibnamefont{Korepin}},\ }%
  \emph{\Doi{10.2277/0521802628}{\bibinfo {title} {{T}he {O}ne-{D}imensional
  {Hubbard} {M}odel}}}\ (\bibinfo {publisher} {Cambridge University Press},\
  \bibinfo {address} {Cambridge (UK)},\ \bibinfo {year} {2005})%
  \bibAnnoteFile{NoStop}{HUBBARD}%
\bibitem{VeWo85}%
  \BibitemOpen
  \bibfield{author}{%
  \bibinfo {author} {\bibfnamefont{H.~J.}\ \bibnamefont{de~Vega}}\ and\
  \bibinfo {author} {\bibfnamefont{F.}~\bibnamefont{Woynarowich}},\ }%
  \bibfield{journal}{%
  \bibinfo {journal} {Nucl. Phys. B}\ }%
  \textbf{\bibinfo {volume} {251}},\ \bibinfo {pages} {439} (\bibinfo {year}
  {1985})%
  \bibAnnoteFile{NoStop}{VeWo85}%
\bibitem{Vega88}%
  \BibitemOpen
  \bibfield{author}{%
  \bibinfo {author} {\bibfnamefont{H.~J.}\ \bibnamefont{de~Vega}},\ }%
  \bibfield{journal}{%
  \bibinfo {journal} {J. Phys. A: Math. Gen.}\ }%
  \textbf{\bibinfo {volume} {21}},\ \bibinfo {pages} {L1089} (\bibinfo {year}
  {1988})%
  \bibAnnoteFile{NoStop}{Vega88}%
\bibitem{WoEc87b}%
  \BibitemOpen
  \bibfield{author}{%
  \bibinfo {author} {\bibfnamefont{F.}~\bibnamefont{Woynarovich}}\ and\
  \bibinfo {author} {\bibfnamefont{H.-P.}\ \bibnamefont{Eckle}},\ }%
  \bibfield{journal}{%
  \bibinfo {journal} {J. Phys. A}\ }%
  \textbf{\bibinfo {volume} {20}},\ \bibinfo {pages} {L443} (\bibinfo {year}
  {1987})%
  \bibAnnoteFile{NoStop}{WoEc87b}%
\bibitem{Suzu88}%
  \BibitemOpen
  \bibfield{author}{%
  \bibinfo {author} {\bibfnamefont{J.}~\bibnamefont{Suzuki}},\ }%
  \bibfield{journal}{%
  \bibinfo {journal} {J. Phys. A: Math. Gen.}\ }%
  \textbf{\bibinfo {volume} {21}},\ \bibinfo {pages} {L1175} (\bibinfo {year}
  {1988})%
  \bibAnnoteFile{NoStop}{Suzu88}%
\bibitem{BlCN86}%
  \BibitemOpen
  \bibfield{author}{%
  \bibinfo {author} {\bibfnamefont{H.~W.~J.}\ \bibnamefont{Bl{\"o}te}},
  \bibinfo {author} {\bibfnamefont{J.~L.}\ \bibnamefont{Cardy}},\ and\ \bibinfo
  {author} {\bibfnamefont{M.~P.}\ \bibnamefont{Nightingale}},\ }%
  \bibfield{journal}{%
  \bibinfo {journal} {Phys. Rev. Lett.}\ }%
  \textbf{\bibinfo {volume} {56}},\ \bibinfo {pages} {742} (\bibinfo {year}
  {1986}).%
\bibitem{Affl86}%
  \BibitemOpen
  \bibfield{author}{%
  \bibinfo {author} {\bibfnamefont{I.}~\bibnamefont{Affleck}},\ }%
  \bibfield{journal}{%
  \bibinfo {journal} {Phys. Rev. Lett.}\ }%
  \textbf{\bibinfo {volume} {56}},\ \bibinfo {pages} {746} (\bibinfo {year}
  {1986}).%
\bibitem{YuFo92}%
  \BibitemOpen
  \bibfield{author}{%
  \bibinfo {author} {\bibfnamefont{N.}~\bibnamefont{Yu}}\ and\ \bibinfo
  {author} {\bibfnamefont{M.}~\bibnamefont{Fowler}},\ }%
  \bibfield{journal}{%
  \bibinfo {journal} {Phys. Rev. B}\ }%
  \textbf{\bibinfo {volume} {46}},\ \bibinfo {pages} {14583} (\bibinfo {year}
  {1992})%
  \bibAnnoteFile{NoStop}{YuFo92}%
\bibitem{JaSa06}%
  \BibitemOpen
  \bibfield{author}{%
  \bibinfo {author} {\bibfnamefont{J.~L.}\ \bibnamefont{Jacobsen}}\ and\
  \bibinfo {author} {\bibfnamefont{H.}~\bibnamefont{Saleur}},\ }%
  \bibfield{journal}{%
  \bibinfo {journal} {Nucl. Phys. B}\ }%
  \textbf{\bibinfo {volume} {743}},\ \bibinfo {pages} {207} (\bibinfo {year}
  {2006}),\ \Eprint{http://arxiv.org/abs/cond-mat/0512058}{cond-mat/0512058}%
  \bibAnnoteFile{NoStop}{JaSa06}%
\bibitem{Card86a}%
  \BibitemOpen
  \bibfield{author}{%
  \bibinfo {author} {\bibfnamefont{J.~L.}\ \bibnamefont{Cardy}},\ }%
  \bibfield{journal}{%
  \bibinfo {journal} {Nucl. Phys. B}\ }%
  \textbf{\bibinfo {volume} {270}},\ \bibinfo {pages} {186} (\bibinfo {year}
  {1986})%
  \bibAnnoteFile{NoStop}{Card86a}%
\bibitem{Ikhl11}%
  \BibitemOpen
  \bibfield{author}{%
  \bibinfo {author} {\bibfnamefont{Y.}~\bibnamefont{Ikhlef}},\ }%
  \bibfield{journal}{%
  \bibinfo {journal} {Mod. Phys. Lett. B}\ }%
  \textbf{\bibinfo {volume} {25}},\ \bibinfo {pages} {291} (\bibinfo {year}
  {2011}),\ \Eprint{http://arxiv.org/abs/1012.2380}{arXiv:1012.2380}%
  \bibAnnoteFile{NoStop}{Ikhl11}%
\bibitem{PoZv93}%
  \BibitemOpen
  \bibfield{author}{%
  \bibinfo {author} {\bibfnamefont{V.~Y.}\ \bibnamefont{Popkov}}\ and\ \bibinfo
  {author} {\bibfnamefont{A.~A.}\ \bibnamefont{Zvyagin}},\ }%
  \bibfield{journal}{%
  \bibinfo {journal} {Phys. Lett. A}\ }%
  \textbf{\bibinfo {volume} {175}},\ \bibinfo {pages} {295} (\bibinfo {year}
  {1993})%
  \bibAnnoteFile{NoStop}{PoZv93}%
\bibitem{FrRo97}%
  \BibitemOpen
  \bibfield{author}{%
  \bibinfo {author} {\bibfnamefont{H.}~\bibnamefont{Frahm}}\ and\ \bibinfo
  {author} {\bibfnamefont{C.}~\bibnamefont{R{\"o}denbeck}},\ }%
  \bibfield{journal}{%
  \bibinfo {journal} {J. Phys. A}\ }%
  \textbf{\bibinfo {volume} {30}},\ \bibinfo {pages} {4467} (\bibinfo {year}
  {1997}),\ \Eprint{http://arxiv.org/abs/cond-mat/9702083}{cond-mat/9702083}%
  \bibAnnoteFile{NoStop}{FrRo97}%
\bibitem{Hame85}%
  \BibitemOpen
  \bibfield{author}{%
  \bibinfo {author} {\bibfnamefont{C.~J.}\ \bibnamefont{Hamer}},\ }%
  \bibfield{journal}{%
  \bibinfo {journal} {J. Phys. A}\ }%
  \textbf{\bibinfo {volume} {18}},\ \bibinfo {pages} {L1133} (\bibinfo {year}
  {1985})%
  \bibAnnoteFile{NoStop}{Hame85}%
\bibitem{WoEc87a}%
  \BibitemOpen
  \bibfield{author}{%
  \bibinfo {author} {\bibfnamefont{F.}~\bibnamefont{Woynarovich}}\ and\
  \bibinfo {author} {\bibfnamefont{H.-P.}\ \bibnamefont{Eckle}},\ }%
  \bibfield{journal}{%
  \bibinfo {journal} {J. Phys. A}\ }%
  \textbf{\bibinfo {volume} {20}},\ \bibinfo {pages} {L97} (\bibinfo {year}
  {1987})%
  \bibAnnoteFile{NoStop}{WoEc87a}%
\bibitem{AlBB87}%
  \BibitemOpen
  \bibfield{author}{%
  \bibinfo {author} {\bibfnamefont{F.~C.}\ \bibnamefont{Alcaraz}}, \bibinfo
  {author} {\bibfnamefont{M.~N.}\ \bibnamefont{Barber}},\ and\ \bibinfo
  {author} {\bibfnamefont{M.~T.}\ \bibnamefont{Batchelor}},\ }%
  \bibfield{journal}{%
  \bibinfo {journal} {Phys. Rev. Lett.}\ }%
  \textbf{\bibinfo {volume} {58}},\ \bibinfo {pages} {771} (\bibinfo {year}
  {1987})%
  \bibAnnoteFile{NoStop}{AlBB87}%
\bibitem{FrPT98}%
  \BibitemOpen
  \bibfield{author}{%
  \bibinfo {author} {\bibfnamefont{H.}~\bibnamefont{Frahm}}, \bibinfo {author}
  {\bibfnamefont{M.~P.}\ \bibnamefont{Pfannm\"uller}},\ and\ \bibinfo {author}
  {\bibfnamefont{A.~M.}\ \bibnamefont{Tsvelik}},\ }%
  \bibfield{journal}{%
  \bibinfo {journal} {Phys. Rev. Lett.}\ }%
  \textbf{\bibinfo {volume} {81}},\ \bibinfo {pages} {2116} (\bibinfo {year}
  {1998}),\ \Eprint{http://arxiv.org/abs/cond-mat/9803145}{cond-mat/9803145}%
  \bibAnnoteFile{NoStop}{FrPT98}%
\bibitem{Frahm99}%
  \BibitemOpen
  \bibfield{author}{%
  \bibinfo {author} {\bibfnamefont{H.}~\bibnamefont{Frahm}},\ }%
  \bibfield{journal}{%
  \bibinfo {journal} {Nucl. Phys. B}\ }%
  \textbf{\bibinfo {volume} {559}},\ \bibinfo {pages} {613} (\bibinfo {year}
  {1999}),\ \Eprint{http://arxiv.org/abs/cond-mat/9904157}{cond-mat/9904157}%
  \bibAnnoteFile{NoStop}{Frahm99}%
\bibitem{AlMa89a}%
  \BibitemOpen
  \bibfield{author}{%
  \bibinfo {author} {\bibfnamefont{F.~C.}\ \bibnamefont{Alcaraz}}\ and\
  \bibinfo {author} {\bibfnamefont{M.~J.}\ \bibnamefont{Martins}},\ }%
  \bibfield{journal}{%
  \bibinfo {journal} {Phys. Rev. Lett.}\ }%
  \textbf{\bibinfo {volume} {63}},\ \bibinfo {pages} {708} (\bibinfo {year}
  {1989})%
  \bibAnnoteFile{NoStop}{AlMa89a}%
\bibitem{AlMa89}%
  \BibitemOpen
  \bibfield{author}{%
  \bibinfo {author} {\bibfnamefont{F.~C.}\ \bibnamefont{Alcaraz}}\ and\
  \bibinfo {author} {\bibfnamefont{M.~J.}\ \bibnamefont{Martins}},\ }%
  \bibfield{journal}{%
  \bibinfo {journal} {J. Phys. A}\ }%
  \textbf{\bibinfo {volume} {22}},\ \bibinfo {pages} {1829} (\bibinfo {year}
  {1989})%
  \bibAnnoteFile{NoStop}{AlMa89}%
\bibitem{FrYF90}%
  \BibitemOpen
  \bibfield{author}{%
  \bibinfo {author} {\bibfnamefont{H.}~\bibnamefont{Frahm}}, \bibinfo {author}
  {\bibfnamefont{N.-C.}\ \bibnamefont{Yu}},\ and\ \bibinfo {author}
  {\bibfnamefont{M.}~\bibnamefont{Fowler}},\ }%
  \bibfield{journal}{%
  \bibinfo {journal} {Nucl. Phys. B}\ }%
  \textbf{\bibinfo {volume} {336}},\ \bibinfo {pages} {396} (\bibinfo {year}
  {1990})%
  \bibAnnoteFile{NoStop}{FrYF90}%
\bibitem{FrYu90}%
  \BibitemOpen
  \bibfield{author}{%
  \bibinfo {author} {\bibfnamefont{H.}~\bibnamefont{Frahm}}\ and\ \bibinfo
  {author} {\bibfnamefont{N.-C.}\ \bibnamefont{Yu}},\ }%
  \bibfield{journal}{%
  \bibinfo {journal} {J. Phys. A}\ }%
  \textbf{\bibinfo {volume} {23}},\ \bibinfo {pages} {2115} (\bibinfo {year}
  {1990})%
  \bibAnnoteFile{NoStop}{FrYu90}%
\end{thebibliography}
%


\newpage
%
\begin{table}
  \caption{\label{one1A}Finite-size sequences (\ref{estimA1}) of the anomalous
    dimension $X_{1,0}^{0,0}(\gamma)$ corresponding to the ground state of the
    antiferromagnetic mixed superspin chain for $\gamma=2\pi/5, 2\pi/9$ and $b
    =  0.1,0.2,0.3,0.4$ from the Bethe ansatz.  The expected exact conformal
    dimension is $X_{1,0}^{0,0}(\gamma)=\frac{1}{4}$, independent of $\gamma$
    and $b$.
  }
\begin{tabular}{|c||c|c|c|c|}
  \hline
$X_{1,0}^{0,0}(2\pi/5)$ & $b=0.1$ & $b=0.2$ & $b=0.3$ & $ b=0.4$ \\ \hline \hline
64 & 0.250048612966262   &    0.250057352809966   &  0.250079908701413  &     0.250153486545075 \\
68 & 0.250044800002668   &    0.250052960280743   &  0.250074088242237  &     0.250143781486991 \\
72 & 0.250041480552865   &    0.250049125512265   &  0.250068974707988  &     0.250135085547961 \\
76 & 0.250038568285938   &    0.250045752223167   &  0.250064450636514  &     0.250127260002517 \\
80 & 0.250035995187684   &    0.250042765513805   &  0.250060424128183  &     0.250120188361586 \\
84 &  0.250033708784690      &  0.250040103975741 &     0.250056819907415   &     0.250113771814941\\
88 &  0.250031664084155      &  0.250037719523553  &    0.250053577683175    &    0.250107929518090\\
92 &  0.250029826347013     &   0.250035574034939  &    0.250050646736442     &   0.250102591235422\\
96 &  0.250028167338885    &    0.250033632461034   &   0.250047987323579     &   0.250097698864218\\
100&   0.250026663737575  &      0.250031869300816  &    0.250045563476697    &    0.250093198867507\\
Extrap. & 0.250002(1)  & 0.250001(1)& 0.24999984(2)& 0.2500001(3)  \\ \hline
\hline 
$X_{1,0}^{0,0}(2\pi/9)$ & $b=0.1$ & $b=0.2$ & $b=0.3$ & $ b=0.4$ \\ \hline \hline
64 & 0.253132058479334   &    0.253431444329609   &  0.254106625377260  &     0.255863205280181 \\
68 & 0.253025177990501   &    0.253314682476050   &  0.253967922029803  &     0.255670948431118 \\
72 & 0.252927764888214   &    0.253208226471161   &  0.253841349125435  &     0.255494860072799 \\
76 & 0.252838520004755   &    0.253110668379048   &  0.253725265274230  &     0.255332854789668 \\
80 & 0.252756378517378   &    0.253020852585427   &  0.253618322313239  &     0.255183198042987  \\
84 & 0.252680460193235   &    0.252937821882064  &   0.253519399812617  &     0.255044436405680\\
88 & 0.252610028646216   &    0.252860776548557  &   0.253427560977248  &     0.254915339947881\\
92 & 0.252544463265816   &    0.252789041850631  &   0.253342011784391  &     0.254794862710576 \\
96 & 0.252483237655954   &    0.252722043984628  &   0.253262079561925  &     0.254682108820548\\
100&  0.252425899534380  &     0.252659291643748 &    0.253187183973606 &      0.254576304302893\\
Extrap.& 0.25003(2)  & 0.24995(3)& 0.25004(2)& 0.24997(2)    \\ \hline
\end{tabular}
\end{table}
%
\begin{table}
  \caption{\label{two1A}Finite-size sequences (\ref{estimA1}) of the anomalous
    dimension $X_{0,0}^{\half,\half}(\gamma)$ of the antiferromagnetic mixed
    superspin chain for $\gamma=2\pi/5, 2\pi/9$ and $b = 0.1,0.2,0.3,0.4$ from
    the Bethe ansatz. The expected exact conformal dimensions are
    $X_{0,0}^{\half,\half}(2\pi/5)= 5/12$ and $X_{0,0}^{\half,\half}(2\pi/9)=
    9/28$ independent of $b$.   
  }
\begin{tabular}{|c||c|c|c|c|}
  \hline
$X_{0,0}^{\half,\half}(2\pi/5)$ & $b=0.1$ & $b=0.2$ 
       & $b=0.3$ & $ b=0.4$ \\ \hline \hline
64 &  0.416670417631978  &     0.416656218465817 &    0.416610197227320  &     0.416354881252023 \\
68 &  0.416669987565812  &     0.416657408090032 &    0.416616640599484  &     0.416390524909717 \\
72 &  0.416669627133090  &     0.416658405958250 &    0.416622040898389  &     0.416420383057850 \\
76 &  0.416669322128664  &     0.416659250096786 &    0.416626611560886  &     0.41644564432787 \\
80  & 0.416669062413019  &     0.416659971230323 &    0.416630513869716  &     0.416467205721020 \\
84 & 0.416668838812804  &     0.416660591916285 &  0.416633872109781    &   0.416485757527825 \\ 
88 & 0.416668644721943  &     0.416661070729271 &    0.416636783046771  &     0.416501777174014\\
92 & 0.416668475963537  &     0.416661547020412 &    0.416639322642474  &     0.416515806307798\\
96 & 0.416668327514699  &     0.416661965760863 &    0.416641552261224  &     0.416528116518884\\
100&  0.416668196783305 &      0.416662334075060&     0.416643519898230 &      0.416538978974116\\
Extrap.&  0.416667(1) & 0.4166663(1)& 0.416667(2)& 0.416667(3)  \\ \hline 
\hline
$X_{0,0}^{\half,\half}(2\pi/9)$ & $b=0.1$ & $b=0.2$    & $b=0.3$ & $ b=0.4$ \\ \hline \hline
64 & 0.321478885785230    &   0.321480244434995   &  0.321478675589893    &   0.321419894207957\\
68 & 0.321475362972286    &   0.321477017780511   &  0.321476806551588    &   0.321428958054849\\
72 & 0.321472274265546    &   0.321474149284625   &  0.321475003385914    &   0.321436049810230\\
76 & 0.321469545411844    &   0.321471583890475   &  0.321473278305405    &   0.321441630433088\\
80 & 0.321467123806211    &   0.321469276575399   &  0.321471637081771    &   0.321446038562155\\
84 & 0.321464951878455  &     0.321467191544666  &   0.321470080991823  &     0.321449528944129\\
88 & 0.321462997920481  &     0.321465298207456  &   0.321468607979076  &     0.321452293350554\\
92 & 0.321461231497336  &     0.321463572029351  &   0.321467215870929  &     0.321454479967015\\
96 & 0.321459626953652  &     0.321461991599086  &   0.321465900513161  &     0.321456201956970\\
100&  0.321458164206755 &      0.321460540439523 &    0.321464653145202 &      0.321457549599824\\
Extrap.&0.321458(1) & 0.321426(3)& 0.321426(1)& 0.32143(1)  \\ \hline 
\end{tabular}
\end{table}
\begin{table}
\caption{\label{three1A}
  Finite-size sequences (\ref{estimA1}) of the anomalous dimension
  $X_{1,-1}^{\half,\half}(\gamma)$ of the antiferromagnetic mixed superspin
  chain for $\gamma=2\pi/5, 2\pi/9$ and $b = 0.1, 0.2,0.3,0.4$ from the Bethe
  ansatz. The expected exact conformal dimensions are
  $X_{1,-1}^{\half,\half}(2\pi/5)= 49/60$ and $X_{1,-1}^{\half,\half}(2\pi/9)=
  137/252$ independent of $b$. 
}
\begin{tabular}{|c||c|c|c|c|}
  \hline
$X_{1,-1}^{\half,\half}(2\pi/5)$ & $b=0.1$ & $b=0.2$ & $b=0.3$ & $ b=0.4$ \\ \hline \hline
64&  0.802543420915095  &     0.800996714941326 &    0.797451714287750  &     0.788002733672379\\
68&  0.803109142135808  &     0.801619792314736 &    0.798202751578452  &     0.789061829077160\\
72&  0.803621371338520  &     0.802184427361242 &    0.798884681127158  &     0.790029738004235\\
76&  0.804087771606711  &     0.802698923084324 &    0.799507125805111  &     0.790918241220402\\
80&  0.804514571755953  &     0.803170041748389 &    0.800077960470532  &     0.791737149187162 \\
84&  0.804906894958283  &     0.803603354252052 &    0.800603695001582  &     0.792494698387001\\
88&  0.805268993033042  &     0.804003490361879 &    0.801089763004215  &     0.793197855830765\\
92&  0.805604428993397  &     0.804374337112643 &    0.801540740173121  &     0.793852555009990\\
96&  0.805916212570146  &     0.804719178147089 &    0.801960503234090  &     0.794463882917438\\
100&  0.806206904992800 &      0.805040815482502 &    0.802352368121435  &     0.79503622240196\\
Extrap. & 0.81661(2) & 0.81667(1) & 0.8166(2) & 0.8174(2)  \\ \hline
  \hline
$X_{1,-1}^{\half,\half}(2\pi/9)$ & $b=0.1$ & $b=0.2$ & $b=0.3$ & $ b=0.4$ \\ \hline \hline
64&  0.469649001673299  &     0.466575740909034 &    0.460184330193937  &     0.445977608618501\\
68&  0.470774831559801  &     0.467741761599685 &    0.461432012191324  &     0.447396493043942\\
72&  0.471822545482396  &     0.468827088511210 &     0.462593839845175 &      0.448719147269481 \\
76&  0.472801422875943  &     0.469841288401414 &    0.463679948377037  &     0.449956804126546\\
80&  0.473719226661201  &     0.470792368025533 &    0.464698825642782  &     0.451118889793702\\
84&  0.474582512357382  &     0.471687089632269 &    0.465657646470560  &     0.452213383888303 \\
88&  0.475396863517694  &     0.472531215460570 &    0.466562525609641  &     0.453247098947592 \\
92&  0.476167071653444  &     0.473329690796934 &    0.467418715541751  &     0.454225894135166\\
96&  0.476897277726334  &     0.474086790324252 &    0.468230758833534  &     0.455154844481359 \\
100&  0.477591081020453 &      0.474806231909338 &    0.469002608145475 &      0.456038369643072\\
Extrap. & 0.5469(2) & 0.546(1) & 0.5492(2) & 0.547(1)  \\ \hline
\end{tabular}
\end{table}
\begin{table}
\caption{\label{four1A}
  Finite-size sequences (\ref{estimA1}) of the anomalous dimension
  $X_{1,0}^{1,0}(\gamma)$ of the antiferromagnetic mixed superspin chain for
  $\gamma=2\pi/5, 2\pi/9$ and $b = 0.1, 0.2,0.3,0.4$ from the Bethe
  ansatz. The expected exact conformal dimensions are $X_{1,0}^{1,0}(2\pi/5)=
  31/24$ and $X_{1,0}^{1,0}(2\pi/9)= 95/56$ independent of $b$.
}
\begin{tabular}{|c||c|c|c|c|}
  \hline
$X_{1,0}^{1,0}(2\pi/5)$ & $b=0.1$ & $b=0.2$ & $b=0.3$ & $ b=0.4$ \\ \hline \hline
64 &  1.29169391271122 &       1.29188030171550 &    1.29246176734499  &      1.29549189382126\\
68 &  1.29169747618030 &       1.29186417985947 &    1.29238361004235  &      1.29508481229779\\
72 &  1.29169998846533 &       1.29185008220992 &    1.29231721713292  &      1.29474153535161\\
76 &  1.29170171985766 &       1.29183766315244 &    1.29226028317640  &      1.29444925164532\\
80 &  1.29170286782319 &       1.29182665069729 &    1.29221104956300  &      1.29419822711471\\
84 &  1.29170357644233 &       1.29181682617240 &    1.29216815127140  &      1.29398095434414\\
88 &  1.29170395121063 &       1.29180801570822 &    1.29213051785738  &      1.29379156808294\\
92 &  1.29170407360045 &       1.29180007506879 &    1.29209729685151  &      1.29362543257076  \\
96 &  1.29170400266840 &       1.29179288604481 &    1.29206780649186  &      1.29347884081817\\
100 &  1.29170378640977 &       1.29178635018844 &    1.29204149259774 &       1.29334880116668\\
Extrap. & 1.2917(1) & 1.29165(2) & 1.291665(2) & 1.291667(1)  \\ \hline
  \hline
$X_{1,0}^{1,0}(2\pi/9)$ & $b=0.1$ & $b=0.2$ & $b=0.3$ & $ b=0.4$ \\ \hline \hline
64 &  1.65831148570030  &      1.65491024458575 &    1.64745726183318 &       1.63027640189379\\
68 &  1.65967750746320  &      1.65637015514941 &    1.64909748064187 &       1.63208407458946\\
72 &  1.66091746837352  &      1.65769829971046 &    1.65059835126374 &       1.63378198262557\\
76 &  1.66204931417987  &      1.65891300418380 &    1.65197797284606 &       1.63537771386844 \\
80 &  1.66308763279172  &      1.66002924590730 &    1.65325137518418 &       1.63687886760066 \\
84 &  1.66404441718139  &      1.66105939574434 &    1.65443114647623 &       1.63829274704172\\
88 &  1.66492962800063  &      1.66201377134831 &    1.65552791636658 &       1.63962620939402\\
92 &  1.66575161458998  &      1.66290105323150 &    1.65655072391135 &       1.64088560250871\\
96 &  1.66651743360543  &      1.66372860292841 &    1.65750730871397 &       1.64207675144390\\
100 &  1.66723309416254 &       1.66450270787275 &    1.65840433835020 &      1.64320497592494 \\
Extrap. & 1.69638(2) & 1.69638(1) & 1.6969(3) & 1.661(3)  \\ \hline
\end{tabular}
\end{table}
\begin{table}
\caption{\label{five1A}Finite-size sequences (\ref{estimA1}) of the anomalous
  dimension $X_{1,1}^{\half,\half}(\gamma)$ of the antiferromagnetic mixed superspin
  chain for $\gamma=2\pi/5, 2\pi/9$ and  $b = 0.1, 0.2, 0.3, 0.4 $ from the
  Bethe ansatz. The expected exact conformal dimensions are
  $X_{1,1}^{\half,\half}(2\pi/5)=61/60$ and
  $X_{1,1}^{\half,\half}(2\pi/9)=277/252$ independent of $b$. 
}
\begin{tabular}{|c||c|c|c|c|}
  \hline
$X_{1,1}^{\half,\half}(2\pi/5)$ & $b=0.1$ & $b=0.2$ & $b=0.3$ & $ b=0.4$ \\ \hline \hline
64&   1.01660809460145  &      1.01674796672621  &   1.01720030948687   &     1.01970307551162\\
68&   1.01661480621747  &      1.01673872337252  &   1.01713948238592   &     1.01935718467132\\
72&   1.01662042714845  &      1.01673097085398  &   1.01708849236842   &     1.01906718883947\\
76&   1.01662518207921  &      1.01672440643128  &   1.01704532699284   &     1.01882166496922\\
80&   1.01662923940361  &      1.01671879759808  &   1.01700846418273   &     1.01861196515601\\
84&   1.01663273029972  &      1.01671396921624  &   1.01697673295315   &     1.01843144543536\\
88&   1.01663575356290  &      1.01670978164280  &   1.01694922441500   &     1.01827493321853\\
92&   1.01663839163587  &      1.01670612582146  &   1.01692522084310   &     1.01813835363556\\
96&   1.01664070491891  &      1.01670291651396  &   1.01690415103221   &     1.01801845973993\\
100 &   1.01664274594897&        1.01670008430629 &    1.01688555504759 &       1.01791264141798\\
Extrap.  & 1.01665(1) &  1.01665(2) & 1.016663(1) & 1.06665(3)   \\ \hline
  \hline
$X_{1,1}^{\half,\half}(2\pi/9)$ & $b=0.1$ & $b=0.2$ & $b=0.3$ & $ b=0.4$ \\ \hline \hline
64 &  1.09896258489785   &     1.09906445762706 &    1.09941621118507   &     1.10154807320311\\
68 &  1.09898211749043   &     1.09907068232367 &    1.09937791947098   &     1.10125137082147\\
72 &  1.09899899574197   &     1.09907651035724 &    1.09934670412561   &     1.10100445331082\\
76 &  1.09901370764206   &     1.09908195585909 &    1.09932102076494   &     1.10079694624626\\
80 &  1.09902663078237   &     1.09908704065466 &    1.09929971588858   &     1.10062102508163\\
84 &  1.09903806201443   &     1.09909178836045 &    1.09928191439564   &     1.10047070214516\\
88 &  1.09904823720116   &     1.09909622419276 &    1.09926694360684   &     1.10034133469567\\
92 &  1.09905734493280   &     1.09910037169640 &    1.09925428104614   &     1.10022927695166 \\
96 &  1.09906553989567   &     1.09910425310586 &    1.09924351573734   &     1.10013163872575\\
100&   1.09907294839689  &      1.09910789126798&     1.09923432056007  &      1.10004610173584\\
Extrap.  &1.09922(1) & 1.09921(2) & 1.0991(1) & 1.0991(2)  \\ \hline
\end{tabular}
\end{table}
\begin{table}
\caption{\label{six1A}Finite-size sequences (\ref{estimA1}) of the anomalous
  dimension  $X_{1,0}^{1,1}(\gamma)$  of the antiferromagnetic mixed superspin
  chain for $\gamma=2\pi/5, 2\pi/9$ and $b = 0.1,0.2,0.3,0.4$
  from the Bethe ansatz. The expected exact
  conformal   dimensions are  $X_{1,0}^{1,1}(2\pi/5)=23/12$ and
  $X_{1,0}^{1,1}(2\pi/9)=43/28$ independent of $b$.
}
\begin{tabular}{|c||c|c|c|c|}
  \hline
$X_{1,0}^{1,1}(2\pi/5)$ & $0.1$ & $0.2$ & $0.3$ & $ 0.4$ \\ \hline \hline
64&   1.91633350061679  &      1.91671340547377 &    1.91794265757312  &      1.92474331627926\\
68&   1.91637144225241  &      1.91670796028526 &    1.91779690210299  &      1.92382383023129\\
72&   1.91640325233943  &      1.91670340851577 &    1.91767474571176  &      1.92305262480562 \\
76&   1.91643018323036  &      1.91669956611603 &    1.91757135907367  &      1.92239949136765\\
80&   1.91645318416930  &      1.91669629383031 &    1.91748308691517  &      1.92184153275738\\
84&   1.91647298557160  &      1.91669348333802 &    1.91740712239614  &      1.92136113809079\\
88&   1.91649015396588  &      1.91669105320029 &    1.91734128093283  &      1.9209445839460\\
92&   1.91650513545003  &      1.91668893792311 &    1.91728384016498  &      1.92058105193391\\
96&   1.91651828702507  &      1.91668708483102 &    1.91723343201604  &      1.92026191802231\\
100&   1.91652989610853 &       1.91668545343676 &    1.91718895267849 &       1.91998024230886\\
Extrap.& 1.91666(1) &  1.91665(1) & 1.91666(2) & 1.9166(1)   \\ \hline 
  \hline
$X_{1,0}^{1,1}(2\pi/9)$ & $0.1$ & $0.2$ & $0.3$ & $ 0.4$ \\ \hline \hline
64&   1.54595893269632  &      1.54714023224294 &    1.55001371490657  &      1.55969299725040\\
68&   1.54564437619687  &      1.54677034969013 &    1.54949371360320  &      1.55852248544641\\
72&   1.54535487074355  &      1.54643219432769 &    1.54902475966904  &      1.55749654171679\\
76&   1.54508734785315  &      1.54612150411845 &    1.54859900060520  &      1.55658895839048 \\
80&   1.54483923019746  &      1.54583477645593 &    1.54821017751172  &      1.55577950742735\\
84&   1.54460833610583  &      1.54556910385993 &    1.54785323101280  &      1.55505233513379 \\
88&   1.54439280622750  &      1.54532204968426 &    1.54752402474063  &      1.55439485052312\\
92&   1.54419104585411  &      1.54509155364219 &    1.54721913528093  &      1.55379692588342\\
96&   1.54400168081111  &      1.54487586080783 &    1.54693570616330  &      1.55325032866461\\
100&   1.54382351731674 &       1.54467346577192&     1.54667132942653 &       1.55274829423512\\
Extrap.& 1.5356(1) & 1.5358(1) & 1.5351(2) & 1.535(1)   \\ \hline 
\end{tabular}
\end{table}

\begin{table}
\caption{\label{one1B}Finite-size sequences (\ref{estimA2}) of the anomalous
  dimension $X_{1,0}^{0,0}(\gamma)$ corresponding to the ground state 
  of the ferromagnetic mixed superspin
  chain for $\gamma=2\pi/5, 2\pi/9$ and $b = 0.1,0.2,0.3,0.4$
  from the Bethe   ansatz.  The expected exact conformal
  dimension is  $X_{1,0}^{0,0}(\gamma)=\frac{1}{4}$ independent of $\gamma$
  and $b$. 
}
\begin{tabular}{|c||c|c|c|c|}
  \hline
$X_{1,0}^{0,0}(2\pi/5)$ & $b=0.1$ & $b=0.2$ & $b=0.3$ & $ b=0.4$ \\ \hline \hline
4 & 0.249618388905792  & 0.249864078062197 & 0.250380674609993 & 0.251433496876099 \\
8 & 0.249912388797508  & 0.249965813355377 & 0.250079922137901 & 0.250320222469547 \\
12 & 0.249962057861862 & 0.249985305677197 & 0.250035162436907 & 0.250140964674203 \\
16 & 0.249978921372708 & 0.249991918899564 & 0.250019836847325 & 0.250079269513615 \\
20 & 0.249986607712271 & 0.249994907534576 & 0.250012748900962 & 0.250050790847099 \\
24 & 0.249990744279497 &0.249996502689051  & 0.250008886931439 & 0.250035317308795 \\
28 & 0.249993222825207 &0.249997451946100  & 0.250006549862109 & 0.250025978300525 \\
32 & 0.249994824338185 &0.249998061879352  & 0.250005028056580 & 0.250019910102813 \\
36 & 0.249995918536672 &0.249998476736720  & 0.250003981517046 & 0.250015745271696 \\
40 & 0.249996699224027 &0.249998771167758  & 0.250003231052877 & 0.250012763515263 \\
Extrap. & 0.25000003(2)  & 0.25000005(1)& 0.25000007(2)& 0.2499996(3)  \\ \hline
  \hline
$X_{1,0}^{0,0}(2\pi/9)$ & $b=0.1$ & $b=0.2$ & $b=0.3$ & $ b=0.4$ \\ \hline \hline
4 &  0.244044806842760 &      0.244461720024188 &    0.245172840974522 &      0.246203546383614 \\
8 &  0.248700725305516   &    0.248802997575032   &  0.248979563899314     &  0.249240366426739 \\
12 &  0.249434549177625  &     0.249479482670048  &   0.249557418184965    &   0.249673343839989 \\
16 &  0.249684085421290  &     0.249709260050335  &   0.249752996660027    &   0.249818218822082 \\
20 &  0.249798433771513  &     0.249814515658493  &   0.249842476615247    &   0.249884223298235 \\
24 &  0.249860253886827  &     0.249871410635206  &   0.249890816324928    &   0.249919808908815 \\
28 &  0.249897430776020  &     0.249905622406331  &   0.249919874671653    &   0.249941175992748  \\
32 &  0.249921520599999  &     0.249927789931196  &   0.249938699258291    &   0.249955008463470 \\
36 & 0.249938018641971  &     0.249942970904318  &   0.249951589132296    &   0.249964475692600 \\
40 &  0.249949810553310  &     0.249953821065760  &   0.249960801110223    &   0.249971239248729 \\
Extrap.& 0.2500003(2)  & 0.25000007(3)& 0.25000002(2)& 0.250000031(1)    \\ \hline
\end{tabular}
\end{table}
%
\begin{table}
\caption{\label{two1B}Finite-size sequences (\ref{estimA2}) of the anomalous
  dimension $X_{0,0}^{\half,-\half}(\gamma)$ of the ferromagnetic mixed superspin
  chain for $\gamma=2\pi/5, 2\pi/9$ and $b = 0.1,0.2,0.3,0.4$
  from the Bethe ansatz. The expected exact conformal
  dimensions are $X_{0,0}^{\half,-\half}(2\pi/5)= 5/12$ and
  $X_{0,0}^{\half,-\half}(2\pi/9)= 9/28$ independent of $b$.
}
\begin{tabular}{|c||c|c|c|c|}
  \hline
$X_{0,0}^{\half,-\half}(2\pi/5)$ & $b=0.1$ & $b=0.2$ 
       & $b=0.3$ & $ b=0.4$ \\ \hline \hline
4 & 0.420153202195165  & 0.418311223187811 & 0.414094943222730 & 0.40447340188129 \\
8 & 0.417533635282876  & 0.417139647726513 & 0.416282739919060 & 0.414402513758005 \\
12 & 0.417050902689717 & 0.416880022540026 & 0.416510598502611 & 0.415713015644990 \\
16 & 0.416882479860486 & 0.416787136134762 & 0.416581379806527 & 0.416139003433824 \\
20 & 0.416804662357638 & 0.416743862247532 & 0.416612753609402 & 0.416331359972793 \\
24 & 0.416762438843948 & 0.416720296295040 & 0.416629457157763 & 0.416434667265353 \\
28 & 0.416736999096700 & 0.416706071365318 & 0.416639421617616 & 0.416496576350315 \\
32 & 0.416720497319091 & 0.416696834653756 & 0.416645848671003 & 0.416536610323140 \\
36 & 0.416709188622272 & 0.416690500639723 & 0.416650237899716 & 0.416563993281584 \\
40 & 0.416701102495937 & 0.416685969815732 & 0.416653369568434 & 0.416583549056764 \\
Extrap.&  0.41666668(2) & 0.41666662(3)& 0.4166665(3)& 0.4166661(3)  \\ \hline 
  \hline
$X_{0,0}^{\half,-\half}(2\pi/9)$ & $b=0.1$ & $b=0.2$ 
       & $b=0.3$ & $ b=0.4$ \\ \hline \hline
4 &  0.319913713673340  & 0.320018967155828 & 0.320190591772206 & 0.320420310281376 \\
8 & 0.321157712483394   & 0.321191122084441 & 0.321248225946738 & 0.321331244861727 \\
12 &  0.321314654780622 & 0.321329741251375 & 0.321355797251027 & 0.321394295060097 \\
16 &  0.321365627553391 & 0.321374157960506 & 0.321388943280466 & 0.321410910201736 \\
20 & 0.321388611392069  & 0.321394083529534 & 0.321403583460742 & 0.321417733723489 \\
24 &  0.321400941690716 & 0.321404746407637 & 0.321411357486342 & 0.321421218332086 \\
28 &  0.321408324757022 & 0.321411122201603 & 0.321415985350110 & 0.321423245212332 \\
32 &  0.321413096052956 & 0.321415238901914 & 0.321418965356214 & 0.321424531307902 \\
36 &  0.321416358100397 & 0.321418051745520 & 0.321420997553802 & 0.321425399388384 \\
40 & 0.321418686741470  & 0.321420059101585 & 0.321422446146568 & 0.321426013997661 \\
Extrap.&0.321425(2) & 0.321428(34)& 0.3214285(3)& 0.3214286(3)  \\ \hline 
\end{tabular}
\end{table}
\begin{table}
\caption{\label{three1B}
  Finite-size sequences (\ref{estimA2}) of the anomalous dimension
  $X_{1,1}^{\half,-\half}(\gamma)$  of the ferromagnetic mixed superspin
  chain for $\gamma=2\pi/5, 2\pi/9$ and $b = 0.1, 0.2,0.3,0.4$
  from the Bethe ansatz. The expected exact conformal
  dimensions are $X_{1,1}^{\half,-\half}(2\pi/5)= 49/60$ and
  $X_{1,1}^{\half,-\half}(2\pi/9)= 137/252$ independent of $b$.
}
\begin{tabular}{|c||c|c|c|c|}
  \hline
$X_{1,1}^{\half,-\half}(2\pi/5)$ & $b=0.1$ & $b=0.2$ & $b=0.3$ & $ b=0.4$ \\ \hline \hline
4 &  0.804938854579059  & 0.814319917467641 & 0.834792393417294 & 0.879143521859337 \\
8 &  0.813596692907556  & 0.815893063874643 & 0.820852468461199 & 0.831532156138152 \\
12 &  0.815277718645391 & 0.816292762124510 & 0.818482319585240 & 0.823183312552782 \\
16 &  0.815878039474781 & 0.816447384142183 & 0.817674984015266 & 0.820308238730800 \\
20 &  0.816159032430526 & 0.816522772350630 & 0.817306868984783 & 0.818988000334640 \\
24 &  0.816312767449947 & 0.816565061146465 & 0.817108836773558 & 0.818274369988834 \\
28 &  0.816405927448350 & 0.816591124286526 & 0.816990242152023 & 0.817845542170740 \\
32 &  0.816466614117071 & 0.816608311410614 & 0.816913659290931 & 0.817567916228810 \\
36 &  0.816508337929270 & 0.816620237845708 & 0.816861360234222 & 0.817377944643833 \\
40 &  0.816538249251599 & 0.816628850050168 & 0.816824067600668 & 0.817242267787161 \\
Extrap. & 0.81666693(2) & 0.8166668(2) & 0.8166675(2) & 0.816673(2)  \\ \hline
  \hline
$X_{1,1}^{\half,-\half}(2\pi/9)$ & $b=0.1$ & $b=0.2$ & $b=0.3$ & $ b=0.4$ \\ \hline \hline
4 &  0.533880505305017 & 0.534967698841839 & 0.536853524001006 & 0.539660072909485 \\
8 &  0.541383434410729 & 0.541648424113656 & 0.542108818742814  & 0.542795515425274 \\
12 &  0.542654427057378 & 0.542771434479493 & 0.542974956206111 & 0.543279030322532 \\
16 &  0.543092408233721 & 0.543158077439541 & 0.543272348894322 & 0.543443184803063 \\
20 &  0.543294026061150 & 0.543336010860762 & 0.543409082836059 & 0.543518357900393 \\
24 &  0.543403261934016 & 0.543432401612150 & 0.543483122740523 & 0.543558985879656 \\
28 &  0.543469032017435 & 0.543490433589493 & 0.543527687861490 & 0.543583414067994 \\
32 &  0.543511681223300 & 0.543528063059823 & 0.543556580857472 & 0.543599241290840 \\
36 &  0.543540903977126 & 0.543553845883840 & 0.543576375481245 & 0.543610079844243 \\
40 &  0.543561798440967 & 0.543572279952902 & 0.543590527601463 & 0.543617826543895 \\
Extrap. & 0.543652(2) & 0.0543653(1) & 0.5436508(2) & 0.5436508(3)  \\ \hline
\end{tabular}
\end{table}
\begin{table}
\caption{\label{four1B}
  Finite-size sequences (\ref{estimA2}) of the anomalous dimension
  $X_{1,1}^{\half,\half}(\gamma)$  of the ferromagnetic mixed superspin
  chain for $\gamma=2\pi/5, 2\pi/9$ and $b = 0.1, 0.2,0.3,0.4$
  from the Bethe ansatz. The expected exact conformal
  dimensions are $X_{1,1}^{\half,\half}(2\pi/5)= 41/40$ and
  $X_{1,1}^{\half,\half}(2\pi/9)= 97/72$ independent of $b$.
}
\begin{tabular}{|c||c|c|c|c|}
  \hline
$X_{1,1}^{\half,\half}(2\pi/5)$ & $b=0.1$ & $b=0.2$ & $b=0.3$ & $ b=0.4$ \\ \hline \hline
4 &  0.994997978852449  & 1.00876393795618 & 1.03824887162911 & 1.10011863417448 \\
8 &  1.01750044752832   & 1.02109273301685 & 1.02880785279794 & 1.04523038678615 \\
12 &  1.02166727960952  & 1.02327839396988 & 1.02674367461960 & 1.03414003184710 \\
16 &  1.02312543799154  & 1.02403462682038 & 1.02599129860889 & 1.03017264439620 \\
20 &  1.02380030916302  & 1.02438306807198 & 1.02563757651283 & 1.02831998897873 \\
24 &  1.02416689237986  & 1.02457191882455 & 1.02544395528163 & 1.02730916557915 \\
28 &  1.02438792592022  & 1.02468564407009 & 1.02532670183939 & 1.02669814116381 \\
32 &  1.02453138316540  & 1.02475939722319 & 1.02525039524152 & 1.02630094270188 \\
36 &  1.02462973613628  & 1.02480993532828 & 1.02519798652844 & 1.02602834133088 \\
40 &  1.02470008713720  & 1.02484607136855 & 1.02516045226876 & 1.02583320955474 \\
Extrap. & 1.0250002(2) & 1.0250001(1) & 1.0249997(3) & 1.0249994(2)  \\ \hline
  \hline
$X_{1,1}^{\half,\half}(2\pi/9)$ & $b=0.1$ & $b=0.2$ & $b=0.3$ & $ b=0.4$ \\ \hline \hline
4 &  1.29876758647094   &     1.30325656988360  &   1.31102766447121   &  1.32254599554157\\
8 &  1.33452774930199   &     1.33573236225976  &   1.33782984630232   &   1.34096834546194\\
12&   1.34154352562186  &      1.34208650591172 &    1.34303195767985  &   1.34444680084643\\
16&   1.34402121289879  &      1.34432814492269 &    1.34486256347996  &   1.34566227355483\\
20&   1.34517162315680  &      1.34536850730852 &    1.34571130659309  &   1.34622425907196\\
24&   1.34579746591406  &      1.34593436028266 &    1.34617270652002  &   1.34652935177740\\
28&   1.34617514127825  &      1.34627579182547 &    1.34645103251638  &   1.34671324708849\\
32&   1.34642039180554  &      1.34649748974364 &    1.34663172250430  &   1.34683257472766\\
36&   1.34658859119652  &      1.34664952840261 &    1.34675562343607  &   1.34691437258345\\
40&   1.34670893138310  &      1.34675830227098 &    1.34684425932718  &   1.34697287559906\\
Extrap. & 1.3472222(1) &1.3472222(2) & 1.3472223(2) & 1.3742224(3)  \\ \hline
\end{tabular}
\end{table}
\begin{table}
\caption{\label{five1B}Finite-size sequences (\ref{estimA2}) of the anomalous
  dimension $X_{1,0}^{1,-1}(\gamma)$  of the ferromagnetic mixed superspin
  chain for $\gamma=2\pi/5, 2\pi/9$ and $b = 0.1, 0.2,
  0.3, 0.4 $ from the Bethe ansatz. The expected exact conformal
  dimensions are
  $X_{1,0}^{1,-1}(2\pi/5)=23/12$ and $X_{1,0}^{1,-1}(2\pi/9)=43/28$
  independent of $b$.
}
\begin{tabular}{|c||c|c|c|c|}
  \hline
$X_{1,0}^{1,-1}(2\pi/5)$ & $b=0.1$ & $b=0.2$ & $b=0.3$ & $ b=0.4$ \\ \hline \hline
4 &  1.92275836268774 &  1.93708511027515 & 1.96438958339881 &  2.00429835568618 \\
8 &  1.91908565206600  & 1.92092744160645 & 1.92450401713200 &  1.93029738987036 \\
12 &  1.91781098647046  & 1.91846609255480 & 1.91979319308352 & 1.92224186047280 \\
16 &  1.91732324822074  & 1.91765863919210 & 1.91835375244184 & 1.91971517305235 \\
20 &  1.91709049473449  & 1.91729520769754 & 1.91772483742181 & 1.91859227129214 \\
24 &  1.91696227916640  & 1.91710065979594 & 1.91739323998151 & 1.91799427030424  \\
28 &  1.91688438802683  & 1.91698436898660 & 1.91719675456519 & 1.91763773565356 \\
32 &  1.91683360960376  & 1.91690931467571 & 1.91707063708865 & 1.91740796311456 \\
36 &  1.91679869850490  & 1.91685805584467 & 1.91698481982676 & 1.91725118033239 \\
40 &  1.91677368043983  & 1.91682149275017 & 1.91692376310090 & 1.91713941274508 \\
Extrap.  & 1.9166665(3) &  1.9166664(2) & 1.9166667(4) & 1.916667(2)   \\ \hline
  \hline
$X_{1,0}^{1,-1}(2\pi/9)$ & $b=0.1$ & $b=0.2$ & $b=0.3$ & $ b=0.4$ \\ \hline \hline
4 &  1.89267410304585  & 1.87067227593596 &  1.82808451415017 & 1.75087323907445 \\
8 &  1.64192241218324  & 1.63438221960052 & 1.62092224163197  & 1.59994006363493 \\
12 &  1.58476516588047  & 1.58126414904826 & 1.57510198900755 & 1.56571877644079 \\
16 &  1.56369613535041  & 1.56169796597256 & 1.55819786351628 & 1.55290966874609 \\
20 &  1.55374177481035  & 1.55245446377004 & 1.55020450327766 & 1.54681712875058 \\
24 &  1.54827895082732  & 1.54738179557628 & 1.54581561424246 & 1.54346217373252 \\
28 &  1.54496582300292  & 1.54430527807228 & 1.54315297687471 & 1.54142343412900 \\
32 &  1.54280766022497  & 1.54230123042885 & 1.54141818784558 & 1.54009376604051 \\
36 &  1.54132445289797  & 1.54092393191263 & 1.54022577993160 & 1.53917919299817 \\
40 &  1.54026173042878  & 1.53993708895457 & 1.53937133167230 & 1.53852352062602 \\
Extrap.  & 1.53571(1) & 1.53573(2) & 1.53571(2) & 1.53572(2)  \\ \hline
\end{tabular}
\end{table}
\begin{table}
\caption{\label{six1B}Finite-size sequences (\ref{estimA2}) of the anomalous
  dimension  $X_{0,0}^{\half,\half}(\gamma)$  of the ferromagnetic mixed superspin
  chain for $\gamma=2\pi/5, 2\pi/9$ and $b = 0.1,0.2,0.3,0.4$
  from the Bethe ansatz. The expected exact
  conformal   dimensions are  $X_{0,0}^{\half,\half}(2\pi/5)=5/8$ and
  $X_{0,0}^{\half,\half}(2\pi/9)=9/8$ independent of $b$.
}
\begin{tabular}{|c||c|c|c|c|}
  \hline
$X_{0,0}^{\half,\half}(2\pi/5)$ & $0.1$ & $0.2$ & $0.3$ & $ 0.4$ \\ \hline \hline
4 &  0.630213720410703   & 0.628167445380927 &  0.623556940842337 & 0.613210392494294 \\
8 &  0.626298472836777   & 0.625855321761323 &  0.624894142565653 & 0.622799370606034 \\
12 &  0.625576535867603  & 0.625384174503429 &  0.624968726489793 & 0.624073859686108 \\
16 &  0.625324185193522  & 0.625216855887439 &  0.624985345732743 & 0.624488157113924 \\
20 &  0.625207443504917  & 0.625139008388433 &  0.624991474730969 & 0.624675026917663 \\
24 &  0.625144044650655  & 0.625096616112297 &  0.624994398748262 & 0.624775292259350 \\
28 &  0.625105822737553  & 0.625071019767160 &  0.624996025401022 & 0.624835333745489 \\
32 &  0.625081017557765  & 0.625054392599770 &  0.624997026550134 & 0.624874137620650 \\
36 &  0.625064012291477  & 0.625042986503641 &  0.624997688262107 & 0.624900666695744 \\
40 &  0.625051849144170  & 0.625034824633739 &  0.624998149205000 & 0.624919605647811 \\
Extrap.& 0.62499996(1) &  0.6250002(2) & 0.624997(1) & 0.62499995(1)   \\ \hline 
  \hline
$X_{0,0}^{\half,\half}(2\pi/9)$ & $0.1$ & $0.2$ & $0.3$ & $ 0.4$ \\ \hline \hline
4 &  1.14652880714786  &      1.14785606161029  &   1.15022589201266   &  1.15391189127984\\
8 &  1.12985951474016  &      1.13011226978579  &   1.13055659229965   &  1.13123131520870\\
12&   1.12712669630877 &       1.12723438773761 &    1.12742271780656  &  1.12770646795383\\
16&   1.12619019285676 &       1.12624989511655 &    1.12635410217068  &  1.12651064369857\\
20&   1.12575996440568 &       1.12579791958604 &    1.12586410849908  &  1.12596339789932\\
24&   1.12552709642448 &       1.12555335909051 &    1.12559913518060  &  1.12566775024573\\
28&   1.12538696538072 &       1.12540621828467 &    1.12543976663790  &  1.12549002936557\\
32&   1.12529612700170 &       1.12531084678639 &    1.12533649097860  &  1.12537489994630\\
36&   1.12523389945651 &       1.12524551879330 &    1.12526575843882  &  1.12529606621631\\
40&   1.12518941396948 &       1.12519881883003 &    1.12521519996600  &  1.12523972647179\\
Extrap.& 1.1249994(3) &  1.249996(3) & 1.1249995(2) & 1.249997(4)   \\ \hline 
\end{tabular}
\end{table}

\begin{table}
  \caption{\label{tab:logestC}
    Estimate of the amplitude $A(\gamma,b)$ in (\ref{fslog}) for various
    values of the deformation parameter $\gamma$ on the self-dual line $\gamma
    b=\pi/4$.  Also given is our conjectured value from Eq.~(\ref{ampC}).
  }
\begin{tabular}{|c|lllllll|}
\hline
 $L$ \textbackslash $\gamma/\pi$ 
      & $0.45$  & $0.40$  & $0.35$ & $0.30$  & $0.25$  & $0.20$ & $0.15 $ \\
\hline\hline
  16 & 0.172989  & 0.227420 & 0.712062 & 1.34413 & 2.12657 & 3.24556 & 5.21619\\
  32 & 0.0633073 & 0.349476 & 0.901673 & 1.53538 & 2.35209 & 3.54601 & 5.59051\\
  64 & 0.0713748 & 0.473826 & 0.995320 & 1.60634 & 2.42813 & 3.64798 & 5.70463\\
 128 & 0.116176  & 0.554262 & 1.03345 & 1.62997 & 2.45182 & 3.68013 & 5.73713 \\
 256 & 0.181819  & 0.592325 & 1.04769 & 1.63789 & 2.45947 & 3.69057 & 5.74664 \\
 512 & 0.233158  & 0.607484 & 1.05292 & 1.64080 & 2.46234 & 3.69445 & 5.74995 \\
1024 & 0.258706  & 0.613073 & 1.05493 & 1.64209 & 2.46371 & 3.69626 & 5.75155 \\
2048 & 0.268682  & 0.615098 & 1.05580 & 1.64278 & 2.46452 & 3.69729 & 5.75262 \\
\hline
conject. & 5/18    & 5/8      & 15/14   & 5/3     & 5/2     & 15/4    & 35/6\\
\hline
\end{tabular}
\end{table}

\end{document}